\documentclass[12pt]{article}

\usepackage{subfigure}
\usepackage{booktabs}
\usepackage{multirow}
\usepackage{longtable}
\usepackage{setspace,relsize} 
\usepackage{amsmath,amssymb}
\usepackage{graphicx}
\usepackage{caption}
\usepackage{color}
\usepackage{dcolumn}
\usepackage{bm}
\usepackage{float}
\usepackage{hyperref} 
\usepackage{apacite}
\usepackage{newtxmath}
\usepackage{natbib}
\usepackage{hyperref}
\usepackage{graphicx}
\usepackage{soul}
\usepackage{mwe}
\usepackage[T1]{fontenc}
\usepackage{upquote}

\usepackage{xcolor}

\hypersetup{ colorlinks = true, citecolor = black, urlcolor = blue}
\usepackage{blindtext}
\usepackage{colortbl}
\usepackage[figuresright]{rotating}
\definecolor{mygray2}{gray}{.7}
\definecolor{background-color}{gray}{0.98}
\title{The Variational Bayesian Inference for Network Autoregression Models}

\author{Wei-Ting Lai\\
\small{Department of Statistics, National Cheng Kung University, Tainan, Taiwan}\\
Ray-Bing Chen\\
\small{Department of Statistics, National Cheng Kung University, Tainan, Taiwan}\\
\small{
Institute of Data Science, National Cheng Kung University, Tainan, Taiwan}\\
Ying Chen\\
\small{Department of Mathematics, National University of Singapore, Singapore}\\ 
\small{Risk Management Institute, National University of Singapore, Singapore}\\
\small{Institute of Data Science, National University of Singapore, Singapore}\\
Thorsten Koch\\
 \small{Chair of Software and Algorithms for Discrete Optimization,}\\
 \small{ Technische Universität Berlin, Berlin, Germany}\\
 \small{ Department of Applied Algorithmic Intelligence Methods,}\\ 
 \small{Zuse Institute Berlin, Berlin, Germany}\\
}

\date{}
\begin{document}
\maketitle

\noindent \textbf{Abstract: }We develop a variational Bayesian (VB) approach for estimating large-scale dynamic network models in the network autoregression framework. The VB approach allows for the automatic identification of the dynamic structure of such a model and obtains a direct approximation of the posterior density. Compared to Markov Chain Monte Carlo (MCMC) based sampling approaches, the VB approach achieves enhanced computational efficiency without sacrificing estimation accuracy. In the simulation study conducted here, the proposed VB approach detects various types of proper active structures for dynamic network models. Compared to the alternative approach, the proposed method achieves similar or better accuracy, and its computational time is halved. In a real data analysis scenario of day-ahead natural gas flow prediction in the German gas transmission network with 51 nodes between October 2013 and September 2015, the VB approach delivers promising forecasting accuracy along with clearly detected structures in terms of dynamic dependence. \\


\renewcommand{\baselinestretch}{1.5}
\normalsize
\newpage

\section{\sffamily \Large Introduction}

Networks have emerged and become available in various fields, such as energy transmission, logistics and transportation, and financial systems. Networks are dynamic in terms of their temporal dependence, and they often have large scales. The understanding and inference of network dynamics have profound implications for operations and decision making in modern industries. Tremendous growth and heterogeneity in both nodes/edges and dependence over time are the key characteristics of such networks. However, conventional statistical methods either assume that networks are static or consider only low-dimensional temporal data. This creates a need for efficient computational approaches that are able to reveal the essential dependence structures in high dimensions and simultaneously deliver accurate inferences with a low computational cost.  

Industrial networks contain series of temporal-spatial data collected over time. While the nodes/edges are often fixed or possess trivial changes, the lead-lag temporal dependence issue can no longer be ignored in network inference. Graph theory has been widely used for unraveling structural information in large-scale network analysis. For example, \cite{fan2009network} and \cite{guo2011joint} proposed a sparse graphic network. \cite{liu2012high} proposed the semiparametric Gaussian copula graphical model.
Despite their efficiency, these graphical models assume static networks, and the evolution of the network dependence is not considered in their estimation processes. 

The temporal dependence of a network can be represented in the vector autoregressive (VAR) modeling framework. In the VAR framework, each node is considered as one time series, and the network dependence is measured as the lead-lag cross-correlations among multiple time series. Both the theoretical properties and empirical performance of VAR have been well studied with respect to multivariate data; see \cite{lutkepohl2007} and \cite{banbura2010}. However, the application of VAR for large-scale network analysis is still challenging. Given a network with $m$ models, which correspond to $m$ time series, and supposing that the dynamics depend on the last $p$ lags, to the result is that there are $pm^2$ unknown coefficients in the VAR model. When the number of nodes $m$ becomes large or the temporal dependence $p$ increases, VAR is overparameterized and this leads to low estimation accuracy or even infeasibility with regard to model inference.  

A unique feature in industrial networks is that their lead-lag temporal dependences, such as concurrent dependence among their nodes, are less dense than the networks themselves. Individual networks are also much sparser than other social networks. It is conceivable that large-scale industrial networks, are driven by a few essential and cohesive connections among their nodes to facilitate network evolution. This motivates the modeling of large-scale dynamic networks using sparse VAR. In particular, penalties are imposed on the parameter space of the VAR framework with various possible types of structural assumptions. For the purpose of both estimation and interpretability, structural sparsity can be enabled in elements, groups and lags. Lag sparsity investigates the effect of time-lagged information, while group sparsity highlights the impacts of certain nodes on others. In addition to the universal effect on a group of lag coefficients, a sparse element illustrates a single effect. \cite{basu2015} investigated the theoretical properties of $\ell_1$-regularized estimates, where multiple time series were assumed to be stable Gaussian processes. \cite{melnyk2016} established bounds on the non-asymptotic estimation error of the Lasso-type estimator for structured VAR parameters. \cite{nicholson2014} proposed several structures for VAR and Lasso, group Lasso and sparse group penatly functions to achieve sparsity in the elements and groups of a network; see also \cite{hsu2008subset}, \cite{song2011large}, and Chen et al. \citeyearpar{chen2020modeling}. Moreover, VAR models can be easily extended to the above three kinds of sparsity by building up a hierarchical lag structure in the autoregression model via the inclusion of high-dimensional exogenous variables.

The estimation of the structured of a VAR framework faces two challenges. First, the sparse structure needs to be specified to avoid the overfitting of the high-dimensional models. Second, the framework should be able to adapt to different kinds of stochastic behaviors, as empirical data are likely non-Gaussian. Bayesian methods are natural choices because they deliver stable performances without prespecified assumptions about the structure and distribution of the model. Stochastic search variable selection (SSVS), for example, is the most commonly used Bayesian variable selection approach. It introduces latent indicators embedded in the priors and stochastically searches subsets by generating posterior samples with the Markov Chain Monte Carlo (MCMC) algorithm; see \cite{george1993variable}. \cite{Geweke1996} proposed the component-wise Gibbs method for the purpose of improving computational efficiency. By using the spike-and-slab prior, it avoids computing inverse matrices and thus reduces the computational cost. However, the Gibbs sampler requests either random or systematic updating of the coefficients. Chen et al. \citeyearpar{chen2011stochastic} proposed the stochastic matching pursuit (SMP) algorithm, which updates the coefficients of each step for obtaining the best fit based on the current residual vector. In terms of structural selection, \cite{2010} introduced Bayesian-constrained variable selection. Chen et al. \citeyearpar{chen2016bayesian} proposed the groupwise Gibbs sampler. As a proof of concept, \cite{chu2019bayesian} implemented a Bayesian variable selection approach in the VAR framework and named it the VAGSA, the vector autoregression-based Gibbs sampler algorithm. For the high-dimensional VAR model, \cite{kastner2020sparse} proposed a large Bayesian vector autoregression approach with a Dirichlet-Laplace prior and factor stochastic volatility (FSV), and they applied it on high-dimensional US economic data.

Nevertheless, these MCMC algorithms are known to be computationally expensive for sequentially generating posterior samples. Variational inference, as an alternative, shows great potential in terms of improving computational speed without sacrificing much accuracy. It obtains an approximation of the target posterior density using the Kullbak-Leibler divergence, based on which an EM-type algorithm is devised a reduced computational cost. \cite{titsias2011spike} and \cite{carbonetto2012scalable} introduced variational Bayesian approaches with spike-and-slab priors for dealing with variable selection problems in linear regression models. \cite{cai2020bivas} proposed a variational Bayesian method for sparse group selection in linear models and extended it to multiple response models. 

In our study, we propose a variational Bayesian (VB) approach for estimating a large-scale dynamic network model. The serial dependence in a given network is represented in a vector autoregression framework with three possible types of structural assumptions and various nesting types. Here, we also call this model a network autoregression (NAR) model. We derive variational inferences and develop the corresponding algorithms. The VB approach allows for the automatic identification of the dynamic structure of data and obtains an approximation of the posterior density directly. Compared to MCMC-based sampling approaches, such as the VAGSA in \cite{chu2019bayesian}, the VB approach achieves enhanced numerical performance with similar accuracy. A simulation study shows that compared with existing methods, the VB approach not only detects the proper active structures in various dynamic network models but also halves the computational time, with similar or better accuracies. In a real data analysis, we predict day-ahead natural gas flows for a German network with 51 nodes over 2 years from Oct 1, 2013, to Sep 30, 2015. Germany's gas transport system is essential to the European energy supply. The adequate, high-precision estimation of supply and demand is a crucial issue for efficient control and operations in gas transmission. The VB approach delivers a clear dynamic dependence structure, providing interpretability and insights for understanding and managing the gas transmission network. To the best of our knowledge, this is the first attempt to derive variational inference for a large-scale dynamic network analysis in a structured NAR/VAR framework.

This paper is organized as follows. Section \hyperlink{sec2}{2} introduces the dynamic network model in the VAR framework. Several types of structural assumptions are also demonstrated. Section \hyperlink{sec3}{3} presents the proposed variational Bayesian algorithms for large-scale dynamic network inference. Section \hyperlink{sec4}{4} investigates the finite-sample performance of the proposed VB approach. Section \hyperlink{sec5}{5} reports the network inference for day-ahead gas flow forecasting with 51 high-pressure nodes in the natural gas transmission network in Germany. Section \hyperlink{sec6}{6} provides a brief conclusion of our work. \

 \hypertarget{sec2}{}\section{\sffamily \Large Model}

Let $\boldsymbol{Y}_t\in \mathbb {R}^{1\times m}$ denote a vectorized time series of networks with $m$ nodes at time $t$ over the time period $[1,T]$. Without loss of generality, $\boldsymbol{Y}_t$ is demeaned. We consider a dynamic network model in a vector autoregression framework $\boldsymbol{Y}_t$ which is assumed to depend on the past values of the network at lags $1$ to $p$, i.e.,
\begin{eqnarray*} \label{model}
\boldsymbol{Y}_t=\boldsymbol{Y}_{t-1}B_1+\cdots+\boldsymbol{Y}_{t-p}B_p+\epsilon_t,~~~~t=p+1,p+2,\dots,T.
\end{eqnarray*}
Here, we let $B_\ell$ be an $m \times m$ coefficient matrix for lag-$\ell$, $\ell=1,2,\cdots,p$, which is used to measure the lead-lag temporal dependence in the network. The number of coefficients in $\boldsymbol{B}_\ell$ grows quadratically with the number of nodes $m$. Given $m=51$ in the gas transformation network, there are $51^2 = 2601$ unknown coefficients for each $B_\ell$. Obviously, the diagonal elements of $B_\ell$ indicate the serial dependence of each node on its own lag-$\ell$ value and the dependences of off-diagonal elements on other {nodes\textquotesingle} lag-$\ell$ values. The term $\left\{\epsilon_t\right\}_{t=p+1}^{T}$ is a sequence of serially uncorrelated $1 \times m$ random vectors, with a mean vector of zero and a covariance matrix $\Sigma$. Then, the dynamic network model can be represented in matrix form as follows: 
\begin{eqnarray}
\boldsymbol{Y}&=\boldsymbol{X}\boldsymbol{B}+\boldsymbol{\epsilon},
\end{eqnarray}
where $\boldsymbol{Y}\in \mathbb {R}^{(T-p) \times m}$ is the response matrix, $\boldsymbol{X}=(\boldsymbol{X}_1,...,\boldsymbol{X}_\ell)$ is a $(T-p) \times m$ matrix with $\boldsymbol{X}_\ell = \left(\boldsymbol{Y}'_{p+1-\ell}\right.$ $\left.,\boldsymbol{Y'}_{p+2-\ell},\dots,\boldsymbol{Y}'_{T-\ell}\right)'$, $\ell=1,2,\cdots,p$, $\boldsymbol{B} =\left(B'_1,B'_2,\dots,B'_p\right)'$, and $
\boldsymbol{\epsilon} = \left(\epsilon'_{p+1},\epsilon'_2,\cdots,\epsilon'_T\right)'$.

\begin{figure}[t!]
   \centering
    \includegraphics[width=17cm]{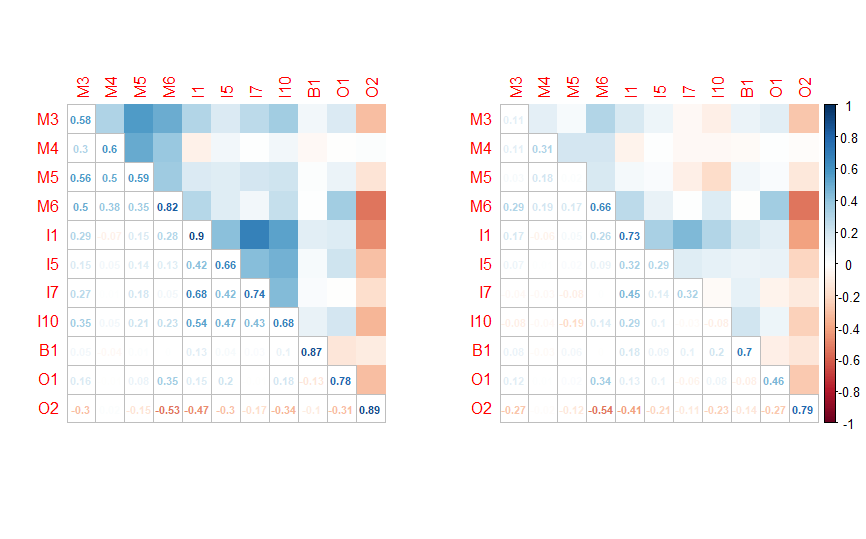}
    \caption{The cross-correlations of 11 nodes arbitrarily selected from the German natural gas transmission network.}
    \label{fig:crosscorrelation}
\end{figure}

In this paper, we consider a structured NAR/VAR framework, i.e., dynamic dependence is sparsely motivated by a large-scale industrial network, such as the German gas transformation network. Figure \ref{fig:crosscorrelation} displays the lag-1 and lag-3 cross-correlations of 11 nodes arbitrarily selected from the German natural gas transmission network. The 11 nodes belong to 4 different types: municipal (labeled with M), industrial (I), border (B), and others (O). The left-hand side of Figure \ref{fig:crosscorrelation} is the lag-1 cross-correlation matrix. The right-hand side of Figure \ref{fig:crosscorrelation} is the cross-correlation matrix for lag-3. This shows the coexistence of strong serial dependence and sparsity in elements, groups, and lags. According to Figure \ref{fig:crosscorrelation}, due to their cross-correlation values, nodes sharing the same type may possess similar patterns. Thus, the dynamics of the network are not driven by each node individually, every group of nodes, or each lagged network in the past.

While element sparsity and lag sparsity are clear, there are different kinds of group sparsity. Following \cite{song2011large}, we categorize the various structures into three types and discuss them as follows.

\begin{figure}[t!]
 \centering
        \begin{minipage}{0.8\textwidth}
            \begin{subfigure}
            \centering
           \includegraphics[width=14cm]{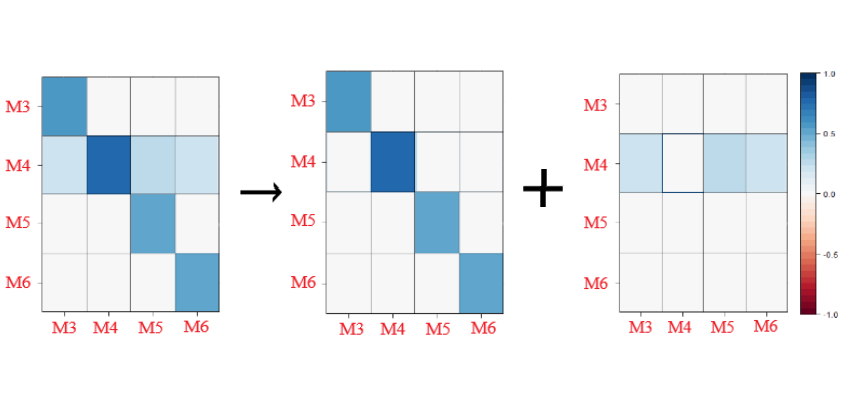}
            \caption*{}
            \end{subfigure}
        \end{minipage}
        \par
{\Large
  \begin{minipage}{.9\textwidth}
 \begin{eqnarray*}
 \begin{array}{c}
 \left(\begin{array}{cccc}
 \bullet&0&0&0\\
  \bullet&\bullet&\bullet&\bullet\\
 0&0&\bullet&0\\
 0&0&0&\bullet\\
 \end{array}
 \right)\\
  (\mbox{Total})
 \end{array}
 \rightarrow
 \begin{array}{c}
 \left(\begin{array}{cccc}
 \bullet& & & \\
 &\bullet&&\\
 &&\bullet&\\
 &&&\bullet\\
 \end{array}
 \right)\\
 (\mbox{Own})
 \end{array}
 +
 \begin{array}{c}
 \left(\begin{array}{cccc}
  &0 &0&0\\
 \bullet&&\bullet&\bullet\\
 0&0& &0\\
 0&0&0& \\

 \end{array}
 \right)\\
  (\mbox{Others})
 \end{array}
 \end{eqnarray*}
 \end{minipage}}
\caption{An illustration of the coefficient matrix for the universal grouping structure.}
 \label{fig: universal grouping}
 \end{figure}

\begin{itemize}
 \item UG Structure: The universal grouping (UG) structure in the coefficient matrix $B_\ell$ means that the off-diagonal coefficients have the same sparsity pattern across the different columns. For example,  municipal nodes M3, M4, M5, and M6 may have similar patterns, and we can treat them as a group. Therefore, if one node affects another node, the others are all influenced by this node. As shown in Figure \ref{fig: universal grouping}, node M4 affects other nodes, and nodes M3, M5 and M6 do not affect other nodes. As such, the coefficient matrix can be separated into a diagonal matrix reflecting the dynamic dependence of each node on its own, and a sparse matrix showing the temporal dependence of each node on others. In network analysis, the columns that have the same row sparsity are grouped together. 
\end{itemize}

\begin{itemize}
 \item SG Structure: For the segmentation grouping (SG) structure in $B_\ell$, the nodes in the same segment interact with each other but are independent from the other nodes. This is associated with the empirical observation that there are different types of nodes in the network. Figure \ref{fig: segmentized grouping} illustrates the SG structure among municipal and industrial nodes. The network is divided into disjoint segments of the municipal and industrial types. The estimation process can be conducted segment by segment.
 
 \begin{figure}[h]
 \centering
        \begin{minipage}{0.8\textwidth}
            \begin{subfigure}
            \centering
            \includegraphics[width=14cm]{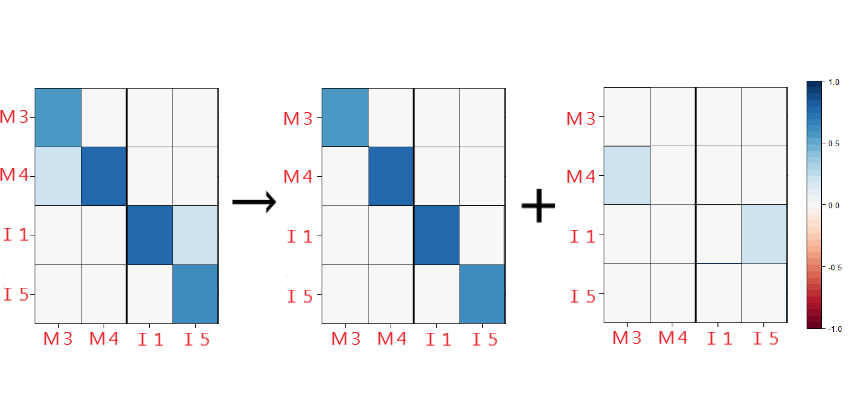}
            \caption*{}
            \end{subfigure}
        \end{minipage}
        \par
{\Large
  \begin{minipage}{.9\textwidth}
 \begin{eqnarray*}
 \begin{array}{c}
 \left(\begin{array}{cc|cc}
 \bullet&0&0&0\\
 \bullet&\bullet&0&0\\
 0&0&\bullet&\bullet\\
0&0&0&\bullet\\
 \end{array}
 \right)\\
  (\mbox{Total})
 \end{array}
 \rightarrow
 \begin{array}{c}
  \left(\begin{array}{cccc}
 \bullet&&&\\
 &\bullet&&\\
 &&\bullet&\\
 &&&\bullet\\
 \end{array}
 \right)\\
  (\mbox{Own})
 \end{array}
 +
 \begin{array}{c}
  \left(\begin{array}{cc|cc}
 &0&0&0\\
 \bullet&&0\\
 0&0&&\bullet\\
 0&0&0&\\
 \end{array}
 \right)\\
  (\mbox{Others})
 \end{array}
 \end{eqnarray*}
  \caption{An illustration of the coefficient matrix for the segmentation grouping structure.} \label{fig: segmentized grouping}
 \end{minipage}}
 \end{figure}
 \end{itemize}

\begin{itemize}
\item NG Structure: Consider the ``no grouping'' (NG) structure in $B_\ell$. Each node has its own impact on the other nodes over time. This is a scenario where no regular pattern is identified among the nodes. Figure \ref{fig: no grouping} shows an example with different types of nodes. In fact, there are no similar patterns among them. In this case, we separate them during the inference procedure.

 \begin{figure}[h!]
 \centering
        \begin{minipage}{0.8\textwidth}
            \begin{subfigure}
            \centering
            \includegraphics[width=14cm]{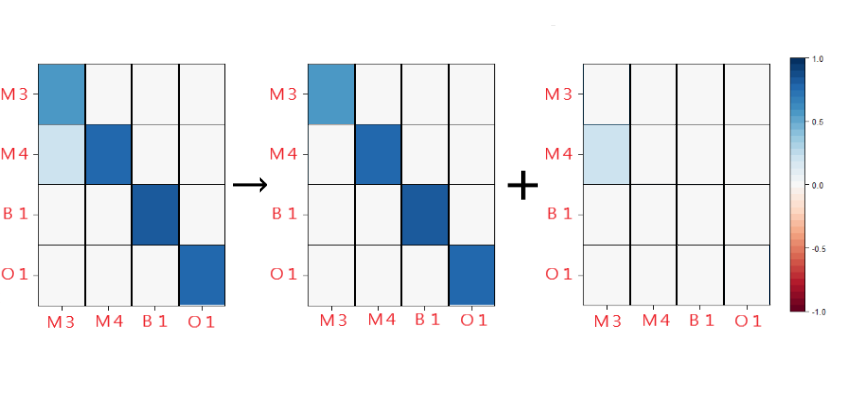}
            \caption*{}
            \end{subfigure}
        \end{minipage}
        \par
{\Large
  \begin{minipage}{1\textwidth}
 \begin{eqnarray*}
 \begin{array}{c}
 \left(
 \begin{array}{c|c|c|c}
\bullet &0&0&0\\
 \bullet&\bullet&0&0\\
 0&0&\bullet&0\\
 0&0&0&\bullet\\
 \end{array}
 \right)\\
  (\mbox{Total})
 \end{array}
 \rightarrow
  \begin{array}{c}
  \left(
  \begin{array}{cccc}
 \bullet&&&\\
 &\bullet&&\\
 &&\bullet&\\
 &&&\bullet\\
 \end{array}
 \right)\\
  (\mbox{Own})
 \end{array}
  +
 \begin{array}{c}
  \left(
  \begin{array}{c|c|c|c}
 &0&0&0\\
 \bullet&&0&0\\
 0&0&&0\\
 0&0&0&\\
 \end{array}
 \right)\\
  (\mbox{Others})
 \end{array}
 \end{eqnarray*}
  \end{minipage}}
 \caption{An illustration of the coefficient matrix for the ``no grouping'' structure.}
 \label{fig: no grouping}
 \end{figure}  
\end{itemize}

It is easy to see that the first and third types of structures, UG and NG, are two special cases of the SG structure when the numbers of segments are $1$ and $m$, respectively. While the three types of structures are defined for each coefficient matrix, they may appear in different time lags. The identification of a proper structure can help not only increase the estimation accuracy but also enhance computational time of the algorithm.

\section{\sffamily \Large Variational Approximation Algorithm}

We introduce the variational Bayesian approach for NAR/VAR structure inference in high-dimensional scenarios, and it is expected to automatically choose the proper structure and perform estimation at a low computational cost. Here, we derive only the VB method for the segmentation grouping structure because it nests the other two types of structures as special cases of segmentation grouping. Thus, it is a derivation under a general setup. To simplify the procedure, we assume that all coefficient matrices share the same segmentation grouping structure. Denote $S=\left\{ s_1,s_2,\cdots,s_g\right\}$ as an overall index set for the columns in each $B_\ell$, where $s_k$ is the index set of the $k$th segment with size $0 < |s_k|< m$ and $\sum_{k=1}^g|s_k|=m$, and $g$ is the total number of segments (groups). In this section, we start by introducing the Bayesian structure selection approach, and then we mention the details of the proposed VB method.

\subsection{The Bayesian structure selection algorithm}

For the Bayesian hierarchical model used to select segmentation structures, \cite{chu2019bayesian} introduced two indicators $\gamma_{\ell,i}$ and $\eta_{\ell, i, k}$ to identify the active structures. These indicators denote the active nonzero coefficients of the $k$th segment in the $i$th row in $B_\ell$ with respect to the lag values of itself and others. For its own lags, $\gamma_{\ell,i}=1$ denotes that the element of the $i$th row and the $i$th column in the coefficient matrix for lag-$\ell$, $B_{\ell,i,i}$, is nonzero, and $\gamma_{\ell,i}=0$ indicates a coefficient of zero, i.e., $B_{\ell,i,i}=0$. In addition, $\eta_{\ell,i,k}=1$ indicates that the $i$th row in $B_{\ell}$ and the columns of the $k$th segment for the off-diagonal elements in $B_\ell$ are all nonzero, i.e. $B_{l,i,\widetilde{s}_k} \neq \boldsymbol{0}$, and $\eta_{l,i,k} = 0$ otherwise. Consider the Bayesian structure selection approach for determining the segmentation structure in the NAR/VAR model. Following the literature, the prior distributions of the indicators $\gamma_{\ell,i}$ and $\eta_{\ell,i,k}$ are chosen to be independent Bernoulli distributions with $P\left(\gamma_{\ell,i}= 1\right) =\pi_{1}$ and $P\left(\eta_{\ell,i,k}= 1\right) =\pi_{2}$, i.e., $Ber(\pi_{1})$ and $Ber(\pi_{2})$, respectively. The priors of the elements in matrix $B_\ell$ are dependent on $\boldsymbol{\gamma_\ell}$ and $\boldsymbol{\eta_\ell}$, which are the vectors of the indicators in lag-$\ell$, as follows:
 \begin{align}
B_{\ell,i,i}|\gamma_{\ell,i},\sigma_{B}^2 &\sim \gamma_{\ell,i}N(0,\sigma_{B}^2)+(1-\gamma_{\ell,i})\delta_0,\label{coef1}\\
B_{\ell,i,\widetilde{s}_k}|\eta_{\ell,i,k},\sigma_{B}^2 &\sim \eta_{\ell,i,k}MN_{1 \times |\widetilde{s}_k|}(\boldsymbol{0}, I ,\sigma_{B}^2{I}_{|\widetilde{s}_k|})+(1-\eta_{\ell,i,k})\delta_{\boldsymbol{0}} ,\label{coef2}
\end{align}
where $\widetilde{s}_k$ is the index set of the $k$th segment except $i$, i.e., $ i \in s_k, \widetilde{s}_k=s_k\setminus \left\{i\right\}$, and $|\widetilde{s}_k|$ denotes the number of elements contained in $\widetilde{s}_k$. In addition, $MN_{1 \times |\widetilde{s}_k|}(\boldsymbol{0},\boldsymbol{I},\sigma_{B}^2I_{|\widetilde{s}_k|})$ is a $1 \times |\widetilde{s}_k|$ multivariate normal distribution with a mean vector of $\boldsymbol{0}$ and a covariance matrix, $\sigma_{B}^2{I}_{|\widetilde{s}_k|}$, and $\delta_0$ and $\delta_{\mathbf{0}}$ are a point mass at $0$ and a zero vector $\mathbf{0}$, respectively. Therefore, the coefficient prior is a mixture prior of the normal distribution and a point mass. Last, $\left(B_{\ell,i,i},\gamma_{\ell,i}\right)$ and $\left(B_{\ell,i,\widetilde{s}_k},\eta_{\ell,i,k}\right)$ for $k=1,2,\cdots,g$, $i=1,2,\dots,m$ and $l=1,2,\cdots,p$ are assumed to be independent. \

Based on these prior assumptions regarding the coefficients and the additional inverse-Wishart prior for the covariance matrix of the NAR/VAR model, coefficient inference can be performed by the vector autoregression Gibbs sampler \citep[VAGSA,][]{chu2019bayesian}. Basically, in the VAGSA, we need to iteratively generate the posterior samples of the two indicators and coefficients for the further inference. Similar to other MCMC algorithms, the VAGSA is computationally expensive, especially when the number of nodes increases. Instead of generating the posterior samples as an approximation of the posterior distribution, the variational Bayesian approach is adopted in our study to directly obtain the approximation of the posterior density function. Here, an EM-type algorithm is used to solve the corresponding optimization problem, and this is expected to significantly improve the computational efficiency of our approach.

\subsection{The Variational inference procedure}

 Before introducing the variational Bayesian method, we first reparametrize the NAR/VAR model. Recall that in the VAGSA, the coefficient prior is a mixture distribution with a normal distribution and a delta function at point zero. When an indicator is equal to zero, we can simply set the corresponding coefficient to zero. To simplify the model structure, we reparametrize the coefficient as the product of the coefficient and the indicator. That is, $B_{\ell,i,i} = \gamma_{\ell,i} \widetilde{B}_{\ell,i,i}$ and $B_{\ell,i,\widetilde{s}_k}=\eta_{\ell,i,k}\widetilde{B}_{\ell,i,\widetilde{s}_k}$. Here, the priors of the indicators, $\gamma_{l,i}$ and $\eta_{l,i, k}$, are still independent Bernoulli distributions with probabilities $\pi_1$ and $\pi_2$, respectively. In addition, the prior of $\widetilde{B}_{\ell,i,i}$ is chosen as a normal distribution with mean zero and variance $\sigma^2_{\beta}$, and the prior of $\widetilde{B}_{\ell,i,\widetilde{s}_k}$ comes from the multivariate normal distribution with a mean zero vector and a covariance matrix $\sigma^2_{\beta}\boldsymbol{I}_{|\widetilde{s}_k|}$. Among them, the priors of $\widetilde{B}_{\ell,i,i}$ and $\widetilde{B}_{\ell,i,\widetilde{s}_k}$ are independent of $\gamma_{\ell,i}$ and $\eta_{\ell,i,k}$. Thus, $\gamma_{\ell,i}\widetilde{B}_{\ell,i,i}$ and $\eta_{\ell,i,k}\widetilde{B}_{\ell,i,\widetilde{s}_k}$ have the same effects as those shown in Eqs. (\ref{coef1}) and (\ref{coef2}).

Instead of generating the posterior samples directly, the variational Bayesian approach identifies the best approximate distribution of the true posterior for the further Bayesian inference. According to the ordinary variational Bayesian approach \citep{bishop2006pattern}, the Kullback-Leibler divergence (KL divergence) is used to measure the dissimilarity of the true posterior distribution from the approximated posterior distribution. Let $\theta=\left\{ \pi_1,\pi_2,\Sigma,\sigma^2_{B}\right\}$ be the set of parameters and $\left\{\boldsymbol{\eta},\boldsymbol{\gamma},\boldsymbol{\widetilde{B}} \right\}$ be the set of the indicator variables and coefficient matrices, where $\boldsymbol{\eta}$ is the vector of $\eta_{\ell,i,k}$, $\boldsymbol{\gamma}$ is the vector of $\gamma_{\ell,i}$,  and $\boldsymbol{\widetilde{B}} = \left(\widetilde{B}'_1,\widetilde{B}'_2,\dots,\widetilde{B}'_p\right)' $. Given the prior assumptions, the posterior density function of $\boldsymbol{\eta},\boldsymbol{\gamma},\boldsymbol{\widetilde{B}}$ is proportional to the the following joint density function:  
\begin{align}
 P(\boldsymbol{Y},\boldsymbol{\eta},\boldsymbol{\gamma},\boldsymbol{\widetilde{B}}|\boldsymbol{X},\theta) =& 
      P(\boldsymbol{Y}|\boldsymbol{\eta},\boldsymbol{\gamma},\boldsymbol{\widetilde{B}},\boldsymbol{X},\theta)P(\boldsymbol{\eta},\boldsymbol{\gamma},\boldsymbol{\widetilde{B}}|\boldsymbol{X},\theta) \notag\\     
      =& MN_{T\times m}(\boldsymbol{XB},\boldsymbol{I},\Sigma)\prod_\ell^p\prod_i^mN(0,\sigma^2_{B}){\pi_1}^{\gamma_{\ell,i}}\left(1-\pi_1\right)^{(1-\gamma_{\ell,i})} \notag\\
      &\prod_k^gMN_{1 \times |\widetilde{s}_k|}(\boldsymbol{0},\boldsymbol{I}, \sigma^2_{B}{I}_{|\widetilde{s}_k|}){\pi_2}^{\eta_{\ell,i,k}}\left(1-\pi_2\right)^{(1-\eta_{\ell,i,k})},
     \label{Jointpdf}
           \end{align}
 Define $q(\boldsymbol{\eta},\boldsymbol{\gamma},\boldsymbol{\widetilde{B}})$ as an approximate posterior density function of $P(\widetilde{\boldsymbol{B}},\boldsymbol{\eta},\boldsymbol{\gamma}|\boldsymbol{Y},\boldsymbol{X})$. 
Since
\begin{align}
KL(q||P)&=\int\sum_{\boldsymbol{\gamma}}\sum_{\boldsymbol{\eta}} q(\boldsymbol{\widetilde{B}},\boldsymbol{\gamma},\boldsymbol{\eta})\log\left(\frac{q(\boldsymbol{\widetilde{B}},\boldsymbol{\gamma},\boldsymbol{\eta})}{P(\boldsymbol{\widetilde{B}},\boldsymbol{\gamma},\boldsymbol{\eta}|\boldsymbol{Y},\boldsymbol{X};\theta)}\right)d\boldsymbol{\widetilde{B}}\notag\\
&=\underset{-L(q)}{\underbrace{\int\sum_{\boldsymbol{\gamma}}\sum_{\boldsymbol{\eta}} q(\boldsymbol{\widetilde{B}},\boldsymbol{\gamma},\boldsymbol{\eta})\log\left(\frac{q(\boldsymbol{\widetilde{B}},\boldsymbol{\gamma},\boldsymbol{\eta})}{P(\boldsymbol{\widetilde{B}},\boldsymbol{\gamma},\boldsymbol{\eta},\boldsymbol{Y}|\boldsymbol{X};\theta)}\right)d\boldsymbol{\widetilde{B}}}}\notag\\
&+\underset{\log P(\boldsymbol{Y}|\boldsymbol{X};\boldsymbol{\theta})}{\underbrace{\int\sum_{\boldsymbol{\gamma}}\sum_{\boldsymbol{\eta}}q(\boldsymbol{\widetilde{B}},\boldsymbol{\gamma},\boldsymbol{\eta})\log P(\boldsymbol{Y}|\boldsymbol{X};\theta)d\boldsymbol{\widetilde{B}}}},
\label{KLLq}
\end{align}
we have
\begin{align}
    \log P(\boldsymbol{Y}|\boldsymbol{X};\boldsymbol{\theta})=L(q)+KL(q||P).
\end{align} 
According to Eq. (\ref{KLLq}), the marginal likelihood $\log P(\boldsymbol{Y}|\boldsymbol{X}; \theta)$ is independent of $q$ and can be treated as a fixed constant. Thus, $L(q)$ can be defined as the lower bound of the KL divergence between $q$ and $P$. Minimizing the KL divergence with respect to $q$, is equivalent to maximizing the lower bound $L(q)$ of $q$. 
Here, one key is to specify the approximation density $q$ via a factorization structure. Following  \cite{titsias2011spike} and \cite{cai2020bivas}, due to the independence assumption among the disjoint groups and their own lags, the following hierarchically factorized distribution is chosen as an approximate density function: $q(\boldsymbol{\eta},\boldsymbol{\gamma},\widetilde{\boldsymbol{B}})$, i.e.,
\begin{align*}\label{aprq_orginal}
q(\boldsymbol{\eta},\boldsymbol{\gamma},\widetilde{B})&=
\prod_\ell^p\prod_i^m\prod_k^gq_{\ell,i}(\widetilde{B}_{\ell,i,i},\gamma_{\ell,i})q_{\ell,i,k}(\widetilde{B}_{\ell,i,\widetilde{s}_k},\eta_{\ell,i,k}),
\end{align*}
where $q_{\ell,i}(\widetilde{B}_{\ell,i,i},\gamma_{\ell,i})$ and $q_{\ell,i,k}(\widetilde{B}_{\ell,i,\widetilde{s}_k},\eta_{\ell,i,k})$ are chosen from the corresponding prior distributions. Then, a variational extension of the EM algorithm is adopted in the estimation process by maximizing the corresponding $L(q)$ with respect to the parameters. In the E-step, one would take the expectation of $L(q)$ with respect to $\boldsymbol{\eta}, \boldsymbol{\gamma}$ and $\widetilde{B}$, and then in the M-step, one optimizes $L(q)$ with respect to $\theta$. Iterate the two steps until the lower bound $L(q)$ converges.
 
In the E-step, $q(\boldsymbol{\eta},\boldsymbol{\gamma},\widetilde{B})$ is updated as follows:
\begin{eqnarray}\label{aprq}
q(\boldsymbol{\eta},\boldsymbol{\gamma},\widetilde{B})&=&
\prod_\ell^p\prod_i^m\prod_k^gq_{\ell,i}(\widetilde{B}_{\ell,i,i}|\gamma_{\ell,i})q_{\ell,i}(\gamma_{\ell,i})q_{\ell,i,k}(\widetilde{B}_{\ell,i,\widetilde{s}_k}|\eta_{\ell,i,k})q_{\ell,i,k}(\eta_{\ell,i,k}). \notag \\ \nonumber
&=& 
\prod_\ell^p\prod_i^m\prod_k^g\left(\phi_{1,\ell,i}N\left(\mu_{1,\ell,i,i},\Sigma_{B_{\ell,i,i}}\right)\right)^{\gamma_{\ell,i}}\left(\left(1-\phi_{1,\ell,i}\right)N(0,\sigma^2_{B})\right)^{(1-\gamma_{\ell,i})}\\ \nonumber
& & \left(\phi_{2,\ell,i,k}MN_{1\times|\widetilde{s}_k|}(\boldsymbol{\mu}_{2,\ell,i,\widetilde{s}_k},\boldsymbol{I},\Sigma_{B_{\ell,i,\widetilde{s}_k}})\right)^{\eta_{\ell,i,k}}\left(\left(1-\phi_{2,\ell,i,k}\right)MN_{1\times|\widetilde{s}_k|}(\boldsymbol{0},\boldsymbol{I},\sigma_{B}^2{I}_{|\widetilde{s}_k|})\right)^{(1-\eta_{\ell,i,k})},
\end{eqnarray}
where
\begin{eqnarray*}
\Sigma_{B_{\ell,i,i}}&=&\left(\boldsymbol{X}_\ell^{(i)'}\boldsymbol{X}_\ell^{(i)}(\Sigma^{-1})_{i,i}+\sigma_{B}^2\right)^{-1},\\
\mu_{1,\ell,i,i}&=&\Sigma_{B_{\ell,i}}\left((\Sigma^{-1})_{i,i}\boldsymbol{X}_\ell^{(i)'}\left(Y^{(i)}-\sum_{j \neq l}^p\boldsymbol{X}_jE(B_j^{(i)})-\boldsymbol{X}_{\ell}^{(-i)}E(B_{\ell}^{(-i,i)})\right)\right.\\
& & \left.+E\left( tr\left((\Sigma^{-1})_{-i,i}\boldsymbol{X}_\ell^{(i)'}\left(Y^{(-i)}-\boldsymbol{XB}^{(-i)}\right)\right)\right)\right),
\\
\phi_{1,\ell,i}&=& Inv-logit\left\{logit(\pi_1)-\frac{1}{2}\log(\sigma^2_{B})+\frac{1}{2}\log(det(\Sigma_{B_{\ell,i,i}}))+\frac{(\Sigma_{B_{\ell,i,i}})^{-1}\mu_{1,\ell,i,i}^2}{2}\right\},
\end{eqnarray*}
\begin{eqnarray*}
\Sigma_{B_{\ell,i,\widetilde{s}_k}}&=&\left(\boldsymbol{X}_\ell^{(i)'}\boldsymbol{X}_\ell^{(i)}(\Sigma^{-1})_{\widetilde{s}_k,\widetilde{s}_k}+\sigma_{B}^2\boldsymbol{I}_{|\widetilde{s}_k|}\right)^{-1},\\
\boldsymbol{\mu}_{2,\ell,i,\widetilde{s}_k}&=&\left(\boldsymbol{X}_\ell^{(i)'}\left(Y^{(\widetilde{s}_k)}-\sum_{j \neq l}^p\boldsymbol{X}_jE(B_j^{(\widetilde{s}_k)})-\boldsymbol{X}_{\ell}^{(-i)}E(B_{\ell}^{(-i,\widetilde{s}_k)})\right)(\Sigma^{-1})_{\widetilde{s}_k,\widetilde{s}_k}\right.\\
& &\left. +E\left(tr\left(\boldsymbol{X}_\ell^{(i)'}\left(Y^{(-\widetilde{s}_k)}-\boldsymbol{X}\boldsymbol{B}^{(-\widetilde{s}_k)}\right)(\Sigma^{-1})_{-\widetilde{s}_k,\widetilde{s}_k}\right)\right)\right)\Sigma_{B_{\ell,i,\widetilde{s}_k}},\\
\phi_{2,\ell,i,k}&=&  Inv-logit\left \{ logit(\pi_2)-\frac{1}{2}\log(det(\sigma^2_\beta\boldsymbol{I}_{\widetilde{s}_k}))+\frac{1}{2}\log(det(\Sigma_{B_{\ell,i,\widetilde{s}_k}} ))\right.\\
& &\left.+\frac{1}{2}tr\left((\Sigma_{B_{\ell,i,\widetilde{s}_k}})^{-1}\boldsymbol{\mu}'_{2,\ell,i,\widetilde{s}_k}\boldsymbol{\mu}_{2,\ell,i,\widetilde{s}_k}\right)\right\}.
\label{Sstep}
\end{eqnarray*}
Here, $ \boldsymbol {X} _ \ell ^ {(i)} $ and $ \boldsymbol {X} _ \ell ^ {(-i)} $ denote the $i$th column of $\boldsymbol{X}_\ell$ and matrix $\boldsymbol{X}_\ell$ excluding the $i$th column, respectively. The same definition structure is used for $\boldsymbol{Y}$ and $B$. In addition, $B_\ell^{(-i,i)}$ and $(\Sigma)^{-1}_{-i,i}$ denote the $i$th column of $B_\ell$ and $\Sigma^{-1}$ without $i$th element. $Inv-logit(\cdot)$ is an inverse logistic function.

In the M-step, we take the derivative of $L(q)$ with respect to $\theta$ and then set it as a zero vector. Thus, the components in $\theta$ can be updated as the solutions of the normal equations, i.e.,
\begin{eqnarray*}
\pi_{1} &=&\frac{\sum_{\ell=1}^p\sum_{i=1}^m\phi_{1,\ell,i}}{mp},\\
\pi_{2} &=&\frac{\sum_{\ell=1}^{p}\sum_{i=1}^m\sum_{k=1}^g\phi_{2,\ell,i,k}}{\sum_{k=1}^g|\widetilde{s}_k|mp},\\
\Sigma &=& \frac{E\left(\left(\boldsymbol{Y}-\boldsymbol{X}\boldsymbol{B}\right)'\left(\boldsymbol{Y}-\boldsymbol{X}\boldsymbol{B}\right)\right)}{T},\\
\sigma^2_{B} &=& \frac{\sum_l^p\sum_i^m\phi_{1,l,i}(\Sigma_{B,l,i,i}+\mu^2_{1,l,i})+\sum_l^p\sum_i^m\sum_k^g\phi_{2,l,i,k}tr(\Sigma_{B,l,i,\widetilde{s}_k}+\boldsymbol{\mu}_{2,l,i,\widetilde{s}_k}'\boldsymbol{\mu}_{2,l,i,\widetilde{s}_k})}{\sum_l^p\sum_i^m\left(\phi_{1,l,i}+\sum_k^{g}|\widetilde{s}_k|\phi_{2,l,i,k}\right)}.
\label{Mstep}
\end{eqnarray*}
More details regarding both steps are shown in the supplementary materials.\

Since we iterate the E- and M-steps sequentially, a natural choice of a stopping criterion is that the difference between the values of $L(q)$ in two consecutive iterations is less than a certain threshold. Then one can approximate the posterior inclusion probabilities as follows:
\begin{eqnarray*}
P(\eta_{\ell,i,k}=1|\boldsymbol{Y},\boldsymbol{X},\hat{\theta}) &\approx& q(\eta_{\ell,i,k}=1 | \hat{\theta})=\phi_{2,\ell,i,k},\\
P(\gamma_{\ell,i,i}=1|\boldsymbol{Y},\boldsymbol{X},\hat{\theta}) &\approx& q(\gamma_{\ell,i,i}=1 | \hat{\theta})=\phi_{1,\ell,i}.
\label{postphi}
\end{eqnarray*}
Thus, based on the median probability criterion \citep{barbieri2004optimal}, $B_{\ell,i,i}$ or $B_{\ell,i,\widetilde{s}_k}$ is identified as non zero if $\phi_{1, l, i}$ or $\phi_{2, l,i,k}$ is larger than or equal to $1/2$. Finally, the one-step head prediction $\hat{\boldsymbol{Y}}_{T+1}$ can be obtained by
$\hat{\boldsymbol{Y}}_{T+1}=\boldsymbol{Y}_{T}\hat{B}_1+\boldsymbol{Y}_{T-1}\hat{B}_2+\dots,\boldsymbol{Y}_{T+1-p}\hat{B}_p$, where the $\hat{B}_\ell$ are estimated based on the identified active structures.

The EM-type method is a locally optimal approach and may be sensitive to the initial status. Realistic initial values of $\theta$ may provide poor estimation results due to improper choices of the initial prior probabilities. In our study, the initial values of the prior probabilities, $\pi_1$ and $\pi_2$, are set to be small, e.g., $\pi_1 = \pi_2 = 0.01$. The initial values of the coefficient matrix, $B$, are obtained via the least-squares estimation method, and the initial value of $\Sigma$ is set as the sample variance of $\boldsymbol{Y}$ divided by $2$. The threshold value of the stopping criterion is usually set to a prespecified value, which is less than or equal to $10^{-6}$.

 \hypertarget{sec4}{}\section{\sffamily \Large Simulation}

This section investigates the finite-sample performance of the proposed VB approach compared with that of a known data generation process. We first revisit the simulation studies of VAGSA in \cite{chu2019bayesian}. Given dynamic networks, we care not only about the structure selection ability but also about the computational efficiency of our approach. For medium-sized examples, i.e., $m = 10$ and $20$, we directly compare the performances of the VB method and the VAGSA. When the dimensionality increases to $m=50$, the VAGSA requires an extremely long computational time, and thus, we illustrate only the results of the proposed VB method. 

\hypertarget{sec4.1}{}\subsection{Medium-sized network examples}

We follow the simulation setups in \cite{chu2019bayesian} and generate medium-sized network series with $m = 10$ and $20$ for a fair comparison. We assume that there are multiple lags $p$ in the VAR framework. Three dependence structures are considered in the simulations.
\begin{description}
\item[m10UG and m20UG:] The true models follow the universal grouping structure with $(m, p) = (10, 5)$ and $(20,5)$, respectively. There are $72$ and $145$
nonzero coefficients in two cases.
\item[m10SG and m20SG:] The segmentation structure is considered in these two simulations with $m = 10$ and $20$. Let $S_{m,g}$ denote
a specific group structure with $m$ time series for $g$ disjoint groups. We set $S_{10,3} = \left\{(1, 2, 3), (4, 5, 6), (7, 8, 9, 10)\right\}$ for $(g,m,p) = (3,10, 5)$ and
$S_{20,4} = \left\{(1, 2, 3, 4, 5), (6, 7, 8, \right.$
$\left.9, 10),(11, 12, 13), (14, \dots, 20)\right\}$ for the case where $(g, m, p) = (4,20, 5)$. There are 40 and 109 nonzero coefficients, respectively.
\item[m10NG and m20NG:] In the ``no grouping'' cases, there are 18 and
27 nonzero coefficients for $(m, p) = (10, 5)$ and
$(20,5)$, respectively.
\end{description} 
In addition, the error terms are generated from a multinormal distribution with zero mean. We consider two different covariance matrices when generating the stochastic noises, namely, an identity matrix $ \boldsymbol{I} _m $ indicating that there are no concurrent correlations among the nodes and a symmetric matrix $\Sigma$ defined as follows:
\begin{enumerate}
\item[$\Sigma_{10}$:] The diagonal of $\Sigma_{10}$ is $\left(0.9,0.9,0.9,0.9,0.9,0.9,0.8,0.8,0.8,0.8\right)$, and the off-diagonal correlation coefficients of  $\Sigma_{10}$ are $0.4^{|i'-i|}$ for $i' \neq i$.\
\item[$\Sigma_{20}$:] The diagonal of $\Sigma_{20}$ is $\left(0.9,0.9,0.9,0.8,0.8,0.8,0.9,\cdots,0.9\right)$, and the off-diagonal correlation coefficients of  $\Sigma_{20}$ are $0.4^{|i'-i|}$ for $i' \neq i$.
 \end{enumerate}
For each simulation, we generate data with $T = 301$, where the first $300$ samples are used for model training and the last point $Y_{301}$ is used to demonstrate the prediction ability of the proposed method. Each simulation is replicated $N =100$ times, and the segmentation structure is assumed to be known. We simply fix the number of lags $p$ to 10 and set the threshold value of the stopping criteria to $10^{-8}$ in all simulations. \

As mentioned before, we also implement the VAGSA for the simulation cases for comparison purposes. According to Chu et al. (2019), first, we add the inverse Wishart prior for $\Sigma$ with $m$ degrees of freedom and a scale matrix $I_m$. For the other parameters, the prior probabilities, $\pi_1$ and $\pi_2$, are fixed at 0.5, and for the variance in the mixture prior distribution, $\sigma_B$ is 0.5 for the cases of m10UG, m20UG, m10SG and m20SG and 15 for the other two cases, m10NG and m20NG. When we implement the VAGSA, there are 3000 sweeps in total, and we take the last 1000 samples for inference. For the details of implementing the VAGSA, please refer to \cite{chu2019bayesian}. 

\hypertarget{sec4.2}{}\subsection{Large-scale network examples}
\begin{figure}
    \centering
 
        \begin{minipage}{.34\textwidth}
            \begin{subfigure}
            \centering
            \includegraphics[width=7.5cm]{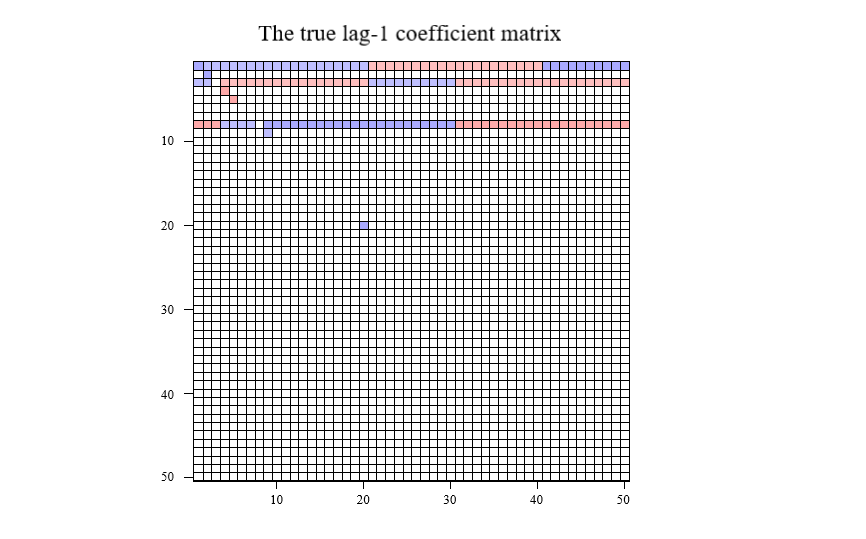}
            \end{subfigure}
            \begin{subfigure}
            \centering
            \includegraphics[width=7.5cm]{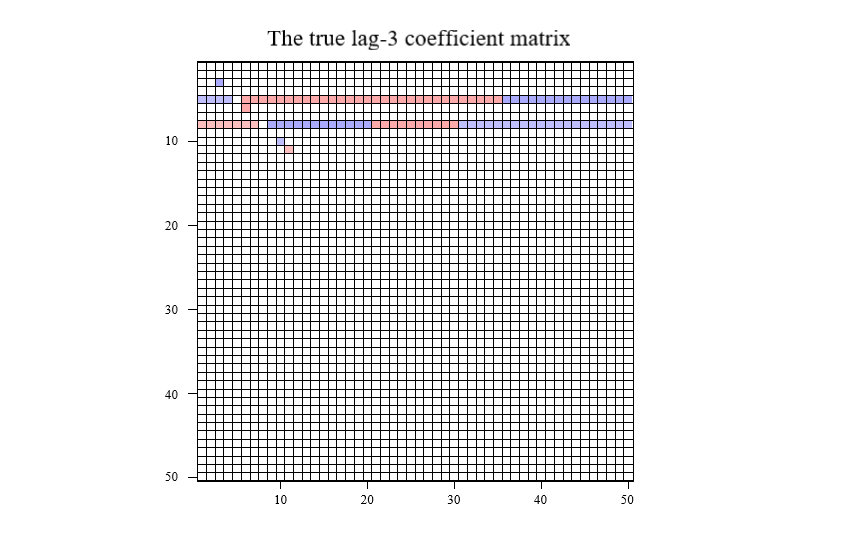}
            \end{subfigure}   
            \begin{subfigure}
            \centering
            \includegraphics[width=7.5cm]{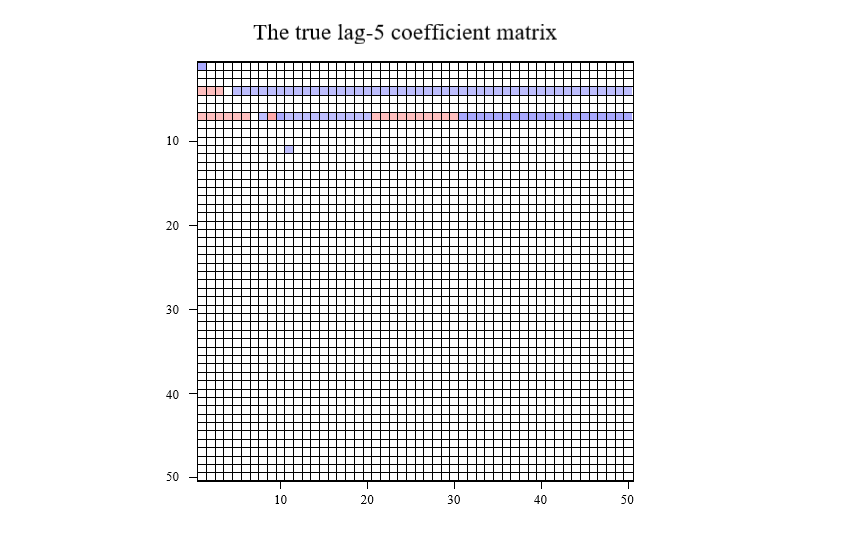}
                         \caption*{(a) m50UG}
            \end{subfigure} 
        \end{minipage}
        \begin{minipage}{.345\textwidth}
            \begin{subfigure}
            \centering
            \includegraphics[width=7.5cm]{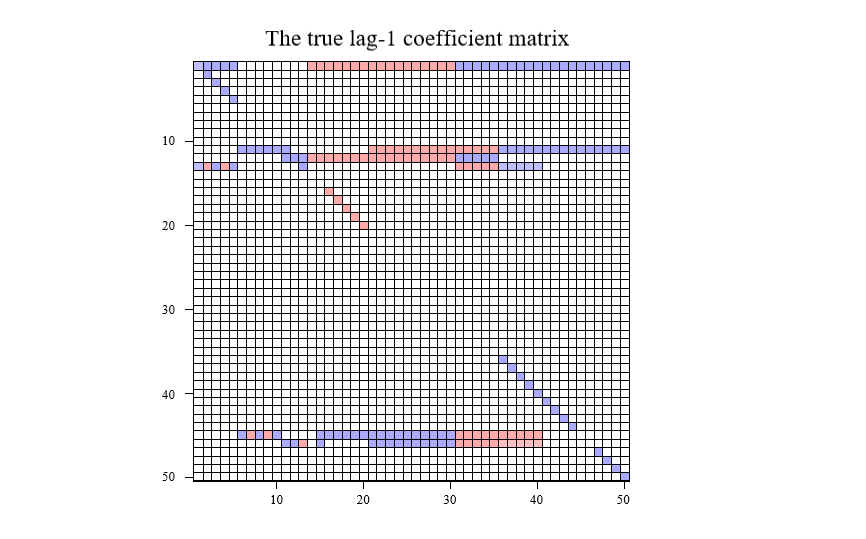}
            \end{subfigure}
            \begin{subfigure}
            \centering
            \includegraphics[width=7.5cm]{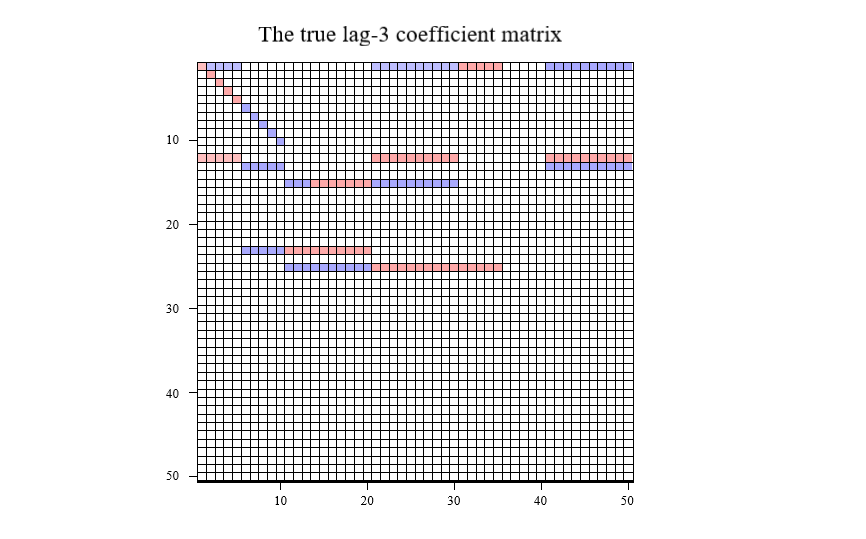}
            \end{subfigure}  
             \begin{subfigure}
            \centering
            \includegraphics[width=7.5cm]{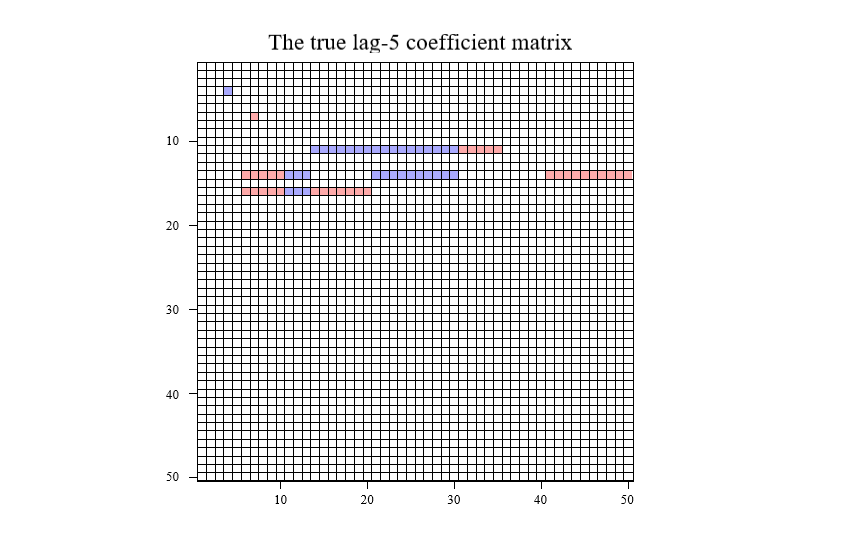}
            \caption*{(b) m50SG}
            \end{subfigure}  
        \end{minipage}
        \hfill
        \begin{minipage}{.3\textwidth}
            \begin{subfigure}
            \centering
            \includegraphics[width=7.5cm]{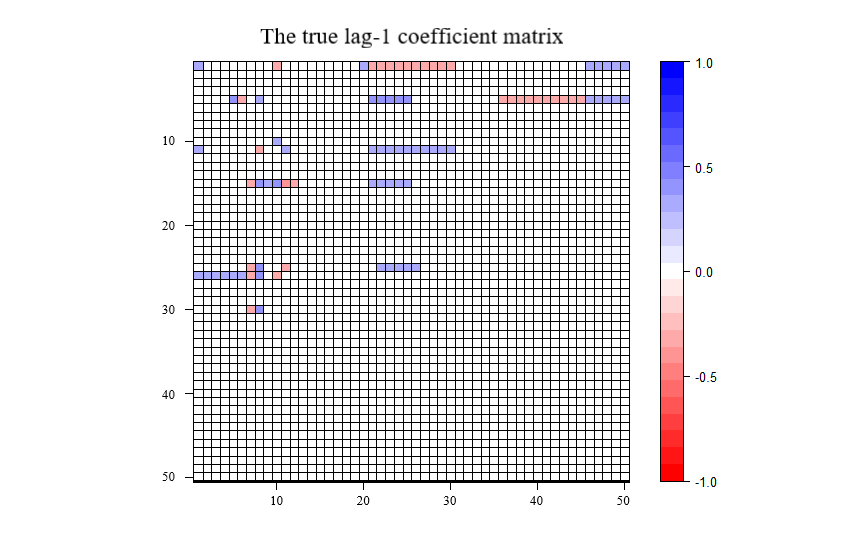}
            \end{subfigure}
            \begin{subfigure}
            \centering
            \includegraphics[width=7.5cm]{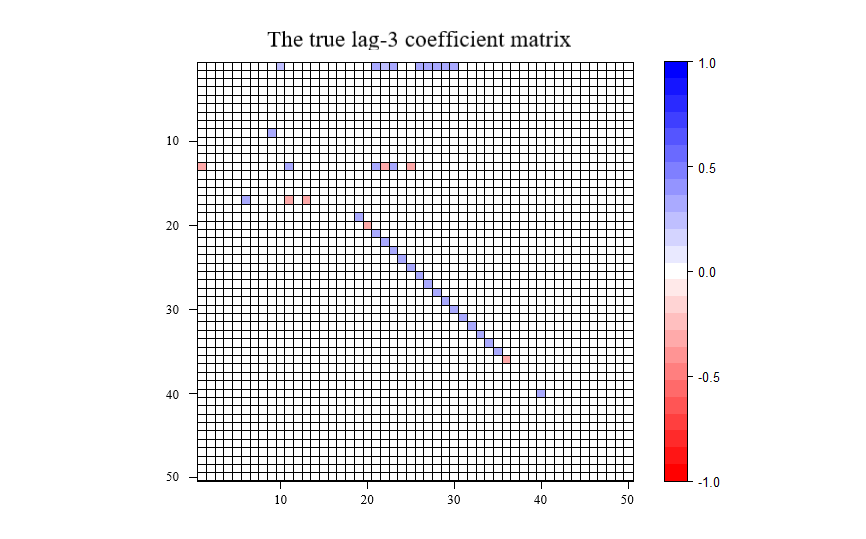}
            \end{subfigure}  
             \begin{subfigure}
            \centering
            \includegraphics[width=7.5cm]{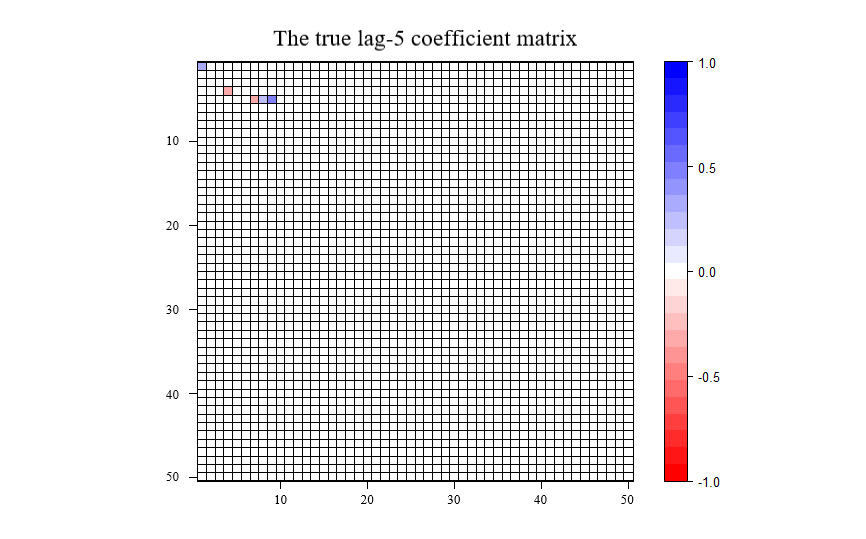}
            \caption*{(c) m50NG}
            \end{subfigure} 
        \end{minipage}
           \caption{The true active coefficient matrices for all UG, SG and NG cases. }
           \label{fig:coefmatrix}
    \end{figure}

To illustrate the feasibility of the proposed VB approach for high-dimensional scenarios, we conduct simulations with $m = 50$ nodes. The setups of the large-scale simulations are similar to those used in Section \hyperlink{sec4.1}{4.1}. Here, we also consider the three different grouping structures, and the details of three cases are shown as follows. 
\begin{description}
\item[m50UG:] There are 355 nonzero coefficients in $B_1$, $B_3$ and $B_5$ for $(m, p) = (50, 5)$.
\item[m50SG:] We set $S_{50,8} = \{(1, 2, 3, 4, 5),$ $(6, 7, 8, 9, 10),$
$(11, 12, 13),$ $(14,\dots, 20),$ $(21,\dots,30),$ $(31,\dots,35),$
$(36,\dots,40),$ $(41,\dots,50)\}$. There are 360 nonzero coefficients in $B_1$, $B_3$ and $B_5$ for $(m, p) = (50,5)$.
\item[m50NG:] There are 
128 nonzero coefficients in $B_1$, $B_3$ and $B_6$ for $(m, p) = (50,5)$.
\end{description}
Figure \ref{fig:coefmatrix} visualizes the true sparsity of the network dependence with the nonzero coefficients marked in bold. Note that only the active coefficient matrices are displayed, while the others, such as lag-2 and lag-4, are omitted as zero everywhere.

Similarly, two different noise covariance matrices are used for data generation, i.e., a $50 \times 50$ identity matrix $\boldsymbol{I}_{50}$ and a symmetric matrix $\Sigma_{50}$, which is defined as follows:
\begin{enumerate}
\item[$\Sigma_{50}$:]  The diagonal elements are $\left(0.9,0.9,0.9,0.8,0.8,0.8,\underset{14}{\underbrace{0.9,\cdots,0.9}},\underset{20}{\underbrace{0.8,\cdots,0.8}}, \underset{10}{\underbrace{0.9,\cdots,0.9}}\right)$, and the correlation coefficients of  $\Sigma_{50}$ are set as $0.4^{|i'-i|}$ for $i' \neq i$.
\end{enumerate}
In this simulation example, we generate data with $T= 701$. The first 700 samples are used for model training, and the last sample $Y_{701}$ is used to illustrate the prediction ability of the model. 

\hypertarget{sec4.3}{}\subsection{Simulation results}
Based on the simulation setups in Sections \hyperlink{sec4.1}{4.1} and \hyperlink{sec4.2}{4.2}, we independently repeat each case 100 times. When we implement the proposed VB approach, the initial setups are basically the same as those use in Section \hyperlink{sec3}{3}, except for the case of m50NG with a covariance matrix of $\Sigma_{50}$. For this case, we set $\pi_1 = \pi_2 =0.5$, instead of 0.01. To illustrate the performances of the proposed method, we consider four measurements. The true positive rate (TPR) and the false positive rate (FPR) are designed to show the accuracy of structure identification. The average model size (AMS) measures how many active elements are identified. Compared to the TPR and the FPR, the AMS gives an overall indicator of identification accuracy. The last measurement is the average of the mean square prediction errors (MSPEs), and this is used to report the prediction ability of the model. The definitions of these four measurements are shown as follows:
\begin{align}
\text{TPR}&=\frac{\text{\# correctly identified active variables}}{\text{\# true active variables}},\nonumber \\
\text{FPR}&=\frac{\text{\# incorrectly identified active variables}}{\text{\# true inactive variables}},\nonumber \\
AMS&=\frac{1}{N}\sum_{j}^N \text{ \# identified active variables in the $j$th replicate} ,\nonumber
\end{align}
where $N$ is the number of replicates and
\begin{align*}
\text{MSPE}=\frac{1}{mN}\sum_{i=1}^m\sum_{j=1}^N\left(Y_{T+1,i,j}-\hat{Y}_{T+1,i,j}\right)^2.
\end{align*}
Last, the average CPU time for a replication is also reported in seconds. Here, we run the R code for the proposed VB method, and we run the VAGSA based on its MATLAB code. All codes are implemented on a PC with an Intel(R) Core(TM) i7-4770 CPU @3.400 GHz and 16.0 GB of RAM.

The results of all cases are summarized in Table \ref{tab:example_10_TPR}. For the medium-sized networks with  $m=10$ and $m=20$, both the VB and VAGSA results are reported. For large-scale networks with $m=50$, only the VB results are reported and the measurements for the VAGSA are denoted by ``-''. Consider the performances of the medium-sized networks. According to Table \ref{tab:example_10_TPR}, both Bayesian methods achieve similar structural identifications and inference accuracies for medium-sized networks. In particular, the TPRs are perfect with values close to $100\%$, implying that all active elements are successfully identified. In terms of AMS, both methods share similar AMS values and are all close to the true model sizes. In addition, the VB method has better performance in terms of the FPR than the alternative VAGSA. For the UG and SG cases, the FDR values of the VB method are less than a quarter of those of the VAGSA. The VAGSA performs slightly better for the four NG cases; however, the differences between the two methods are minor. To check the selection results in detail, we find that the VB approach might not identify a variable with a small coefficient for the NG cases. Finally, consider the prediction ability. The VB method has slightly better accuracies than those of the VAGSA in most cases, as reflected in the MSPE. Most importantly, there is a dramatic improvement in terms of the average CPU time when using the VB method. In general, it only needs approximately $1/7$ of the computational cost required by the VAGSA. For some cases with $m = 10$ and $I_{10}$, the VB approach even saves $34/35
$ CPU time, without sacrificing accuracy. 

\begin{figure}
    \centering
 
        \begin{minipage}{.34\textwidth}
            \begin{subfigure}
            \centering
            \includegraphics[width=7.5cm]{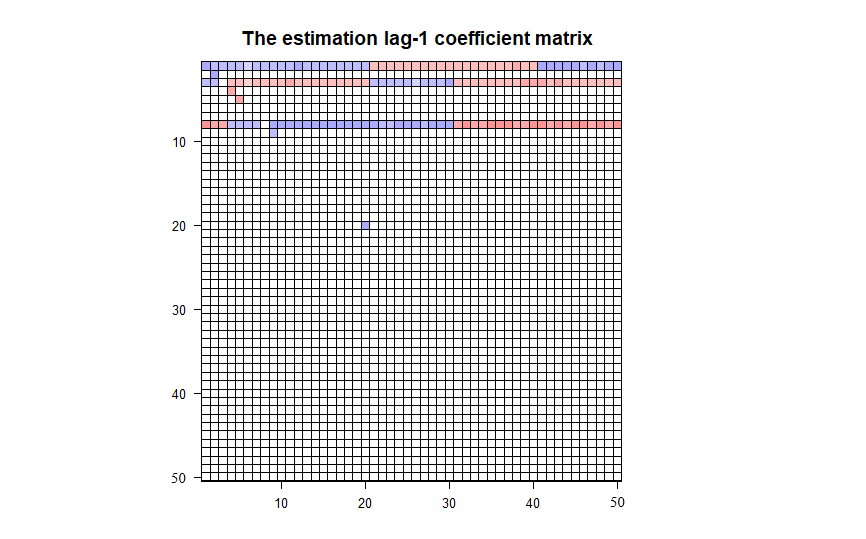}
            \end{subfigure}\\
            \begin{subfigure}
            \centering
            \includegraphics[width=7.5cm]{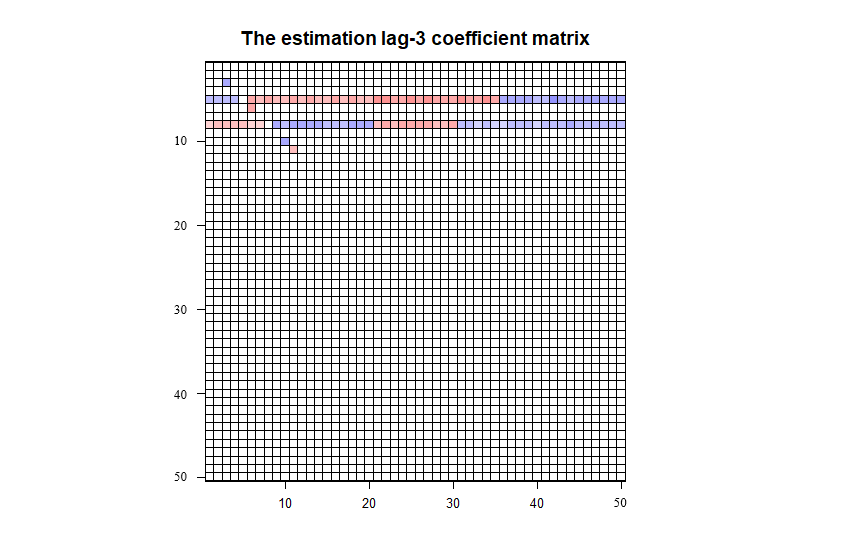}
            \end{subfigure}    
            \begin{subfigure}
            \centering
            \includegraphics[width=7.5cm]{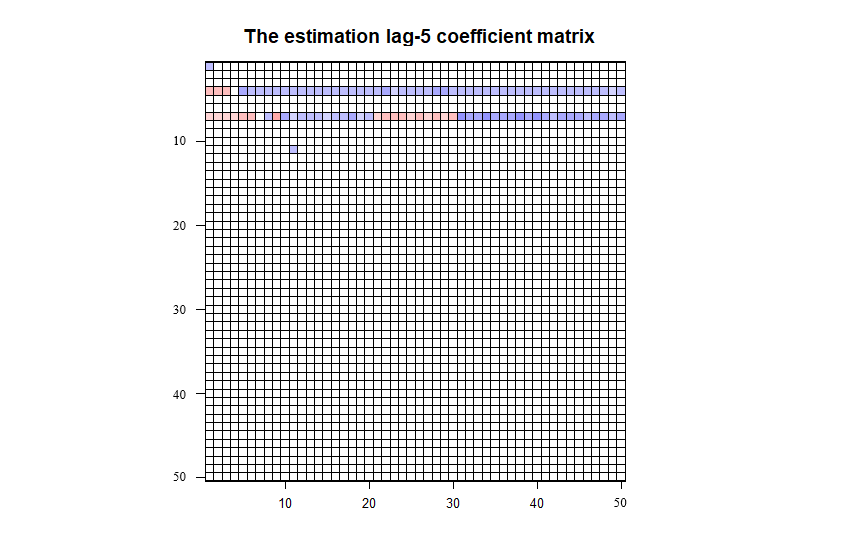}
                         \caption*{(a) m50UG}
            \end{subfigure} 
        \end{minipage}
        \hfill
        \begin{minipage}{.345\textwidth}
            \begin{subfigure}
            \centering
            \includegraphics[width=7.5cm]{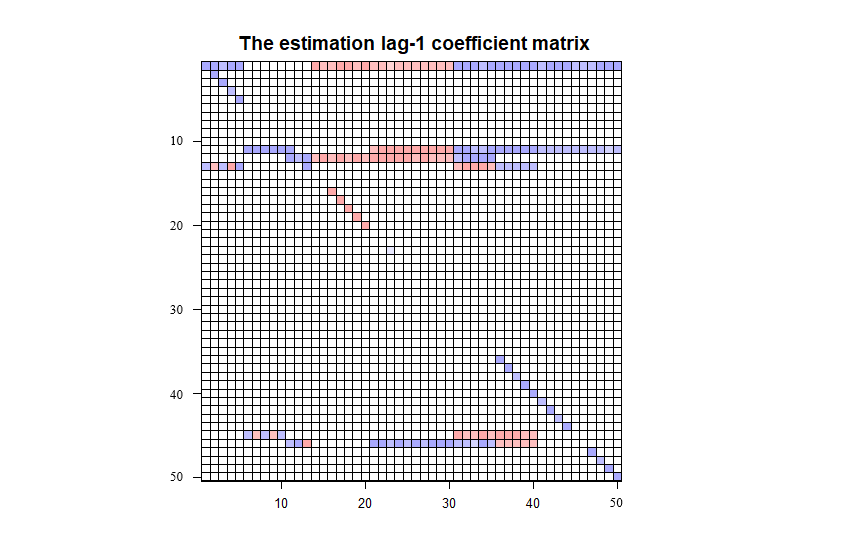}
            \end{subfigure}\\
            \begin{subfigure}
            \centering
            \includegraphics[width=7.55cm]{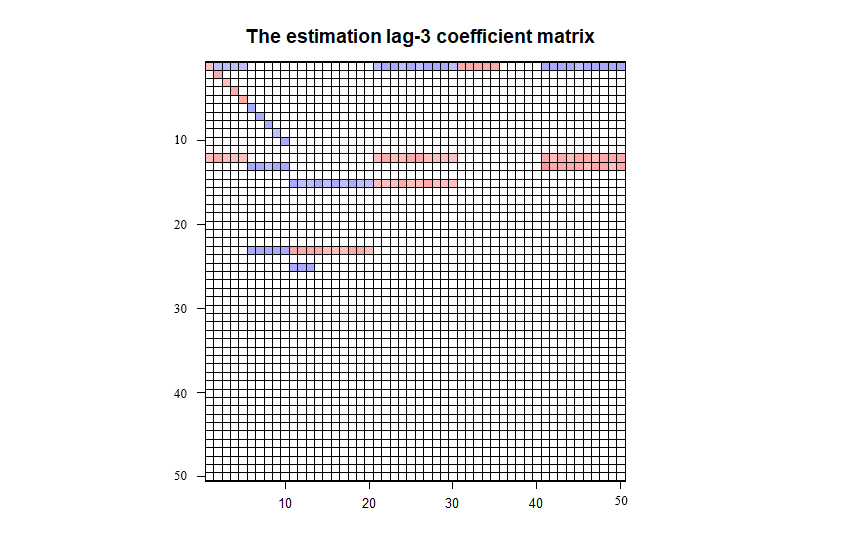}
            \end{subfigure}  
             \begin{subfigure}
            \centering
            \includegraphics[width=7.6cm]{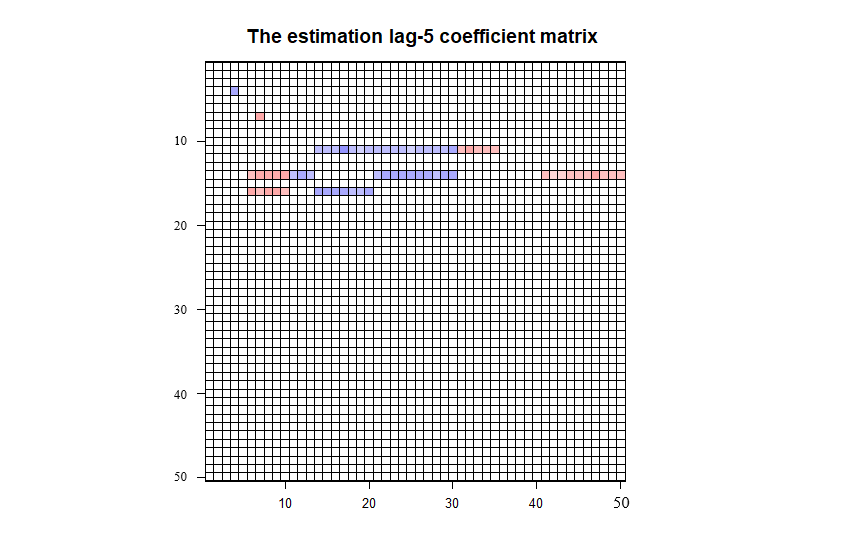}
             \caption*{(b) m50SG}
            \end{subfigure}  
        \end{minipage}
        \hfill
        \begin{minipage}{.3\textwidth}
            \begin{subfigure}
            \centering
            \includegraphics[width=7.5cm]{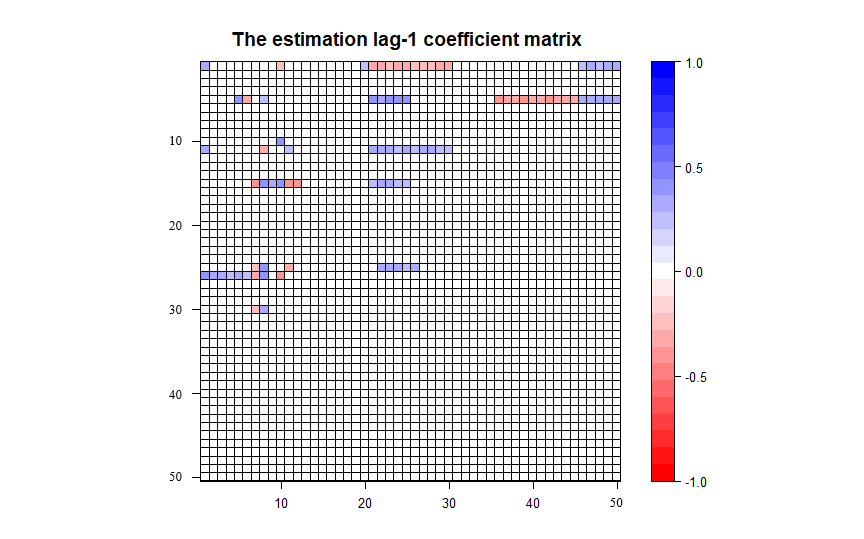}
            \end{subfigure}\\
            \begin{subfigure}
            \centering
            \includegraphics[width=7.5cm]{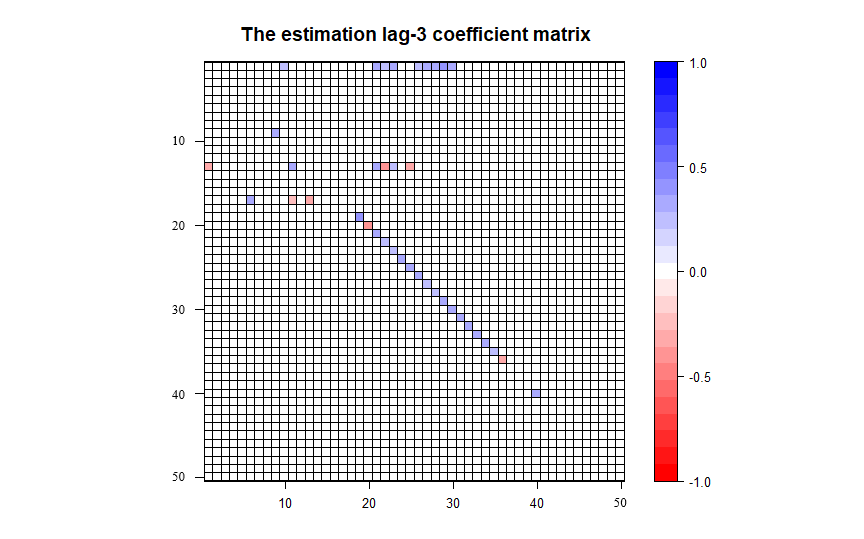}
            \end{subfigure}  
             \begin{subfigure}
            \centering
            \includegraphics[width=7.5cm]{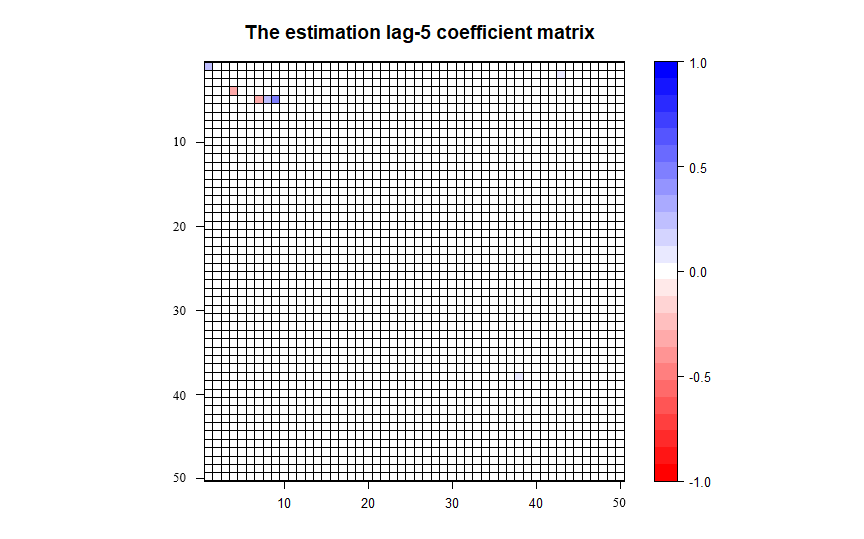}
            \caption*{(c) m50NG}
            \end{subfigure} 
        \end{minipage}
           \caption{Estimated coefficient matrices with $\Sigma=\boldsymbol{I}_{50}$ for all UG, SG and NG cases. }
           \label{fig:coefmatrix_I} 
    \end{figure}

\begin{figure}
    \centering
 
        \begin{minipage}{.34\textwidth}
            \begin{subfigure}
            \centering
            \includegraphics[width=7.5cm]{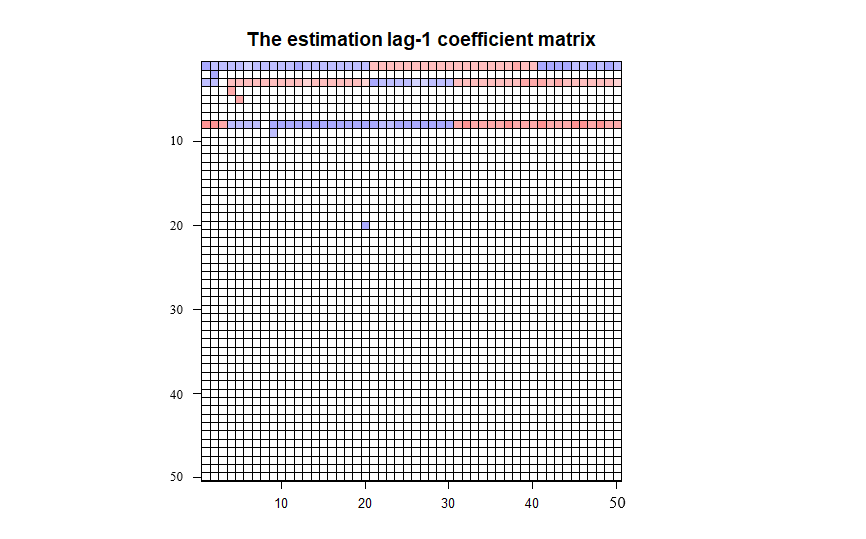}
            \end{subfigure}\\
            \begin{subfigure}
            \centering
            \includegraphics[width=7.5cm]{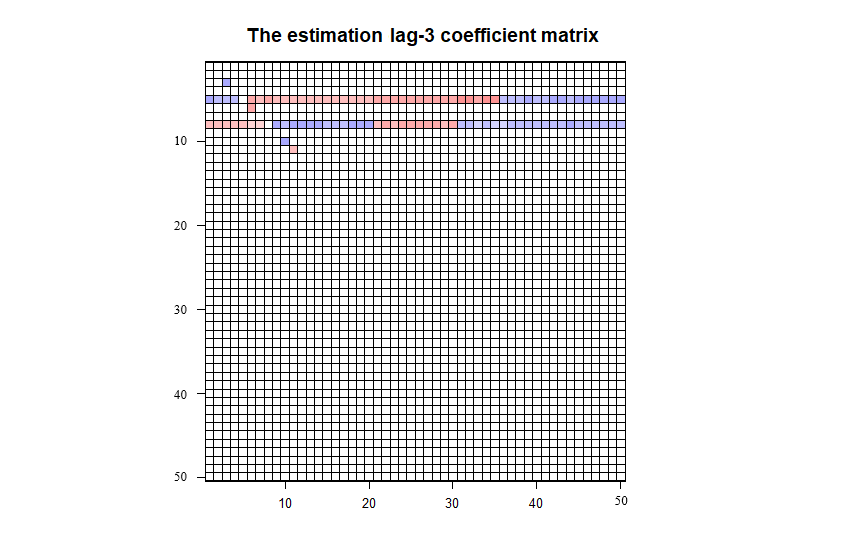}
            \end{subfigure}    
            \begin{subfigure}
            \centering
            \includegraphics[width=7.5cm]{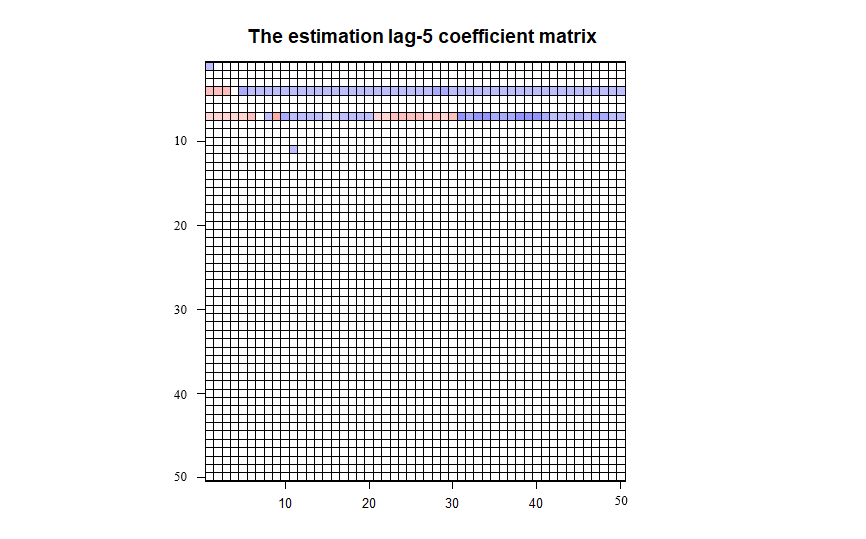}
                         \caption*{(a) m50UG}
            \end{subfigure} 

        \end{minipage}
        \hfill
        \begin{minipage}{.34\textwidth}
            \begin{subfigure}
            \centering
            \includegraphics[width=7.6cm]{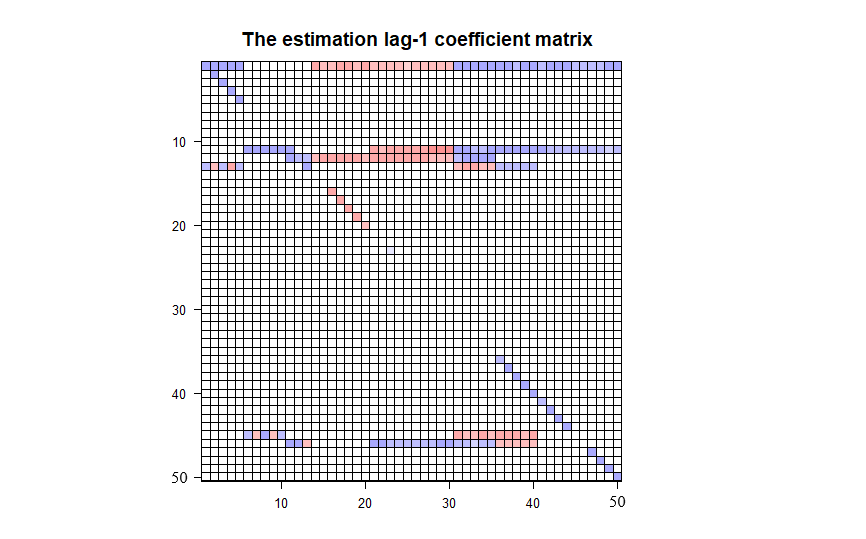}
            \end{subfigure}\\
            \begin{subfigure}
            \centering
            \includegraphics[width=7.6cm]{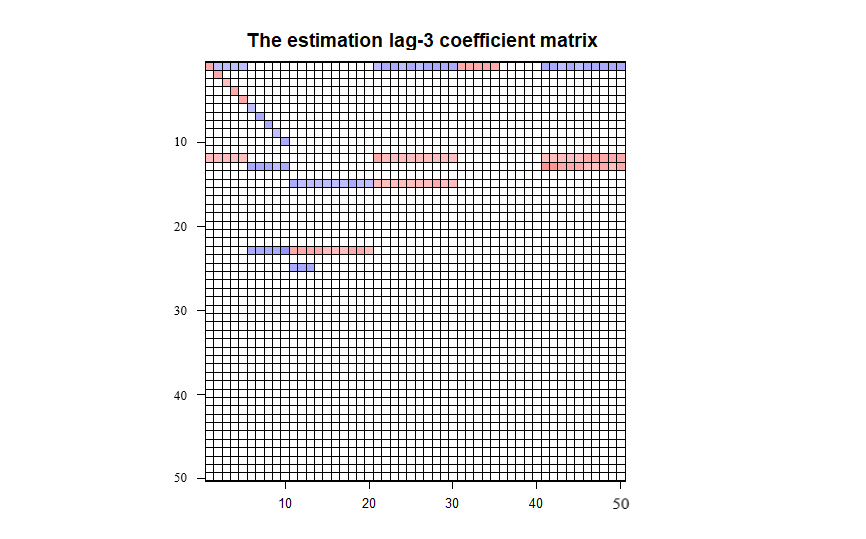}
            \end{subfigure}  
             \begin{subfigure}
            \centering
            \includegraphics[width=7.6cm]{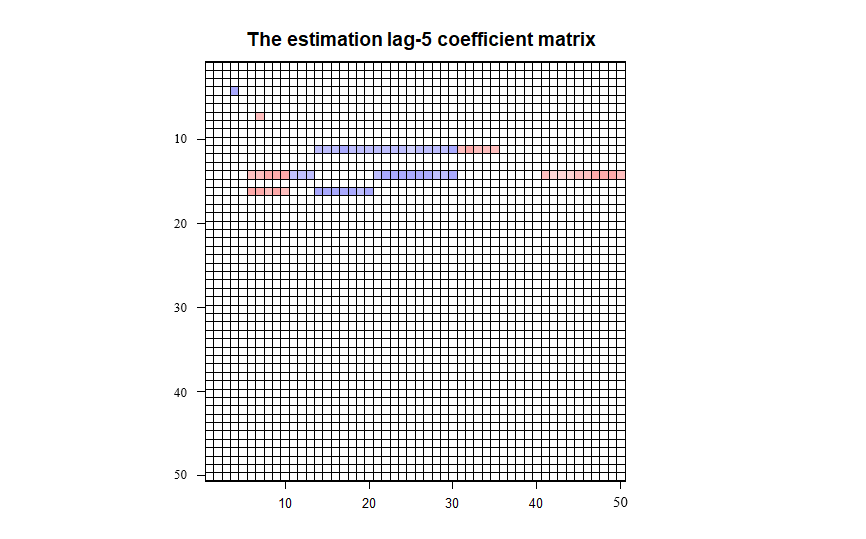}
              \caption*{(b) m50SG}
            \end{subfigure}  
        \end{minipage}
        \hfill
        \begin{minipage}{.3\textwidth}
            \begin{subfigure}
            \centering
            \includegraphics[width=7.5cm]{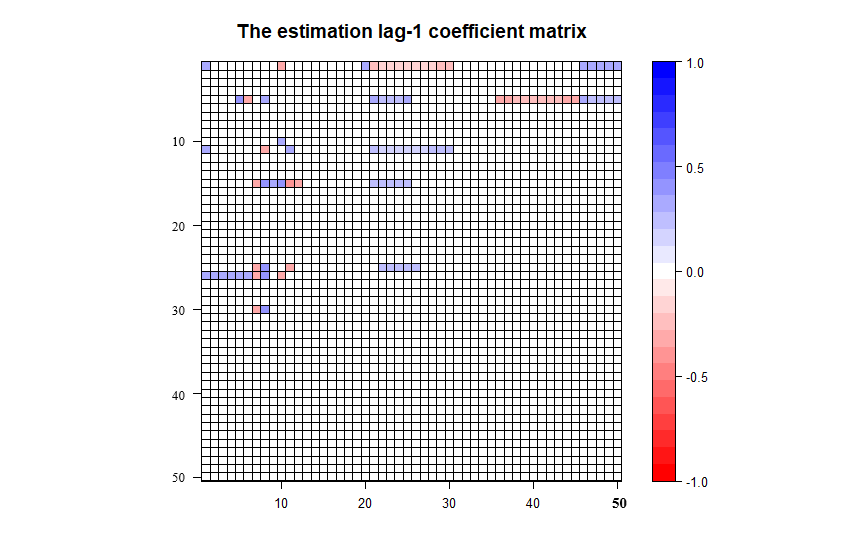}
            \end{subfigure}\\
            \begin{subfigure}
            \centering
            \includegraphics[width=7.5cm]{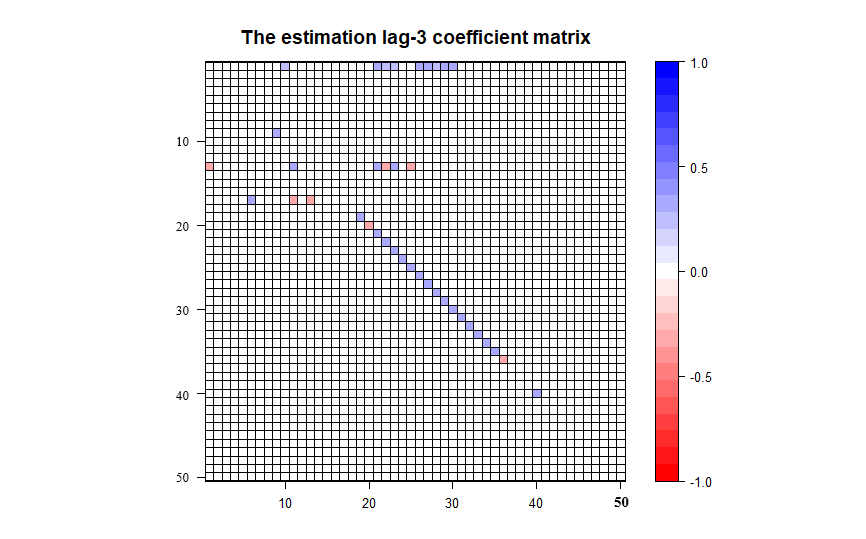}
            \end{subfigure}  
             \begin{subfigure}
            \centering
            \includegraphics[width=7.5cm]{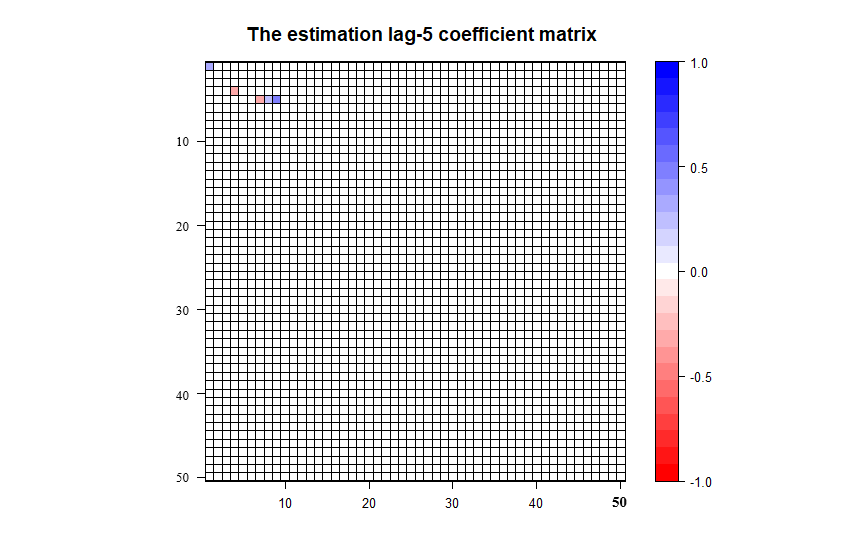}
            \caption*{(c) m50NG}
            \end{subfigure} 
        \end{minipage}
           \caption{Estimated coefficient matrices with $\Sigma=\Sigma_{50}$ for all UG, SG and NG cases. }
           \label{fig:coefmatrix_Sigma}
    \end{figure}

The good performance of the VB method continues for the large-scale networks because it is computationally possible. Figures \ref{fig:coefmatrix_I} and \ref{fig:coefmatrix_Sigma} show the average estimates of the coefficient matrices for two different covariance matrices. These coefficient matrices are all close to the true matrices. Use the UG case as an example. The structure identification results are perfect because the TPR $=100\%$, the FPR $=0\%$ and the AMS is still close to the true value. 
In the NG case with a covariance matrix of, $ \Sigma_ {50} $, the corresponding TPR is $92\%$, which is slight lower than the TPRs in the other cases. This may be because the VB method is used to approximate the true posterior distribution, and thus, active variables with small coefficients might not be detected. Overall, the proposed VB method has good performance in identifying the active structures for large-scale network cases. In addition, the average CPU times show that the case with a higher level structure, namely, UG, requires the least computational time, followed by the SG and NG cases. 

\begin{table}[H]
   \caption{Numerical Comparison between the VB Method and the VAGSA} 
   \label{tab:example_10_TPR}
   \small 
   \centering 
   \scalebox{0.9}{
  \begin{tabular}{l|ccccr} 
  \toprule[\heavyrulewidth]
   \bottomrule [\heavyrulewidth]
   \textbf{VB/VAGSA}&\textbf{TPR(\%)}&\textbf{FPR(\%)}&\textbf{AMS}&\textbf{MSPE}& \textbf{Ave. CPU Time}\\
   \bottomrule [\heavyrulewidth]
   m10UG-$I_{10}$ &100/100 &0.07/0.31  &72.62/74.90 &1.00/1.01& 4/148 \\
   m10UG-$\Sigma_{10}$ & 100/100&0.06/0.28  &72.51/74.59 &1.00/0.83& 5/138\\
   
   m10SG-$I_{10}$&100/100 &0.15/0.63  & 41.35/46.02&1.07/1.09&12/303 \\
   m10SG-$\Sigma_{10}$  &100/100&0.13/0.26&41.17/46.02&0.86/1.09&11/303\\

   m10NG-$I_{10}$&98/97 &0.15/0.11  &19.07/18.58& 1.00/1.00&40/628 \\
   m10NG-$\Sigma_{10}$&99/99 &0.11/0.08  & 18.86/18.55 &0.89/0.88&36/653\\
   \midrule
   m20UG-$I_{20}$&100/100 & 0.03/0.18 & 145.79/151.89  &0.98/0.99&15/236\\
   m20UG-$\Sigma_{20}$& 100/100 & 0.02/0.14 &  145.56/150.54 &0.95/0.95&180/237  \\
   
   m20SG-$I_{20}$& 100/100&0.06/0.26  & 109.25/117.05& 1.10/1.11 &44/652 \\
   m20SG-$\Sigma_{20}$& 100/100 & 0.06/0.21 &  109.22/115.20 &0.90/0.90 &51/672 \\
   
   m20NG-$I_{20}$&97/98 & 0.08/0.15 &29.29/32.39&0.99/0.99&291/2183\\
   m20NG-$\Sigma_{20}$&97/98 & 0.06/0.13 & 28.55/31.43&0.86/0.86  &262/2247 \\
   \midrule
    m50UG-$I_{50}$&100/- & 0.00/- &  355.47/- &1.02/- & 213/- \\
    m50UG-$\Sigma_{50}$& 100/-& 0.00/- &  355.57/- &0.87/-& 239/-  \\
    m50SG-$I_{50}$&100/- & 0.01/- &  361.58/- &1.02/- & 1082/- \\
    m50SG-$\Sigma_{50}$& 100/- & 0.01/- &  361.15/- &0.88/-& 1767/-  \\
    m50NG-$I_{50}$ &100/- & 0.03/- &134.67/-&1.02/- & 7860/-  \\
    m50NG-$\Sigma_{50}$&92/- & 0.03/- & 123.33/-&0.89/-& 58194/-  \\
   \bottomrule [\heavyrulewidth]
   \end{tabular}
   }
\end{table}

 \hypertarget{sec5}{}\section{\sffamily \Large Empirical Study}

\begin{figure}
    \centering
        \begin{minipage}{.45\textwidth}
          \includegraphics[scale=0.25]{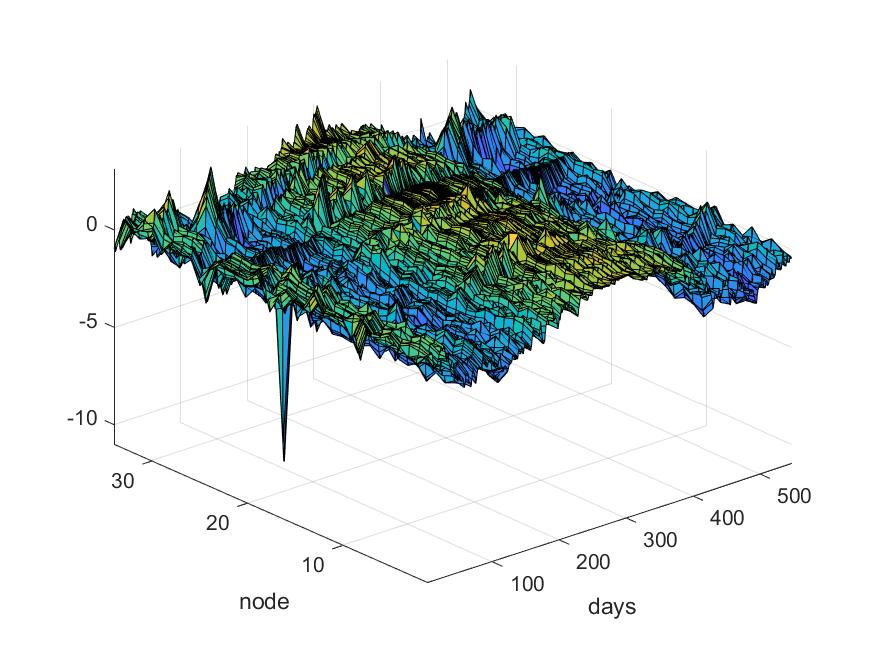}
          \caption*{Municipal (34 nodes)} 
           \end{minipage}
        \hfill
         \begin{minipage}{.45\textwidth}
          \includegraphics[scale=0.25]{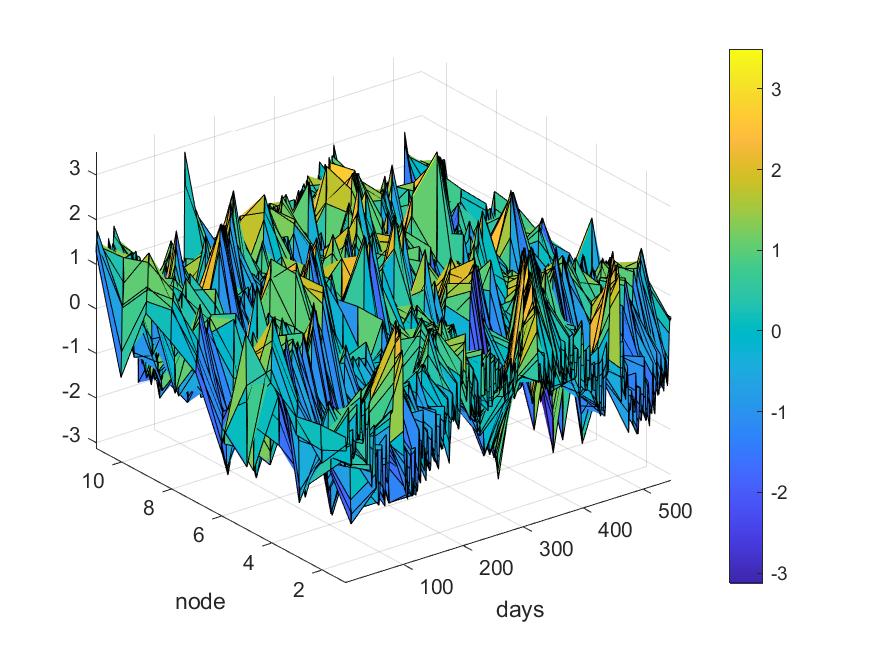}    \caption*{Industrial (11 nodes) }
        \end{minipage}
         \par
         \begin{minipage}{.45\textwidth}
          \includegraphics[scale=0.3]{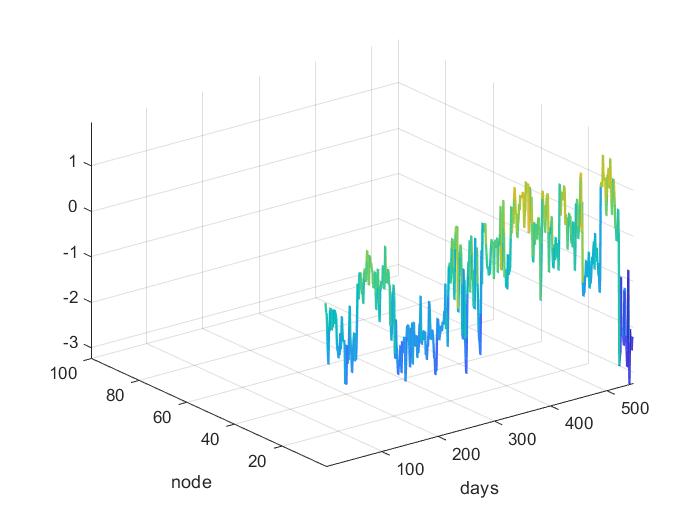}   
          \caption*{Border (1 node) }
        \end{minipage}
        \hfill
          \begin{minipage}{.45\textwidth}
          \includegraphics[scale=0.25]{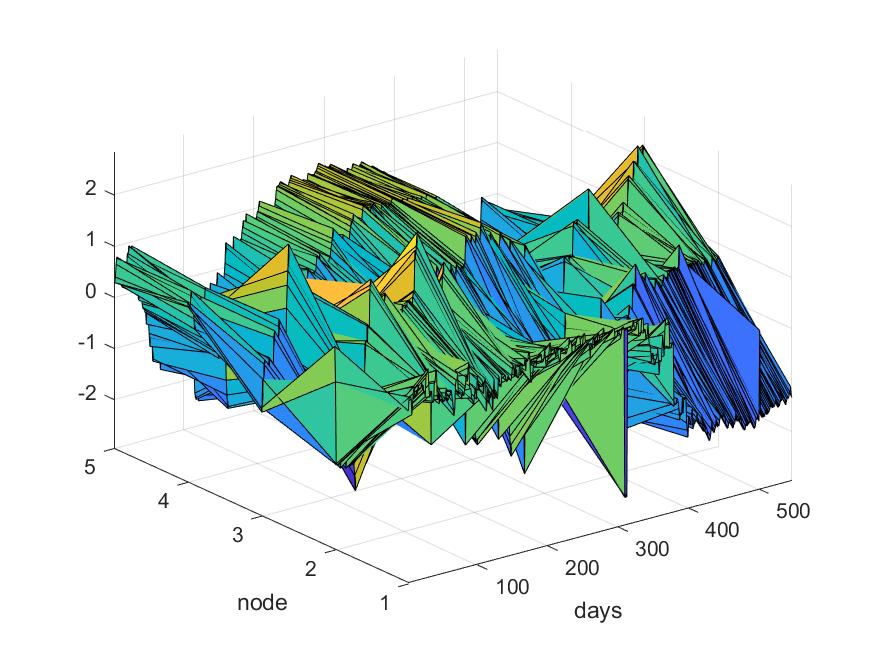}     \caption*{Others (5 nodes) }
           \end{minipage}

           \caption{Normalize daily gas flows values in different types. }
           \label{fig:GMAP}
    \end{figure}

We apply the VB approach to estimate the temporal dependences of the German natural gas flow networks and perform day-ahead forecasting with the NAR/VAR framework. The data cover two years, from 1st October 2013 to 30th September 2015. The gas flows contain both inflows (supply) and outflows (demand) recorded 7 days a week at $51$ distribution nodes belonging to 4 categories with different functions. There are 34 municipal nodes, that provide gas to local residential areas and small business districts. The 11 industry nodes are responsible for factory production. Moreover there is 1 border node, which is important in Germany, because these nodes serve as network transfer points for natural gas imported and exported via Germany. The rest 5 nodes are categorized to ``others'' that serve as switch nodes or perform other functions. Figure \ref{fig:GMAP} displays the time series of the 51 nodes that show different dynamic patterns. For further analysis, we normalize the values of the flows. In addition, note that the data in municipal nodes have been seasonally adjusted via the daily temperature, and 3 nodes, O3, O4, and O5, in the ``others'' category have also been seasonally adjusted.

We adopt the NAR/VAR model with the segmentation structure to analyze the gas network data. Here, the segmented grouping structures are defined based on the types of nodes that are involved. That is, each type of node is treated as a group. In addition, we set the number of lags $p$ as $14$ to incorporate social dependence up to two weeks ahead. We use the data from Oct 1st, 2013, to March 31, 2015, as the in-sample training data. The threshold value of the stopping criterion for the proposed VB method is set to be $10^{-6}$, and the other initial settings are the same as those used we set in the simulation studies in Section \hyperlink{sec4}{4}. After learning the model via the VB method, we perform a one-step ahead forecast. That is, at each daily point, we shift one more day, use the expanded sample to retrain the model coefficients based on the active segmentation structures, and perform a one-day ahead forecast. Note that we repeat the iterative forecast for the remaining $0.5$ years.

 Overall, we obtain 183 forecasts. To illustrate the performance of the proposed VB method, in addition to the mean absolute percentage estimator (MAPE), the normalized MSE (NRMSE) for the 51 nodes is also used, and both measures are defined as follows:
\begin{align}
MAPE&= \frac{1}{183\times51}\sum_{t=547}^{729}\sum_{i=1}^{51}\frac{|Y_{t+1,i}-\hat{Y}_{t+1,i}|}{|Y_{t+1,i}|},\\
NRMSE&=\frac{1}{183\times51}\left(\sum_{t=547}^{729}\sum_{i=1}^{51}(Y_{t+1,i}-\hat{Y}_{t+1,i})^2\right)/\frac{1}{183\times51}\sum_{t=547}^{729}\sum_{i=1}^{51}Y_{t+1,i}.\label{mape}
\end{align}

\begin{figure}
    \centering
        \begin{minipage}{-.4\textwidth}
          \includegraphics[scale=0.45]{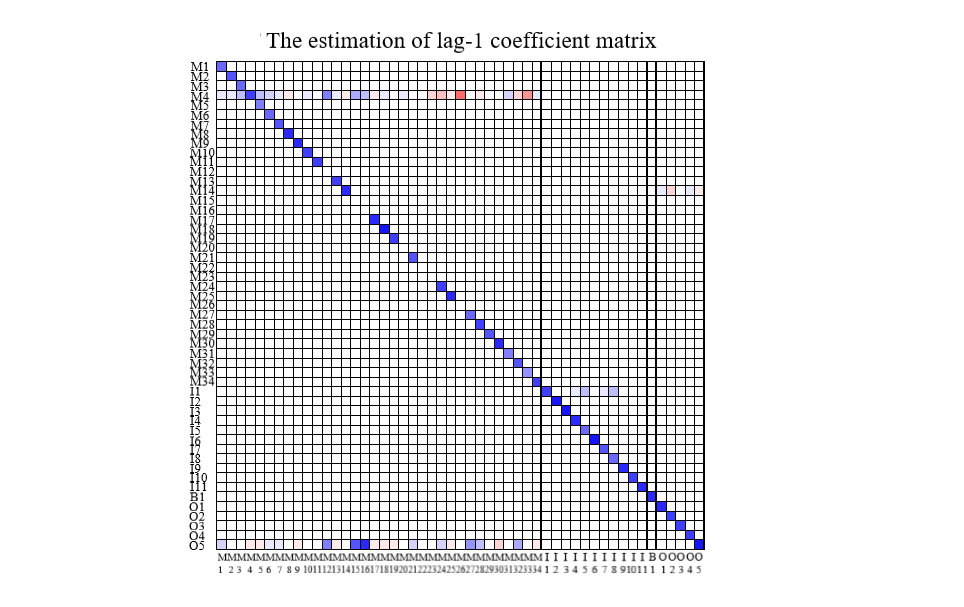}
           \end{minipage}
        \hfill
         \begin{minipage}{.5\textwidth}
          \includegraphics[scale=0.45]{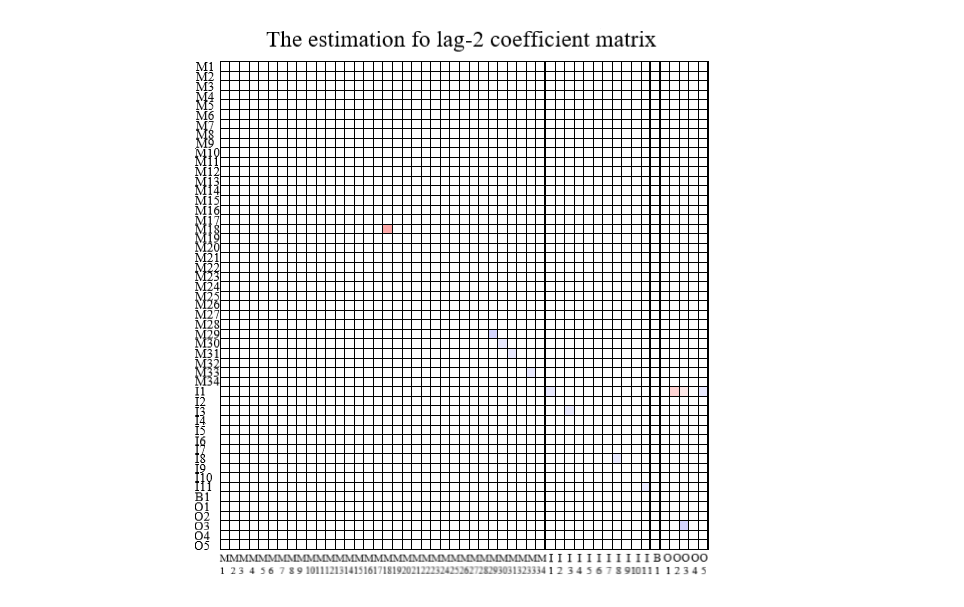}   
        \end{minipage}
                \par
         \begin{minipage}{1\textwidth}
          \includegraphics[scale=0.45]{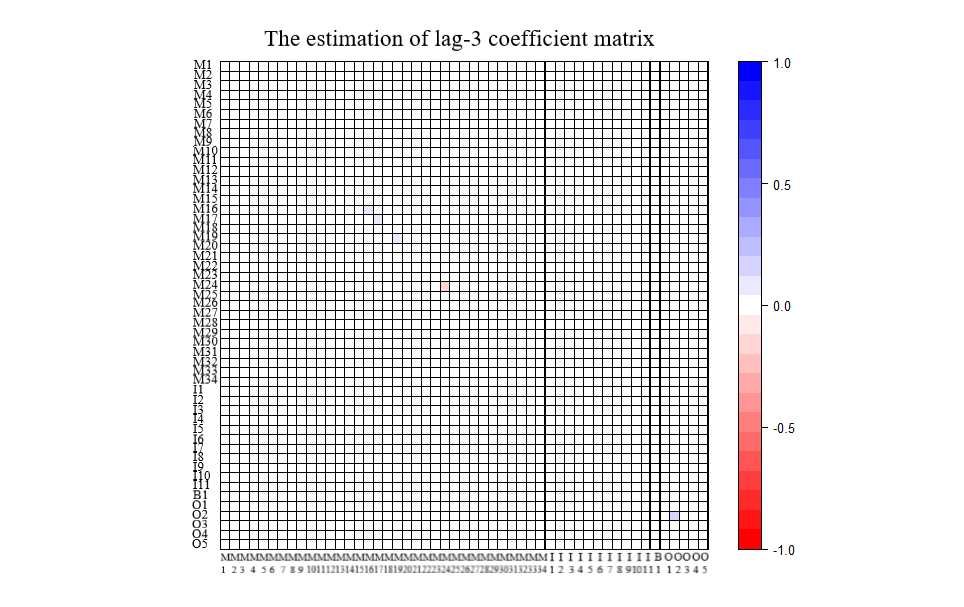}  
        \end{minipage}

\caption{The coefficient matrix for lag-1 to lag-3.}
\label{fig:coefmatrixlag1}
    \end{figure}
Figures \ref{fig:coefmatrixlag1} displays the estimated coefficient matrices for lag-1 to lag-3, and these three coefficient matrices are quite sparse. For lag-1, it can be observed that the $4$th and $14$th rows, corresponding to municipal nodes M4 and M14, affect the other municipal nodes. In the $35$th row, an industry node, I1, influences the other industry nodes but not the other three types. The $51$st row is the status of node O5; the others node only have relations with municipal nodes. On the other hand, the only border node is only related to itself and has no interactions with the others. For lag-2, only the $35$th row I1 influences the nodes in the other category. For lag-3, some nodes have minor influences by themselves. As the lag increases, there are fewer and fewer nodes that may influence others. Furthermore, there is no more serial dependence in the network after lag-10.

\begin{table}[t!]
 \caption{The MAPEs and NRMSEs for the fitted and forecast results classified by type. }
 \label{mape_fitted_type}
 \huge
 \centering
  \scalebox{0.35}{
\begin{tabular}{crrrrrrr}
\multicolumn{7}{c}{\textbf{In sample from 01/10/2014 to 31/03/2015}} \\
\toprule[\heavyrulewidth]
\bottomrule [\heavyrulewidth]
\multicolumn{1}{c}{\textbf{Type}} 

&
\multicolumn{1}{c}{\textbf{Number}} & \multicolumn{1}{c}{\textbf{MAPE (\%)}}  &
\multicolumn{1}{c}{\textbf{Range (\%)}}    &
\multicolumn{1}{c}{\textbf{SD}} & \multicolumn{1}{c}{\textbf{NRMSE}} & \multicolumn{1}{c}{\textbf{Range}} & \multicolumn{1}{c}{\textbf{SD}} \\
\bottomrule [\heavyrulewidth]
\multicolumn{1}{c}{\textbf{Municipal}}&
34 &
6.28&
(3.20, 14.70) &
1.93&
0.09&
(0.04, 0.36)&
0.05
\\
\textbf{Industrial}&
11 &
11.69&
(0.89, 34.22) &
11.74&
0.12&
(0.01, 0.31)&
0.10
\\
\multicolumn{1}{c}{\textbf{Border}}&
{1}   &
{9.36} &
{$-$}&
{$-$}&
{0.11}&
{$-$}&
{$-$}\\
\multicolumn{1}{c}{\textbf{Others}}&
{5}  &
{13.05} &
{(3.78, 36.30)} &
{13.26}&
{0.14}&
{(0.05, 0.34)}&
{0.12}
\\

\bottomrule[\heavyrulewidth]

\multicolumn{7}{c}{\textbf{Out of sample from 01/04/2015 to 30/09/2015}} \\
\toprule[\heavyrulewidth]
\bottomrule [\heavyrulewidth]

\multicolumn{1}{c}{\textbf{Municipal}}&
{34}  &
{7.89} &
{(4.66, 11.85)} &
{1.60}&
{0.10}&
{(0.06, 0.15)}&
{0.02}
\\
\multicolumn{1}{c}{\textbf{Industrial}}&
{11}  &
{15.98} &
{(1.70, 54.57)} &
{19.00}&
{0.12}&
{(0.02, 0.30)}&
{0.10}
\\
\multicolumn{1}{c}{\textbf{Border}}&
{1}   &
{11.77} &
{$-$}&
{$-$}&
{0.14}&
{$-$}&
{$-$}\\
\multicolumn{1}{c}{\textbf{Others}}&
{5}  &
{8.65} &
{(5.52, 13.47)} &
{3.32}&
{0.10}&
{(0.07, 0.12)}&
{0.02}
\\

\bottomrule[\heavyrulewidth]
\end{tabular}
}

\end{table}
 
Table \ref{mape_fitted_type} reports the overall model fitting and forecasting performances via the obtained MAPE and NRMSE values. The details of individual nodes are shown in the Appendix. Basically, the VB method delivers good performances for both the in-sample training period from 1st October 2014 to 31st March 2015 and the out-of-sample forecasting period from 1st April 2015 to 30th September 2015. According to Table \ref{mape_fitted_type}, regardless of whether the in-sample training and out-of-sample forecasting results are examined, the average values of the MAPE are less than $15\%$, and the corresponding NRMSE values are all less than $0.15$. Thus, we can claim that the NAR/VAR model with segmentation structures fits the in-sample training data well and provides good prediction capability for out-of-sample forecasting. According to Table \ref{mape_fitted_typeA1} in the Appendix, most nodes have small mean values and standard deviations with repect to MAPE and NRMSE. For nodes I5, I8 and I10, both the MAPE and NRMSE values are large. A possible reason for this phenomenon may be the violation of the stationarity assumption in the NAR/VAR model because the corresponding trend of the daily data changes frequently. In addition, in the ``others'' type, due to some extreme values for a few special days, node O1 may not have good model fitting results. When we consider the out-of-sample forecasting performance, the corresponding MAPE and NRMSE values share similar patterns to those witnessed in the in-sample training results. Nodes I5, I8, and I10 still have large values for both measurements. However, node O1 does have good MAPE and NRMSE values because there are no extreme values shown in the prediction period. Overall, the out-of-sample forecasting performance is acceptable because the average MAPE and NRMSE values are 9.78\% and 0.11, respectively.

 \hypertarget{sec6}{}\section{\sffamily \Large Conclusion}

In this paper, we focus on Bayesian analysis with NAR/VAR models, especially when there are many time series involved. First, to simplify the model complexity, the model structure assumptions in \cite{song2011large} are adopted. Then, model inference can be performed via a Bayesian structure selection approach. Here, to denote active structures, indicators are added to the NAR/VAR model. Instead of generating the posterior samples of these indicators via an MCMC algorithm, we directly obtain the proper approximation of the posterior density function via a variational Bayesian approach. Based on a factorized approximation assumption for the latent structures, a variational EM-type algorithm is used to obtain the best approximation; then, according to the approximated posterior probabilities of the indicators, the median probability criterion is used to determine the active structures in the corresponding coefficient matrices. The simulation results support the notion that the proposed variational Bayesian approach not only identifies the proper active structures in NAR/VAR models but also significantly reduces the computational cost. Finally, German gas flow network data with 51 nodes are analyzed to illustrate the performance of the proposed method. The analytical results of the proposed VB method yield the trends of nodes, which may be useful for assisting operators with performing appropriate operations.

There are several possible future research directions. The first concerns the parameter turning strategy for the proposed VB method. In our experience, the performance may be sensitive to the chosen initial parameters, especially for the prior probabilities $\pi_1$ and $\pi_2$. According to \cite{ormerod2017variational} and \cite{zhang2019novel}, we suggest setting $\pi_i$ as small probabilities, as this would tend to select a compact model. However, sometimes the model may not identify active variables with small coefficients. For example, in the simulated NG case with m = 50 and a covariance matrix of $\Sigma_{50}$, we set $\pi_1 = \pi_2 =0.5,$ instead of 0.01. Thus, choosing the proper initial setups should be an considered issue for the proposed VB method. The cross-validation approach should be examined as a possible strategy, and other possibilities should be related to proper information criteria. The second research direction is to take the group sparsity assumption into account. That is, the variables within an active segmentation still satisfy the elementwise sparsity assumption. Following Chen et al. \citeyearpar{chang2016bayesian}, we need to add new indicators for the variables within each segmentation. The idea of the variational Bayesian approach in \cite{cai2020bivas} could be modified for analyses using NAR models. 
In our real example, the segmentation structures are defined based on the types of nodes present. However, these may not be the best segmentation structures based on these prespecified gas station types. Following \cite{chu2019bayesian}, we may apply a data-driven clustering approach to identify other segmentation structures for the corresponding NAR models. Thus, it would be an interesting research direction to integrate the clustering algorithm with the variational Bayesian approach. That is, we can identify the proper segmentation structures and determine the active structures simultaneously.

\bibliographystyle{apacite}
\bibliography{NARmodel}

\begin{thebibliography}{}

\bibitem [\protect \citeauthoryear {%
Ba{\'n}bura%
, Giannone%
\BCBL {}\ \BBA {} Reichlin%
}{%
Ba{\'n}bura%
\ \protect \BOthers {.}}{%
{\protect \APACyear {2010}}%
}]{%
banbura2010}
\APACinsertmetastar {%
banbura2010}%
\begin{APACrefauthors}%
Ba{\'n}bura, M.%
, Giannone, D.%
\BCBL {}\ \BBA {} Reichlin, L.%
\end{APACrefauthors}%
\unskip\
\newblock
\APACrefYearMonthDay{2010}{}{}.
\newblock
{\BBOQ}\APACrefatitle {Large {Bayesian} vector auto regressions} {Large
  {Bayesian} vector auto regressions}.{\BBCQ}
\newblock
\APACjournalVolNumPages{Journal of Applied Econometrics}{25}{1}{71--92}.
\PrintBackRefs{\CurrentBib}

\bibitem [\protect \citeauthoryear {%
Barbieri%
\ \BBA {} Berger%
}{%
Barbieri%
\ \BBA {} Berger%
}{%
{\protect \APACyear {2004}}%
}]{%
barbieri2004optimal}
\APACinsertmetastar {%
barbieri2004optimal}%
\begin{APACrefauthors}%
Barbieri, M\BPBI M.%
\BCBT {}\ \BBA {} Berger, J\BPBI O.%
\end{APACrefauthors}%
\unskip\
\newblock
\APACrefYearMonthDay{2004}{}{}.
\newblock
{\BBOQ}\APACrefatitle {Optimal Predictive Model Selection} {Optimal predictive
  model selection}.{\BBCQ}
\newblock
\APACjournalVolNumPages{The Annals of Statistics}{32}{}{870--897}.
\PrintBackRefs{\CurrentBib}

\bibitem [\protect \citeauthoryear {%
Basu%
\ \BBA {} Michailidis%
}{%
Basu%
\ \BBA {} Michailidis%
}{%
{\protect \APACyear {2015}}%
}]{%
basu2015}
\APACinsertmetastar {%
basu2015}%
\begin{APACrefauthors}%
Basu, S.%
\BCBT {}\ \BBA {} Michailidis, G.%
\end{APACrefauthors}%
\unskip\
\newblock
\APACrefYearMonthDay{2015}{}{}.
\newblock
{\BBOQ}\APACrefatitle {Regularized estimation in sparse high-dimensional time
  series models} {Regularized estimation in sparse high-dimensional time series
  models}.{\BBCQ}
\newblock
\APACjournalVolNumPages{The Annals of Statistics}{43}{4}{1535--1567}.
\PrintBackRefs{\CurrentBib}

\bibitem [\protect \citeauthoryear {%
Bishop%
}{%
Bishop%
}{%
{\protect \APACyear {2006}}%
}]{%
bishop2006pattern}
\APACinsertmetastar {%
bishop2006pattern}%
\begin{APACrefauthors}%
Bishop, C\BPBI M.%
\end{APACrefauthors}%
\unskip\
\newblock
\APACrefYear{2006}.
\newblock
\APACrefbtitle {Pattern recognition and machine learning} {Pattern recognition
  and machine learning}.
\newblock
\APACaddressPublisher{}{Springer}.
\PrintBackRefs{\CurrentBib}

\bibitem [\protect \citeauthoryear {%
Cai%
\ \protect \BOthers {.}}{%
Cai%
\ \protect \BOthers {.}}{%
{\protect \APACyear {2020}}%
}]{%
cai2020bivas}
\APACinsertmetastar {%
cai2020bivas}%
\begin{APACrefauthors}%
Cai, M.%
, Dai, M.%
, Ming, J.%
, Peng, H.%
, Liu, J.%
\BCBL {}\ \BBA {} Yang, C.%
\end{APACrefauthors}%
\unskip\
\newblock
\APACrefYearMonthDay{2020}{}{}.
\newblock
{\BBOQ}\APACrefatitle {BIVAS: a scalable Bayesian method for bi-level variable
  selection with applications} {Bivas: a scalable bayesian method for bi-level
  variable selection with applications}.{\BBCQ}
\newblock
\APACjournalVolNumPages{Journal of Computational and Graphical
  Statistics}{29}{1}{40--52}.
\PrintBackRefs{\CurrentBib}

\bibitem [\protect \citeauthoryear {%
Carbonetto%
\ \BBA {} Stephens%
}{%
Carbonetto%
\ \BBA {} Stephens%
}{%
{\protect \APACyear {2012}}%
}]{%
carbonetto2012scalable}
\APACinsertmetastar {%
carbonetto2012scalable}%
\begin{APACrefauthors}%
Carbonetto, P.%
\BCBT {}\ \BBA {} Stephens, M.%
\end{APACrefauthors}%
\unskip\
\newblock
\APACrefYearMonthDay{2012}{}{}.
\newblock
{\BBOQ}\APACrefatitle {Scalable variational inference for {Bayesian} variable
  selection in regression, and its accuracy in genetic association studies}
  {Scalable variational inference for {Bayesian} variable selection in
  regression, and its accuracy in genetic association studies}.{\BBCQ}
\newblock
\APACjournalVolNumPages{Bayesian Analysis}{7}{}{73--108}.
\PrintBackRefs{\CurrentBib}

\bibitem [\protect \citeauthoryear {%
Chang%
, Chen%
\BCBL {}\ \BBA {} Chi%
}{%
Chang%
\ \protect \BOthers {.}}{%
{\protect \APACyear {2016}}%
}]{%
chang2016bayesian}
\APACinsertmetastar {%
chang2016bayesian}%
\begin{APACrefauthors}%
Chang, S\BHBI M.%
, Chen, R\BHBI B.%
\BCBL {}\ \BBA {} Chi, Y.%
\end{APACrefauthors}%
\unskip\
\newblock
\APACrefYearMonthDay{2016}{}{}.
\newblock
{\BBOQ}\APACrefatitle {Bayesian variable selections for probit models with
  componentwise {Gibbs} samplers} {Bayesian variable selections for probit
  models with componentwise {Gibbs} samplers}.{\BBCQ}
\newblock
\APACjournalVolNumPages{Communications in Statistics-Simulation and
  Computation}{45}{8}{2752--2766}.
\PrintBackRefs{\CurrentBib}

\bibitem [\protect \citeauthoryear {%
R\BHBI B.~Chen%
, Chu%
, Lai%
\BCBL {}\ \BBA {} Wu%
}{%
R\BHBI B.~Chen%
\ \protect \BOthers {.}}{%
{\protect \APACyear {2011}}%
}]{%
chen2011stochastic}
\APACinsertmetastar {%
chen2011stochastic}%
\begin{APACrefauthors}%
Chen, R\BHBI B.%
, Chu, C\BHBI H.%
, Lai, T\BHBI Y.%
\BCBL {}\ \BBA {} Wu, Y\BPBI N.%
\end{APACrefauthors}%
\unskip\
\newblock
\APACrefYearMonthDay{2011}{}{}.
\newblock
{\BBOQ}\APACrefatitle {Stochastic Matching Pursuit for {Bayesian} Variable
  Selection} {Stochastic matching pursuit for {Bayesian} variable
  selection}.{\BBCQ}
\newblock
\APACjournalVolNumPages{Statistics and Computing}{21}{}{247--259}.
\PrintBackRefs{\CurrentBib}

\bibitem [\protect \citeauthoryear {%
R\BHBI B.~Chen%
, Chu%
, Yuan%
\BCBL {}\ \BBA {} Wu%
}{%
R\BHBI B.~Chen%
\ \protect \BOthers {.}}{%
{\protect \APACyear {2016}}%
}]{%
chen2016bayesian}
\APACinsertmetastar {%
chen2016bayesian}%
\begin{APACrefauthors}%
Chen, R\BHBI B.%
, Chu, C\BHBI H.%
, Yuan, S.%
\BCBL {}\ \BBA {} Wu, Y\BPBI N.%
\end{APACrefauthors}%
\unskip\
\newblock
\APACrefYearMonthDay{2016}{}{}.
\newblock
{\BBOQ}\APACrefatitle {Bayesian Sparse Group Selection} {Bayesian sparse group
  selection}.{\BBCQ}
\newblock
\APACjournalVolNumPages{Journal of Computational and Graphical
  Statistics}{25}{}{665--683}.
\PrintBackRefs{\CurrentBib}

\bibitem [\protect \citeauthoryear {%
Y.~Chen%
, Koch%
, Zakiyeva%
\BCBL {}\ \BBA {} Zhu%
}{%
Y.~Chen%
\ \protect \BOthers {.}}{%
{\protect \APACyear {2020}}%
}]{%
chen2020modeling}
\APACinsertmetastar {%
chen2020modeling}%
\begin{APACrefauthors}%
Chen, Y.%
, Koch, T.%
, Zakiyeva, N.%
\BCBL {}\ \BBA {} Zhu, B.%
\end{APACrefauthors}%
\unskip\
\newblock
\APACrefYearMonthDay{2020}{}{}.
\newblock
{\BBOQ}\APACrefatitle {Modeling and forecasting the dynamics of the natural gas
  transmission network in Germany with the demand and supply balance
  constraint} {Modeling and forecasting the dynamics of the natural gas
  transmission network in germany with the demand and supply balance
  constraint}.{\BBCQ}
\newblock
\APACjournalVolNumPages{Applied Energy}{278}{}{115597}.
\PrintBackRefs{\CurrentBib}

\bibitem [\protect \citeauthoryear {%
Chu%
, Lo~Huang%
, Huang%
\BCBL {}\ \BBA {} Chen%
}{%
Chu%
\ \protect \BOthers {.}}{%
{\protect \APACyear {2019}}%
}]{%
chu2019bayesian}
\APACinsertmetastar {%
chu2019bayesian}%
\begin{APACrefauthors}%
Chu, C\BHBI H.%
, Lo~Huang, M\BHBI N.%
, Huang, S\BHBI F.%
\BCBL {}\ \BBA {} Chen, R\BHBI B.%
\end{APACrefauthors}%
\unskip\
\newblock
\APACrefYearMonthDay{2019}{}{}.
\newblock
{\BBOQ}\APACrefatitle {Bayesian Structure Selection for Vector Autoregression
  Model} {Bayesian structure selection for vector autoregression model}.{\BBCQ}
\newblock
\APACjournalVolNumPages{Journal of Forecasting}{38}{}{422--439}.
\PrintBackRefs{\CurrentBib}

\bibitem [\protect \citeauthoryear {%
Fan%
, Feng%
\BCBL {}\ \BBA {} Wu%
}{%
Fan%
\ \protect \BOthers {.}}{%
{\protect \APACyear {2009}}%
}]{%
fan2009network}
\APACinsertmetastar {%
fan2009network}%
\begin{APACrefauthors}%
Fan, J.%
, Feng, Y.%
\BCBL {}\ \BBA {} Wu, Y.%
\end{APACrefauthors}%
\unskip\
\newblock
\APACrefYearMonthDay{2009}{}{}.
\newblock
{\BBOQ}\APACrefatitle {Network exploration via the adaptive {LASSO} and {SCAD}
  penalties} {Network exploration via the adaptive {LASSO} and {SCAD}
  penalties}.{\BBCQ}
\newblock
\APACjournalVolNumPages{The Annals of Applied Statistics}{3}{2}{521}.
\PrintBackRefs{\CurrentBib}

\bibitem [\protect \citeauthoryear {%
Farcomeni%
}{%
Farcomeni%
}{%
{\protect \APACyear {2010}}%
}]{%
2010}
\APACinsertmetastar {%
2010}%
\begin{APACrefauthors}%
Farcomeni, A.%
\end{APACrefauthors}%
\unskip\
\newblock
\APACrefYearMonthDay{2010}{}{}.
\newblock
{\BBOQ}\APACrefatitle {Bayesian Constrained Variable Selection} {Bayesian
  constrained variable selection}.{\BBCQ}
\newblock
\APACjournalVolNumPages{Statistica Sinica}{20}{}{1043--1062}.
\PrintBackRefs{\CurrentBib}

\bibitem [\protect \citeauthoryear {%
George%
\ \BBA {} McCulloch%
}{%
George%
\ \BBA {} McCulloch%
}{%
{\protect \APACyear {1993}}%
}]{%
george1993variable}
\APACinsertmetastar {%
george1993variable}%
\begin{APACrefauthors}%
George, E\BPBI I.%
\BCBT {}\ \BBA {} McCulloch, R\BPBI E.%
\end{APACrefauthors}%
\unskip\
\newblock
\APACrefYearMonthDay{1993}{}{}.
\newblock
{\BBOQ}\APACrefatitle {Variable selection via {Gibbs} sampling} {Variable
  selection via {Gibbs} sampling}.{\BBCQ}
\newblock
\APACjournalVolNumPages{Journal of the American Statistical
  Association}{88}{423}{881--889}.
\PrintBackRefs{\CurrentBib}

\bibitem [\protect \citeauthoryear {%
Geweke%
}{%
Geweke%
}{%
{\protect \APACyear {1996}}%
}]{%
Geweke1996}
\APACinsertmetastar {%
Geweke1996}%
\begin{APACrefauthors}%
Geweke, J.%
\end{APACrefauthors}%
\unskip\
\newblock
\APACrefYearMonthDay{1996}{}{}.
\newblock
{\BBOQ}\APACrefatitle {Variable Selection and Model Comparison in Regression}
  {Variable selection and model comparison in regression}.{\BBCQ}
\newblock
\APACjournalVolNumPages{In Bayesian Statistics}{5}{}{609–-620}.
\PrintBackRefs{\CurrentBib}

\bibitem [\protect \citeauthoryear {%
Guo%
, Levina%
, Michailidis%
\BCBL {}\ \BBA {} Zhu%
}{%
Guo%
\ \protect \BOthers {.}}{%
{\protect \APACyear {2011}}%
}]{%
guo2011joint}
\APACinsertmetastar {%
guo2011joint}%
\begin{APACrefauthors}%
Guo, J.%
, Levina, E.%
, Michailidis, G.%
\BCBL {}\ \BBA {} Zhu, J.%
\end{APACrefauthors}%
\unskip\
\newblock
\APACrefYearMonthDay{2011}{}{}.
\newblock
{\BBOQ}\APACrefatitle {Joint estimation of multiple graphical models} {Joint
  estimation of multiple graphical models}.{\BBCQ}
\newblock
\APACjournalVolNumPages{Biometrika}{98}{1}{1--15}.
\PrintBackRefs{\CurrentBib}

\bibitem [\protect \citeauthoryear {%
Hsu%
, Hung%
\BCBL {}\ \BBA {} Chang%
}{%
Hsu%
\ \protect \BOthers {.}}{%
{\protect \APACyear {2008}}%
}]{%
hsu2008subset}
\APACinsertmetastar {%
hsu2008subset}%
\begin{APACrefauthors}%
Hsu, N\BHBI J.%
, Hung, H\BHBI L.%
\BCBL {}\ \BBA {} Chang, Y\BHBI M.%
\end{APACrefauthors}%
\unskip\
\newblock
\APACrefYearMonthDay{2008}{}{}.
\newblock
{\BBOQ}\APACrefatitle {Subset selection for vector autoregressive processes
  using {LASSO}} {Subset selection for vector autoregressive processes using
  {LASSO}}.{\BBCQ}
\newblock
\APACjournalVolNumPages{Computational Statistics \& Data
  Analysis}{52}{7}{3645--3657}.
\PrintBackRefs{\CurrentBib}

\bibitem [\protect \citeauthoryear {%
Kastner%
\ \BBA {} Huber%
}{%
Kastner%
\ \BBA {} Huber%
}{%
{\protect \APACyear {2020}}%
}]{%
kastner2020sparse}
\APACinsertmetastar {%
kastner2020sparse}%
\begin{APACrefauthors}%
Kastner, G.%
\BCBT {}\ \BBA {} Huber, F.%
\end{APACrefauthors}%
\unskip\
\newblock
\APACrefYearMonthDay{2020}{}{}.
\newblock
{\BBOQ}\APACrefatitle {Sparse Bayesian vector autoregressions in huge
  dimensions} {Sparse bayesian vector autoregressions in huge
  dimensions}.{\BBCQ}
\newblock
\APACjournalVolNumPages{Journal of Forecasting}{}{}{}.
\PrintBackRefs{\CurrentBib}

\bibitem [\protect \citeauthoryear {%
Liu%
\ \protect \BOthers {.}}{%
Liu%
\ \protect \BOthers {.}}{%
{\protect \APACyear {2012}}%
}]{%
liu2012high}
\APACinsertmetastar {%
liu2012high}%
\begin{APACrefauthors}%
Liu, H.%
, Han, F.%
, Yuan, M.%
, Lafferty, J.%
, Wasserman, L.%
\BCBL {}\ \BOthersPeriod {.}\end{APACrefauthors}%
\unskip\
\newblock
\APACrefYearMonthDay{2012}{}{}.
\newblock
{\BBOQ}\APACrefatitle {High-dimensional semiparametric {Gaussian} copula
  graphical models} {High-dimensional semiparametric {Gaussian} copula
  graphical models}.{\BBCQ}
\newblock
\APACjournalVolNumPages{The Annals of Statistics}{40}{4}{2293--2326}.
\PrintBackRefs{\CurrentBib}

\bibitem [\protect \citeauthoryear {%
L{\"u}tkepohl%
}{%
L{\"u}tkepohl%
}{%
{\protect \APACyear {2007}}%
}]{%
lutkepohl2007}
\APACinsertmetastar {%
lutkepohl2007}%
\begin{APACrefauthors}%
L{\"u}tkepohl, H.%
\end{APACrefauthors}%
\unskip\
\newblock
\APACrefYearMonthDay{2007}{}{}.
\newblock
{\BBOQ}\APACrefatitle {General-to-specific or specific-to-general modelling?
  {An} opinion on current econometric terminology} {General-to-specific or
  specific-to-general modelling? {An} opinion on current econometric
  terminology}.{\BBCQ}
\newblock
\APACjournalVolNumPages{Journal of Econometrics}{136}{1}{319--324}.
\PrintBackRefs{\CurrentBib}

\bibitem [\protect \citeauthoryear {%
Melnyk%
\ \BBA {} Banerjee%
}{%
Melnyk%
\ \BBA {} Banerjee%
}{%
{\protect \APACyear {2016}}%
}]{%
melnyk2016}
\APACinsertmetastar {%
melnyk2016}%
\begin{APACrefauthors}%
Melnyk, I.%
\BCBT {}\ \BBA {} Banerjee, A.%
\end{APACrefauthors}%
\unskip\
\newblock
\APACrefYearMonthDay{2016}{}{}.
\newblock
{\BBOQ}\APACrefatitle {Estimating structured vector autoregressive models}
  {Estimating structured vector autoregressive models}.{\BBCQ}
\newblock
\APACjournalVolNumPages{International Conference on Machine
  Learning}{}{}{830--839}.
\PrintBackRefs{\CurrentBib}

\bibitem [\protect \citeauthoryear {%
Nicholson%
, Matteson%
\BCBL {}\ \BBA {} Bien%
}{%
Nicholson%
\ \protect \BOthers {.}}{%
{\protect \APACyear {2014}}%
}]{%
nicholson2014}
\APACinsertmetastar {%
nicholson2014}%
\begin{APACrefauthors}%
Nicholson, W\BPBI B.%
, Matteson, D\BPBI S.%
\BCBL {}\ \BBA {} Bien, J.%
\end{APACrefauthors}%
\unskip\
\newblock
\APACrefYearMonthDay{2014}{}{}.
\newblock
{\BBOQ}\APACrefatitle {Structured regularization for large vector
  autoregressions} {Structured regularization for large vector
  autoregressions}.{\BBCQ}
\newblock
\APACjournalVolNumPages{Cornell University}{}{}{}.
\PrintBackRefs{\CurrentBib}

\bibitem [\protect \citeauthoryear {%
Ormerod%
, You%
, M{\"u}ller%
\BCBL {}\ \protect \BOthers {.}}{%
Ormerod%
\ \protect \BOthers {.}}{%
{\protect \APACyear {2017}}%
}]{%
ormerod2017variational}
\APACinsertmetastar {%
ormerod2017variational}%
\begin{APACrefauthors}%
Ormerod, J\BPBI T.%
, You, C.%
, M{\"u}ller, S.%
\BCBL {}\ \BOthersPeriod {.}\end{APACrefauthors}%
\unskip\
\newblock
\APACrefYearMonthDay{2017}{}{}.
\newblock
{\BBOQ}\APACrefatitle {A variational {B}ayes approach to variable selection} {A
  variational {B}ayes approach to variable selection}.{\BBCQ}
\newblock
\APACjournalVolNumPages{Electronic Journal of Statistics}{11}{2}{3549--3594}.
\PrintBackRefs{\CurrentBib}

\bibitem [\protect \citeauthoryear {%
Song%
\ \BBA {} Bickel%
}{%
Song%
\ \BBA {} Bickel%
}{%
{\protect \APACyear {2011}}%
}]{%
song2011large}
\APACinsertmetastar {%
song2011large}%
\begin{APACrefauthors}%
Song, S.%
\BCBT {}\ \BBA {} Bickel, P\BPBI J.%
\end{APACrefauthors}%
\unskip\
\newblock
\APACrefYearMonthDay{2011}{}{}.
\newblock
{\BBOQ}\APACrefatitle {Large vector auto regressions} {Large vector auto
  regressions}.{\BBCQ}
\newblock
\APACjournalVolNumPages{arXiv preprint arXiv:1106.3915}{}{}{}.
\PrintBackRefs{\CurrentBib}

\bibitem [\protect \citeauthoryear {%
Titsias%
\ \BBA {} L{\'a}zaro-Gredilla%
}{%
Titsias%
\ \BBA {} L{\'a}zaro-Gredilla%
}{%
{\protect \APACyear {2011}}%
}]{%
titsias2011spike}
\APACinsertmetastar {%
titsias2011spike}%
\begin{APACrefauthors}%
Titsias, M\BPBI K.%
\BCBT {}\ \BBA {} L{\'a}zaro-Gredilla, M.%
\end{APACrefauthors}%
\unskip\
\newblock
\APACrefYearMonthDay{2011}{}{}.
\newblock
{\BBOQ}\APACrefatitle {Spike and slab variational inference for multi-task and
  multiple kernel learning} {Spike and slab variational inference for
  multi-task and multiple kernel learning}.{\BBCQ}
\newblock
\APACjournalVolNumPages{Advances in Neural Information Processing
  Systems}{24}{}{2339-2347}.
\PrintBackRefs{\CurrentBib}

\bibitem [\protect \citeauthoryear {%
Zhang%
, Xu%
\BCBL {}\ \BBA {} Zhang%
}{%
Zhang%
\ \protect \BOthers {.}}{%
{\protect \APACyear {2019}}%
}]{%
zhang2019novel}
\APACinsertmetastar {%
zhang2019novel}%
\begin{APACrefauthors}%
Zhang, C\BHBI X.%
, Xu, S.%
\BCBL {}\ \BBA {} Zhang, J\BHBI S.%
\end{APACrefauthors}%
\unskip\
\newblock
\APACrefYearMonthDay{2019}{}{}.
\newblock
{\BBOQ}\APACrefatitle {A novel variational {B}ayesian method for variable
  selection in logistic regression models} {A novel variational {B}ayesian
  method for variable selection in logistic regression models}.{\BBCQ}
\newblock
\APACjournalVolNumPages{Computational Statistics \& Data
  Analysis}{133}{}{1--19}.
\PrintBackRefs{\CurrentBib}

\end{thebibliography}

\newpage
\appendix
\section{\sffamily \Large Appendix }
\setcounter{table}{0}
\renewcommand{\thetable}{A\arabic{table}}
\begin{table}[h]
 \caption{The MAPEs and NRMSEs for the in-sample training period from 01/10/2014 to 31/03/2015 classified by type. }
 \label{mape_fitted_typeA1}
 \centering
\scalebox{0.8}{
\begin{tabular}{cllllllllllll}
 \toprule[\heavyrulewidth]
 & \multicolumn{1}{c}{\textbf{Municipal:}} &   &  &  & &   &  & &&&& \\ 
  \toprule[\heavyrulewidth] \toprule[\heavyrulewidth]

\textbf{\begin{tabular}[c]{@{}c@{}}MAPE\\ (\%)\end{tabular}} & \multicolumn{1}{c}{\textbf{1}}  &\multicolumn{1}{c}{\textbf{2}}& \multicolumn{1}{c}{\textbf{3}}  & \multicolumn{1}{c}{\textbf{4}}  & \multicolumn{1}{c}{\textbf{5}}  &\multicolumn{1}{c}{\textbf{6}}  &\multicolumn{1}{c}{\textbf{7}} & \multicolumn{1}{c}{\textbf{8} } & \multicolumn{1}{c}{\textbf{9}}&\multicolumn{1}{c}{\textbf{10}} &\multicolumn{1}{c}{\textbf{11}} &\multicolumn{1}{c}{\textbf{12}} \\ \hline
\textbf{mean}                  &\multicolumn{1}{c}{4.53}            & \multicolumn{1}{c}{5.04}           & \multicolumn{1}{c}{8.89}         &\multicolumn{1}{c}{6.04}            & \multicolumn{1}{c}{8.00}           & \multicolumn{1}{c}{4.89}            & \multicolumn{1}{c}{6.00}        &\multicolumn{1}{c}{5.04}           & \multicolumn{1}{c}{5.58}           & \multicolumn{1}{c}{6.285}  &\multicolumn{1}{c}{5.62}               & \multicolumn{1}{c}{6.36} \\
\textbf{sd}        &  \multicolumn{1}{c}{3.59}     & \multicolumn{1}{c}{3.58}            &\multicolumn{1}{c}{7.30}            & \multicolumn{1}{c}{4.66}            &\multicolumn{1}{c}{6.58}            & \multicolumn{1}{c}{3.98}          &\multicolumn{1}{c}{4.81}           &\multicolumn{1}{c}{4.04}    & \multicolumn{1}{c}{4.80}          & \multicolumn{1}{c}{5.30}             & \multicolumn{1}{c}{ 4.39 } & \multicolumn{1}{c}{5.07}               \\
 \bottomrule[\heavyrulewidth]\rowcolor{lightgray}
\textbf{\begin{tabular}[c]{@{}c@{}}NRMSE\\ \end{tabular}}& \multicolumn{1}{c}{0.06}           &\multicolumn{1}{c}{0.06}            & \multicolumn{1}{c}{ 0.11}           & \multicolumn{1}{c}{0.08}         &\multicolumn{1}{c}{0.10 }            & \multicolumn{1}{c}{0.06 }           & \multicolumn{1}{c}{ 0.08}            & \multicolumn{1}{c}{0.07}        &\multicolumn{1}{c}{0.07}           & \multicolumn{1}{c}{0.08}  & \multicolumn{1}{c}{0.07} &\multicolumn{1}{c}{0.08} 
\\
\hline
\textbf{\begin{tabular}[c]{@{}c@{}}MAPE\\ (\%)\end{tabular}} & \multicolumn{1}{c}{\textbf{13}} &\multicolumn{1}{c}{\textbf{14}} & \multicolumn{1}{c}{\textbf{15}} &\multicolumn{1}{c}{\textbf{16}} &\multicolumn{1}{c}{\textbf{17}}& \multicolumn{1}{c}{\textbf{18}}&  \multicolumn{1}{c}{\textbf{19}} &  \multicolumn{1}{c}{\textbf{20}}  &  \multicolumn{1}{c}{\textbf{21}} &  \multicolumn{1}{c}{\textbf{22}}&  \multicolumn{1}{c}{\textbf{23}} &  \multicolumn{1}{c}{\textbf{24}} \\ \hline
\textbf{mean}         & \multicolumn{1}{c}{5.77}         & \multicolumn{1}{c}{4.38}           & \multicolumn{1}{c}{6.53}           &\multicolumn{1}{c}{5.65}          & \multicolumn{1}{c}{6.93}           & \multicolumn{1}{c}{6.93}  & \multicolumn{1}{c}{5.41}          & \multicolumn{1}{c}{7.24}             & \multicolumn{1}{c}{3.20}            & \multicolumn{1}{c}{7.00}  &\multicolumn{1}{c}{7.32}            & \multicolumn{1}{c}{14.70}      \\
\textbf{sd}       & \multicolumn{1}{c}{4.59}           & \multicolumn{1}{c}{3.46}         &\multicolumn{1}{c}{4.89}           &\multicolumn{1}{c}{4.23}           &\multicolumn{1}{c}{5.53}           & \multicolumn{1}{c}{5.70}   &\multicolumn{1}{c}{4.74}          &\multicolumn{1}{c}{5.31}            &\multicolumn{1}{c}{2.44}           &\multicolumn{1}{c}{5.31}  & \multicolumn{1}{c}{5.37}            & \multicolumn{1}{c}{23.15}            \\
 \bottomrule[\heavyrulewidth]\rowcolor{lightgray}
\textbf{\begin{tabular}[c]{@{}c@{}}NRMSE\end{tabular}}& \multicolumn{1}{c}{0.07}            & \multicolumn{1}{c}{0.06}  &\multicolumn{1}{c}{0.09}           & \multicolumn{1}{c}{0.07}         & \multicolumn{1}{c}{0.09}           & \multicolumn{1}{c}{0.09}           &\multicolumn{1}{c}{0.07}          & \multicolumn{1}{c}{0.09}           & \multicolumn{1}{c}{0.04}  & \multicolumn{1}{c}{0.09} &
 \multicolumn{1}{c}{0.10}    & \multicolumn{1}{c}{0.36}
\\
\hline
\textbf{\begin{tabular}[c]{@{}c@{}}MAPE\\ (\%)\end{tabular}}    &  \multicolumn{1}{c}{\textbf{25}} &  \multicolumn{1}{c}{\textbf{26}} &  \multicolumn{1}{c}{\textbf{27}}  &  \multicolumn{1}{c}{\textbf{28}}  &  \multicolumn{1}{c}{\textbf{29}}  & \multicolumn{1}{c}{\textbf{30}}    & \multicolumn{1}{c}{\textbf{31} }  &  \multicolumn{1}{c}{\textbf{32}}&  \multicolumn{1}{c}{\textbf{33}}  &\multicolumn{1}{c}{\textbf{34}}\\  \hline
\textbf{mean}         & \multicolumn{1}{c}{4.50}      &  \multicolumn{1}{c}{5.48}         &  \multicolumn{1}{c}{6.61}           &  \multicolumn{1}{c}{5.27}           &  \multicolumn{1}{c}{8.25}           &  \multicolumn{1}{c}{6.38}          &  \multicolumn{1}{c}{7.89}            &  \multicolumn{1}{c}{4.94}      &  \multicolumn{1}{c}{5.41} &    \multicolumn{1}{c}{5.44}        \\
\textbf{sd}          &\multicolumn{1}{c}{3.54}      &  \multicolumn{1}{c}{4.46}           & \multicolumn{1}{c}{5.42}           &  \multicolumn{1}{c}{4.17}         & \multicolumn{1}{c}{6.78}          &  \multicolumn{1}{c}{5.37}           &  \multicolumn{1}{c}{6.72}            &  \multicolumn{1}{c}{3.50}   &  \multicolumn{1}{c}{3.98}  &  \multicolumn{1}{c}{4.48}    \\
 \bottomrule[\heavyrulewidth]\rowcolor{lightgray}
\textbf{\begin{tabular}[c]{@{}c@{}}NRMSE\end{tabular}}& 
\multicolumn{1}{c}{0.06}         &  \multicolumn{1}{c}{0.07}           &  \multicolumn{1}{c}{0.09}           &  \multicolumn{1}{c}{0.07}           &  \multicolumn{1}{c}{0.10} &  \multicolumn{1}{c}{0.8}      &    \multicolumn{1}{c}{0.10}       &  \multicolumn{1}{c}{0.06}       &  \multicolumn{1}{c}{0.07} 
&  \multicolumn{1}{c}{0.07}
\\
\toprule[\heavyrulewidth]
 & \multicolumn{1}{c}{\textbf{Industry:}} &   &  &  & &   &  &  \\ 
\toprule[\heavyrulewidth]\toprule[\heavyrulewidth]

\textbf{\begin{tabular}[c]{@{}c@{}}MAPE\\ (\%)\end{tabular}}& \multicolumn{1}{c}{\textbf{1}}  & \multicolumn{1}{c}{\textbf{2}}  & \multicolumn{1}{c}{\textbf{3}}  & \multicolumn{1}{c}{\textbf{4}}  & \multicolumn{1}{c}{\textbf{5}} &  \multicolumn{1}{c}{\textbf{6}} &  \multicolumn{1}{c}{\textbf{7}} &   \multicolumn{1}{c}{\textbf{8}} & \multicolumn{1}{c}{\textbf{9}}  & \multicolumn{1}{c}{\textbf{10}}  & \multicolumn{1}{c}{\textbf{11}} & \\ \hline
\textbf{mean}        & \multicolumn{1}{c}{11.04}     & \multicolumn{1}{c}{8.30}           &  \multicolumn{1}{c}{2.80}     & \multicolumn{1}{c}{10.94}           &  \multicolumn{1}{c}{24.35 }           & \multicolumn{1}{c}{1.68}     &\multicolumn{1}{c} {4.71}     & \multicolumn{1}{c} {27.73}          &\multicolumn{1}{c} {1.96}          & \multicolumn{1}{c} {34.22}          & \multicolumn{1}{c} {0.89}       \\
\textbf{sd}          &  \multicolumn{1}{c}{10.20}    &  \multicolumn{1}{c}{9.05}    &  \multicolumn{1}{c}{2.34}         &  \multicolumn{1}{c}{11.57}    &  \multicolumn{1}{c} {33.19 }        &   \multicolumn{1}{c}{1.76}        &    \multicolumn{1}{c} {4.51}   & \multicolumn{1}{c} {39.21}          & \multicolumn{1}{c} {1.63}          & \multicolumn{1}{c} {41.65}          & \multicolumn{1}{c} {0.10}         \\ 
 \bottomrule[\heavyrulewidth]\rowcolor{lightgray}
\textbf{\begin{tabular}[c]{@{}c@{}}\end{tabular}}&
\multicolumn{1}{c}{0.12}          &         \multicolumn{1}{c}{0.10}      &  \multicolumn{1}{c}{0.04}      &  \multicolumn{1}{c}{0.14}           &  \multicolumn{1}{c}{0.22}           & \multicolumn{1}{c}{0.02}     &\multicolumn{1}{c}{0.06}         & \multicolumn{1}{c} {0.22}          &\multicolumn{1}{c} {0.03}       & \multicolumn{1}{c} {0.31}     & \multicolumn{1}{c} {0.01}   
\\
\toprule[\heavyrulewidth]
& \multicolumn{1}{c}{\textbf{Border:}} &   &  &  & &   &  & &&&& \\ 
\toprule[\heavyrulewidth] \toprule[\heavyrulewidth] \textbf{\begin{tabular}[c]{@{}c@{}}MAPE\\ (\%)\end{tabular}} & \multicolumn{1}{c}{\textbf{1}}  &  &  & &   &  & &&&&& \\ \hline
\textbf{mean}                                               & \multicolumn{1}{c}{9.36}           &  &  & &   &  & &&&&&       \\
  
\textbf{sd}                                                 &  \multicolumn{1}{c}{14.23}          &  &  & &   &   &  &&&&&     \\ 
 \bottomrule[\heavyrulewidth]\rowcolor{lightgray}
\textbf{\begin{tabular}[c]{@{}c@{}}NRMSE\end{tabular}}&\multicolumn{1}{c}{0.11}
\\
\toprule[\heavyrulewidth]
    & \multicolumn{1}{c}{\textbf{Others:}} &   &  &  & &   &  & &&&& \\ 
\toprule[\heavyrulewidth] \toprule[\heavyrulewidth]
\textbf{\begin{tabular}[c]{@{}c@{}}MAPE\\ (\%)\end{tabular}} & \multicolumn{1}{c}{\textbf{1}}  &\multicolumn{1}{c}{\textbf{2}}  &\multicolumn{1}{c}{\textbf{3}}  & \multicolumn{1}{c}{\textbf{4}} &    \multicolumn{1}{c}{\textbf{5}} & &&&& \\ \hline
\textbf{mean}        & \multicolumn{1}{c}{36.297}           &\multicolumn{1}{c}{8.507}             &  \multicolumn{1}{c} {10.711
}       &    \multicolumn{1}{c} {5.93}       &     \multicolumn{1}{c} {3.78}        &           &   &&&&        \\
\textbf{sd}       &  \multicolumn{1}{c}{20.69}    &  \multicolumn{1}{c}{10.28}        &     \multicolumn{1}{c} {10.73}      &    \multicolumn{1}{c} {4.46}         &    \multicolumn{1}{c} {3.00}         &          &      &&&&     \\ 
 \bottomrule[\heavyrulewidth]\rowcolor{lightgray}
\textbf{\begin{tabular}[c]{@{}c@{}}NRMSE\end{tabular}}&
\multicolumn{1}{c}{0.34}            &\multicolumn{1}{c}{0.07}           &  \multicolumn{1}{c} {0.13}        &    \multicolumn{1}{c} {0.08}       &     \multicolumn{1}{c} {0.05}               
\\

 \toprule[\heavyrulewidth]
\end{tabular}
}

\end{table}

\newpage
 \begin{table}[h]
 \caption{The MAPEs and RMSEs for the out-of-sample forecasting period  01/01/2015 to 30/09/2015 classified by type.}
 \label{mape_prediction_typeA2}
  \centering
 \scalebox{0.8}{
\begin{tabular}{cllllllllllll}
 \toprule[\heavyrulewidth]
 & \multicolumn{1}{c}{\textbf{Municipal:}} &   &  &  & &   &  & &&&& \\ 
  \toprule[\heavyrulewidth] \toprule[\heavyrulewidth]

\textbf{\begin{tabular}[c]{@{}c@{}}MAPE\\ (\%)\end{tabular}} & \multicolumn{1}{c}{\textbf{1}}  &\multicolumn{1}{c}{\textbf{2}}& \multicolumn{1}{c}{\textbf{3}}  & \multicolumn{1}{c}{\textbf{4}}  & \multicolumn{1}{c}{\textbf{5}}  &\multicolumn{1}{c}{\textbf{6}}  &\multicolumn{1}{c}{\textbf{6}} & \multicolumn{1}{c}{\textbf{8} } & \multicolumn{1}{c}{\textbf{9}}&\multicolumn{1}{c}{\textbf{10}} &\multicolumn{1}{c}{\textbf{11}} &\multicolumn{1}{c}{\textbf{12}} \\ \hline
\textbf{mean}         & \multicolumn{1}{c}{7.56}          &\multicolumn{1}{c}{7.42}            & \multicolumn{1}{c}{10.20}           & \multicolumn{1}{c}{7.51}         &\multicolumn{1}{c}{10.05 }            & \multicolumn{1}{c}{7.23}           & \multicolumn{1}{c}{8.13}            & \multicolumn{1}{c}{7.54}        &\multicolumn{1}{c}{7.66}           & \multicolumn{1}{c}{7.70}           & \multicolumn{1}{c}{7.45}  &\multicolumn{1}{c}{8.77}               \\
\textbf{sd}        &  \multicolumn{1}{c}{5.66}     & \multicolumn{1}{c}{ 5.69 }            &\multicolumn{1}{c}{7.63}            & \multicolumn{1}{c}{5.74}            &\multicolumn{1}{c}{7.23}            & \multicolumn{1}{c}{5.60}          &\multicolumn{1}{c}{6.34 }           &\multicolumn{1}{c}{6.10 }    & \multicolumn{1}{c}{5.85}          & \multicolumn{1}{c}{6.15 }             & \multicolumn{1}{c}{ 5.91 } & \multicolumn{1}{c}{6.78}               \\
  \bottomrule[\heavyrulewidth]\rowcolor{lightgray}
\textbf{\begin{tabular}[c]{@{}c@{}}NRMSE\end{tabular}}&
\multicolumn{1}{c}{0.10}           &\multicolumn{1}{c}{0.10}            & \multicolumn{1}{c}{ 0.13}           & \multicolumn{1}{c}{0.10}         &\multicolumn{1}{c}{0.13 }            & \multicolumn{1}{c}{0.10 }           & \multicolumn{1}{c}{0.11}            & \multicolumn{1}{c}{0.10}        &\multicolumn{1}{c}{0.10}           & \multicolumn{1}{c}{0.10}  & \multicolumn{1}{c}{0.10}
 & \multicolumn{1}{c}{0.11}
\\
\hline

\textbf{\begin{tabular}[c]{@{}c@{}}MAPE\\ (\%)\end{tabular}} & \multicolumn{1}{c}{\textbf{13}} &\multicolumn{1}{c}{\textbf{14}} & \multicolumn{1}{c}{\textbf{15}} &\multicolumn{1}{c}{\textbf{16}} &\multicolumn{1}{c}{\textbf{17}}& \multicolumn{1}{c}{\textbf{18}}&  \multicolumn{1}{c}{\textbf{19}} &  \multicolumn{1}{c}{\textbf{20}}  &  \multicolumn{1}{c}{\textbf{21}} &  \multicolumn{1}{c}{\textbf{22}}&  \multicolumn{1}{c}{\textbf{23}} &  \multicolumn{1}{c}{\textbf{24}} \\ \hline
\textbf{mean}         & \multicolumn{1}{c}{6.59}         & \multicolumn{1}{c}{6.34}           & \multicolumn{1}{c}{9.82}           &\multicolumn{1}{c}{8.93}          & \multicolumn{1}{c}{8.52}           & \multicolumn{1}{c}{8.20}  & \multicolumn{1}{c}{6.50}          & \multicolumn{1}{c}{10.28}             & \multicolumn{1}{c}{5.23}            & \multicolumn{1}{c}{10.49}  &\multicolumn{1}{c}{11.8}            & \multicolumn{1}{c}{8.16}      \\
\textbf{sd}       & \multicolumn{1}{c}{4.91}           & \multicolumn{1}{c}{5.11}         &\multicolumn{1}{c}{7.75}           &\multicolumn{1}{c}{7.37}           &\multicolumn{1}{c}{6.45}           & \multicolumn{1}{c}{6.71}   &\multicolumn{1}{c}{8.35}          &\multicolumn{1}{c}{8.75}            &\multicolumn{1}{c}{4.25}           &\multicolumn{1}{c}{8.88}  & \multicolumn{1}{c}{9.30}            & \multicolumn{1}{c}{6.31}            \\
\bottomrule[\heavyrulewidth]\rowcolor{lightgray}
\textbf{\begin{tabular}[c]{@{}c@{}}NRMSE\end{tabular}}&
\multicolumn{1}{c}{0.09  }            & \multicolumn{1}{c}{0.08}  &\multicolumn{1}{c}{0.12}           & \multicolumn{1}{c}{0.12}         & \multicolumn{1}{c}{0.11}           & \multicolumn{1}{c}{0.10}           &\multicolumn{1}{c}{0.09}          & \multicolumn{1}{c}{0.13}           & \multicolumn{1}{c}{0.07}  & \multicolumn{1}{c}{0.13} &
\multicolumn{1}{c}{0.15}    & \multicolumn{1}{c}{0.12} 
\\
\hline
\textbf{\begin{tabular}[c]{@{}c@{}}MAPE\\ (\%)\end{tabular}}    &  \multicolumn{1}{c}{\textbf{25}} &  \multicolumn{1}{c}{\textbf{261}} &  \multicolumn{1}{c}{\textbf{27}}  &  \multicolumn{1}{c}{\textbf{28}}  &  \multicolumn{1}{c}{\textbf{29}}  & \multicolumn{1}{c}{\textbf{30}}    & \multicolumn{1}{c}{\textbf{31} }  &  \multicolumn{1}{c}{\textbf{32}}&  \multicolumn{1}{c}{\textbf{33}}  &\multicolumn{1}{c}{\textbf{34}}\\  \hline
\textbf{mean}         & \multicolumn{1}{c}{4.65}      &  \multicolumn{1}{c}{7.91}         &  \multicolumn{1}{c}{8.79}           &  \multicolumn{1}{c}{7.67}           &  \multicolumn{1}{c}{8.85}           &  \multicolumn{1}{c}{5.81}          &  \multicolumn{1}{c}{8.06}            &  \multicolumn{1}{c}{6.60}      &  \multicolumn{1}{c}{5.92} &    \multicolumn{1}{c}{5.90}        \\
\textbf{sd}          &\multicolumn{1}{c}{3.79}      &  \multicolumn{1}{c}{5.69}           & \multicolumn{1}{c}{6.44}           &  \multicolumn{1}{c}{5.90}         & \multicolumn{1}{c}{7.54}          &  \multicolumn{1}{c}{4.31}           &  \multicolumn{1}{c}{6.74}            &  \multicolumn{1}{c}{4.95}   &  \multicolumn{1}{c}{4.61}  &  \multicolumn{1}{c}{4.31}    \\ 
\bottomrule[\heavyrulewidth]\rowcolor{lightgray}
\textbf{\begin{tabular}[c]{@{}c@{}}NRMSE\end{tabular}}&
\multicolumn{1}{c}{0.06}         &  \multicolumn{1}{c}{0.10}           &  \multicolumn{1}{c}{0.11}           &  \multicolumn{1}{c}{0.10}           &  \multicolumn{1}{c}{0.11} &  \multicolumn{1}{c}{0.74}      &    \multicolumn{1}{c}{0.10}       &  \multicolumn{1}{c}{0.08}       &  \multicolumn{1}{c}{0.08}     
&  \multicolumn{1}{c}{0.08} 
\\
\toprule[\heavyrulewidth]
 & \multicolumn{1}{c}{\textbf{Industry:}} &   &  &  & &   &  &  \\ 
\toprule[\heavyrulewidth]\toprule[\heavyrulewidth]

\textbf{\begin{tabular}[c]{@{}c@{}}MAPE\\ (\%)\end{tabular}}& \multicolumn{1}{c}{\textbf{1}}  & \multicolumn{1}{c}{\textbf{2}}  & \multicolumn{1}{c}{\textbf{3}}  & \multicolumn{1}{c}{\textbf{4}}  & \multicolumn{1}{c}{\textbf{5}} &  \multicolumn{1}{c}{\textbf{6}} &  \multicolumn{1}{c}{\textbf{7}} &   \multicolumn{1}{c}{\textbf{8}} & \multicolumn{1}{c}{\textbf{9}}  & \multicolumn{1}{c}{\textbf{10}}  & \multicolumn{1}{c}{\textbf{11}} & \\ \hline
\textbf{mean}        & \multicolumn{1}{c}{9.32}     & \multicolumn{1}{c}{8.52}           &  \multicolumn{1}{c}{2.51}     & \multicolumn{1}{c}{9.93 }           &  \multicolumn{1}{c}{32.62 }           & \multicolumn{1}{c}{2.98}     &\multicolumn{1}{c} {6.30}     & \multicolumn{1}{c} {54.57}          &\multicolumn{1}{c} {1.70}          & \multicolumn{1}{c} {45.35}          & \multicolumn{1}{c} {1.94}       \\
\textbf{sd}          &  \multicolumn{1}{c}{8.21}    &  \multicolumn{1}{c}{9.87}    &  \multicolumn{1}{c}{2.80}         &  \multicolumn{1}{c}{8.44}    &  \multicolumn{1}{c} {61.38}        &   \multicolumn{1}{c}{5.15}        &    \multicolumn{1}{c} {5.61}   & \multicolumn{1}{c} {121.58}          & \multicolumn{1}{c} {1.41}          & \multicolumn{1}{c} {80.72}          & \multicolumn{1}{c} {3.70}         \\ 
\bottomrule[\heavyrulewidth]
\rowcolor{lightgray}
\textbf{\begin{tabular}[c]{@{}c@{}}NRMSE\end{tabular}}&
\multicolumn{1}{c}{0.11}          &         \multicolumn{1}{c}{0.10}      &  \multicolumn{1}{c}{0.03}      &  \multicolumn{1}{c}{0.12}           &  \multicolumn{1}{c}{0.23}           & \multicolumn{1}{c}{0.05}     &\multicolumn{1}{c}{0.08}         & \multicolumn{1}{c} {0.27}          &\multicolumn{1}{c} {0.02}       & \multicolumn{1}{c} {0.30}     & \multicolumn{1}{c} {0.04} 
\\
\toprule[\heavyrulewidth]
 & \multicolumn{1}{c}{\textbf{Border:}} &   &  &  & &   &  & &&&& \\ 
\toprule[\heavyrulewidth] \toprule[\heavyrulewidth]
\textbf{\begin{tabular}[c]{@{}c@{}}MAPE\\ (\%)\end{tabular}} & \multicolumn{1}{c}{\textbf{1}}  &  &  & &   &  & &&&&& \\ \hline
\textbf{mean}                                               & \multicolumn{1}{c}{11.77}           &  &  & &   &  & &&&&&       \\
  
\textbf{sd}                                                 &  \multicolumn{1}{c}{11.40}          &  &  & &   &   &  &&&&&     \\ 
\bottomrule[\heavyrulewidth]
\rowcolor{lightgray}
\textbf{\begin{tabular}[c]{@{}c@{}}NRMSE\end{tabular}}&
\multicolumn{1}{c}{0.15}   
\\
\toprule[\heavyrulewidth]
& \multicolumn{1}{c}{\textbf{Others:}} &   &  &  & &   &  & &&&& \\ 
\toprule[\heavyrulewidth] \toprule[\heavyrulewidth]
\textbf{\begin{tabular}[c]{@{}c@{}}MAPE\\ (\%)\end{tabular}} & \multicolumn{1}{c}{\textbf{1}}  &\multicolumn{1}{c}{\textbf{2}}  &\multicolumn{1}{c}{\textbf{3}}  & \multicolumn{1}{c}{\textbf{4}} &    \multicolumn{1}{c}{\textbf{5}} & &&&& \\ \hline
\textbf{mean}        & \multicolumn{1}{c}{5.76}           &\multicolumn{1}{c}{9.73}             &  \multicolumn{1}{c} {13.47}       &    \multicolumn{1}{c} {9.03}       &     \multicolumn{1}{c} {5.53}        &           &   &&&&        \\
\textbf{sd}       &  \multicolumn{1}{c}{5.43}    &  \multicolumn{1}{c}{9.87}        &     \multicolumn{1}{c} {23.81}      &    \multicolumn{1}{c} {6.80}         &    \multicolumn{1}{c} {4.75}         &          &      &&&&     \\ 
 \bottomrule[\heavyrulewidth]
 \rowcolor{lightgray}
\textbf{\begin{tabular}[c]{@{}c@{}}NRMSE\end{tabular}}&
\multicolumn{1}{c}{0.08}            &\multicolumn{1}{c}{0.11}           &  \multicolumn{1}{c} {0.12}        &    \multicolumn{1}{c} {0.11}       &     \multicolumn{1}{c} {0.07}  
\\
\toprule[\heavyrulewidth]
\end{tabular}
}
\end{table}

\newpage
\section{\sffamily \Large Supplementary Document}
\subsection{Variational EM algorithm for th NAR model}
In this extra supplementary document, we provide more details about  the EM algorithm for the VB approach with respect to the NAR/VAR model.
\subsubsection {E-Step}
Let $\theta=\left\{ \pi_1,\pi_2,\Sigma,\sigma^2_{B}\right\}$ be the collection of NAR model parameters and $\left\{\boldsymbol{\eta},\boldsymbol{\gamma},\boldsymbol{B} \right\}$ be the set of latent variables. The joint probability of $\boldsymbol{Y}, \boldsymbol{\eta}, \boldsymbol{\gamma}, \widetilde{B}$ and as follows:
\begin{align*}
 P(\boldsymbol{Y},\boldsymbol{\eta},\boldsymbol{\gamma},\boldsymbol{\widetilde{B}}|\boldsymbol{X},\theta) =& 
      P(\boldsymbol{Y}|\boldsymbol{\eta},\boldsymbol{\gamma},\boldsymbol{\widetilde{B}},\boldsymbol{X},\theta)P(\boldsymbol{\eta},\boldsymbol{\gamma},\boldsymbol{\widetilde{B}}|\boldsymbol{X},\theta) \notag\\     
      =& MN_{T\times m}(\boldsymbol{XB},\boldsymbol{I},\Sigma)\prod_l^p\prod_i^mN(0,\sigma^2_{B}){\pi_1}^{\gamma_{\ell,i}}\left(1-\pi_1\right)^{(1-\gamma_{\ell,i})} \notag\\
      &\prod_k^gMN_{1\times |\widetilde{s}_k|}(\boldsymbol{0},\boldsymbol{I},\sigma^2_{B}{I}_{|\widetilde{s}_k|}){\pi_2}^{\eta_{\ell,i,k}}\left(1-\pi_2\right)^{(1-\eta_{\ell,i,k})}.
\end{align*}
As mentioned before, in this study, we can maximize the lower bound $L(q)$ with respect to the approximate density function $q$. In the reparameterized spike-and-slab prior, each pair of variables $\left\{B_{\ell,i,i},\gamma_{\ell,i}\right\}$ and $\left\{B_{\ell,i,\widetilde{s}_k},\eta_{\ell,i,k}\right\}$ are strongly correlated
since their product is the underlying variable that interacts with the data. Thus, a sensible approximation
must treat each pair $\left\{B_{\ell,i,i},\gamma_{\ell,i}\right\}$ and $\left\{B_{\ell,i,\widetilde{s}_k},\eta_{\ell,i,k}\right\}$ as a unit so that $\left\{B_{\ell,i,i},\gamma_{\ell,i}\right\}$ and $\left\{B_{\ell,i,\widetilde{s}_k},\eta_{\ell,i,k}\right\}$ are placed in the same factor
of the variational distribution. The simplest factorization that achieves this is:
\begin{align}
q(\boldsymbol{\eta},\boldsymbol{\gamma},\widetilde{B})=
\prod_\ell^p\prod_i^m\prod_k^g q_{\ell,i}(\widetilde{B}_{\ell,i,i},\gamma_{\ell,i})q_{\ell,i,k}(\widetilde{B}_{\ell,i,\widetilde{s}_k},\eta_{\ell,i,k}), \notag
\end{align}
and we assume that $q$ can have the following formulation: 
\begin{align}
\label{Eq_posterior}
q(\boldsymbol{\eta},\boldsymbol{\gamma},\widetilde{B})=&
\prod_\ell^p\prod_i^m\prod_k^g q_{\ell,i}(\widetilde{B}_{\ell,i,i},\gamma_{\ell,i})q_{\ell,i,k}(\widetilde{B}_{\ell,i,\widetilde{s}_k},\eta_{\ell,i,k}),\\
=&
\prod_l^p\prod_i^m\prod_k^g q_{\ell,i}(\widetilde{B}_{\ell,i,i}|\gamma_{\ell,i})q_{\ell,i}(\gamma_{\ell,i})q_{\ell,i,k}(\widetilde{B}_{\ell,i,\widetilde{s}_k}|\eta_{\ell,i,k})q_{\ell,i,k}(\eta_{\ell,i,k}), \notag
\end{align}
where $q_{\ell,i}(\widetilde{B}_{\ell,i,i}|\gamma_{\ell,i})$, $q_{\ell,i,k}(\widetilde{B}_{\ell,i,\widetilde{s}_k}|\eta_{\ell,i,k})$, $q_{\ell,i}(\gamma_{\ell,i})$, and $q_{\ell,i,k}(\eta_{\ell,i,k})$ are the approximated posterior distribution of $\widetilde{B}_{\ell,i,i}|\gamma_{\ell,i}$, $\widetilde{B}_{\ell,i,\widetilde{s}_k}|\eta_{\ell,i,k}$, $\gamma_{\ell,i}$, and $\eta_{\ell,i,k}$, respectively, which were obtained from the prior distributions. We have assumed that segments are independent and that the elements with their own lags are also independent. With this assumption, we rewrite the lower bound as
\begin{align*}
L(q)=E_{q(\boldsymbol{\gamma}),q(\boldsymbol{\eta})}\left[E_{q(\boldsymbol{\widetilde{B}}|\boldsymbol{\eta},\boldsymbol{\gamma})}\left[\log P\left(\boldsymbol{Y},\boldsymbol{\eta},\boldsymbol{\gamma},\boldsymbol{\widetilde{B}}|\boldsymbol{X},\theta\right)-\log q\left(\boldsymbol{\eta},\boldsymbol{\gamma},\boldsymbol{\widetilde{B}}\right)\right]\right].
\end{align*}
Hence, we have 
\begin{align*}
&L(q)=\sum_{\gamma}\prod_{\ell}\prod_{i}q(\gamma_{\ell,i})\sum_{\eta}\prod_{\ell}\prod_{i}\prod_{k}q(\eta_{\ell,i,k})\int_{\widetilde{B}}\left(\log P(\boldsymbol{Y},\boldsymbol{\gamma},\boldsymbol{\eta},\boldsymbol{\widetilde{B}}|\boldsymbol{X},\theta)-\log  q(\boldsymbol{\gamma},\boldsymbol{\eta},\boldsymbol{\widetilde{B}})\right)\\
&\prod_{\ell}^{p}\prod_{i}^{m}\prod_{k}^{p}q(\widetilde{B}_{\ell,i,\widetilde{s}_k}|\boldsymbol{\eta}_{\ell,i,k})\prod_{\ell}^{p}\prod_{i}^{m}q(\widetilde{B}_{\ell,i,i}|\gamma_{\ell,i})d\boldsymbol{\widetilde{B}}\\
=&\sum_{\gamma_{\ell,i}}q(\gamma_{\ell,i})\sum_{\boldsymbol{\eta}}\prod_{\ell}\prod_{i}\prod_{k}q(\eta_{\ell,i,k})\int\int q(B_{\ell,i,i}|\gamma_{\ell,i})\left[  \int \log P(\boldsymbol{Y},\boldsymbol{\gamma},\boldsymbol{\eta},\boldsymbol{\widetilde{B}}|\boldsymbol{X},\theta)\right.\\
&\left.\sum_{\gamma_{\ell',i }}\prod_{\ell'\neq \ell}\prod_{ i}q(\gamma_{\ell',i})q(\widetilde{B}_{\ell',i,i}|\gamma_{\ell',i,i})d\widetilde{B}_{\ell',i,i}+\sum_{\gamma_{\ell,i' }}\prod_{\ell}\prod_{ i'\neq i}q(\gamma_{\ell,i'})q(\widetilde{B}_{\ell,i',i'}|\gamma_{\ell,i',i'})d\widetilde{B}_{\ell,i',i'}\right]\\
&d\widetilde{B}_{\ell,i,i}\prod_{\ell}\prod_{i}\prod_{k}q(\widetilde{B}_{\ell,i,\widetilde{s}_k}|\eta_{\ell,i,k})d\widetilde{B}_{\ell,i,\widetilde{s}_k}\\
-&
\sum_{\gamma_{\ell,i}}q(\gamma_{\ell,i})\sum_{\boldsymbol{\eta}}\prod_{\ell}\prod_{i}\prod_{k}q(\eta_{\ell,i,k})\int\int q(B_{\ell,i,i}|\gamma_{\ell,i})\left[\log \left(q(B_{\ell,i,i}|\gamma_{\ell,i})q(\gamma_{l,i})\right)\right.\\
+& \left. \log \left(q(\widetilde{B}_{\ell,i,\widetilde{s}_k}|\eta_{l,i,k})q(\eta_{l,i,k})\right)\right]d\widetilde{B}_{l,i,i}d\widetilde{B}_{l,i,\widetilde{s}_k}+\text{constant}\\
=& E_{q(\boldsymbol{\eta}),q(\boldsymbol{\widetilde{B}}_{\eta}|\boldsymbol{\eta})}\left[ E_{q(\widetilde{B}_{\ell,i,i}|\gamma_{\ell,i}=1)}\left[E_{\ell'\neq \ell,i}\left( \log \left(P(\boldsymbol{Y},\widetilde{\boldsymbol{B}},\boldsymbol{\gamma}_{\ell',i},\gamma_{\ell,i}=1,\boldsymbol{\eta}|\boldsymbol{X},\theta) \right)\right)\right.\right.\\
 +&\left.\left.E_{\ell,i'\neq i}\left(\log \left(P(\boldsymbol{Y},\widetilde{\boldsymbol{B}},\boldsymbol{\gamma}_{\ell,i'},\gamma_{\ell,i}=1,\boldsymbol{\eta}|\boldsymbol{X},\theta) \right) \right)
-\log \left(q(\widetilde{B}_{\ell,i,i}|\gamma_{\ell,i}=1)\right) \right]\right]q(\gamma_{\ell,i}=1)\\
+&E_{q(\boldsymbol{\eta}),q(\boldsymbol{\widetilde{B}}_{\eta}|\boldsymbol{\eta})}\left[ E_{q(\widetilde{B}_{\ell,i,i}|\gamma_{\ell,i}=0)}\left[E_{\ell'\neq \ell,i}\left(\log \left(P(\boldsymbol{Y},\widetilde{\boldsymbol{B}},\boldsymbol{\gamma}_{\ell',i},\gamma_{\ell,i}=0,\boldsymbol{\eta}|\boldsymbol{X},\theta) \right)\right) \right.\right.\\
+&\left.\left.E_{\ell,i'\neq i}\left(\log \left(P(\boldsymbol{Y},\widetilde{\boldsymbol{B}},\boldsymbol{\gamma}_{\ell,i'},\gamma_{\ell,i}=0,\boldsymbol{\eta}|\boldsymbol{X},\theta) \right) \right)-\log \left(q(\widetilde{B}_{\ell,i,i}|\gamma_{\ell,i}=0)\right) \right]\right]q(\gamma_{\ell,i}=0)\\
+&\text{~constant,}
\end{align*}
where $E_{\ell'\neq \ell,i}(\cdot)$ denotes that the expectation is taken with respect to all the $i$th column variables except lag-$\ell$, and $E_{\ell,i' \neq i}(\cdot)$ denotes taking the expectation for all the lag-$\ell$ except the $i$th column. For a fixed $\gamma_{\ell,i}=1$, $L(q)$ can be represented as 
\begin{align*} 
 &E_{q(\boldsymbol{\eta}),q(\boldsymbol{\widetilde{B}}_{\eta}|\boldsymbol{\eta})}\left[ E_{q(\widetilde{B}_{\ell,i,i}|\gamma_{\ell,i}=1)}\left[E_{\ell'\neq l,i}\left(\log \left(P(\boldsymbol{Y},\widetilde{\boldsymbol{B}},\boldsymbol{\gamma}_{\ell',i},\gamma_{\ell,i}=1,\boldsymbol{\eta}|\boldsymbol{X},\theta) \right) \right)\right. \right.\\
&+ \left.\left.E_{\ell,i'\neq i}\left(log\left(P(\boldsymbol{Y},\widetilde{\boldsymbol{B}},\boldsymbol{\gamma}_{\ell,i'},\gamma_{\ell,i}=1,\boldsymbol{\eta}|\boldsymbol{X},\theta) \right) \right) -log\left(q(\widetilde{B}_{\ell,i,i}|\gamma_{\ell,i}=1)\right)\right]\right].
\end{align*}
This formulate can be treated as the negative KL divergence between $E_{\ell'\neq l,i}\left(\log\left(P(\boldsymbol{Y},\widetilde{\boldsymbol{B}},\boldsymbol{\gamma}_{\ell',i}\right.\right.$ $\left.\left.,\gamma_{\ell,i}=1,\boldsymbol{\eta}|\boldsymbol{X},\theta)\right)\right)+E_{\ell,i'\neq i}\left(\log \left(P(\boldsymbol{Y},\widetilde{\boldsymbol{B}},\boldsymbol{\gamma}_{\ell,i'},\gamma_{\ell,i}=1,\boldsymbol{\eta}|\boldsymbol{X},\theta) \right) \right)$ and $q(\widetilde{B}_{\ell,i,i}|\gamma_{\ell,i}=1)$. Thus, when
\begin{align}
\label{Estep_q}
\log \left(q(\widetilde{B}_{\ell,i,i}|\gamma_{\ell,i}=1)\right)&=E_{\ell'\neq \ell,i}\left(\log \left(P(\boldsymbol{Y},\widetilde{\boldsymbol{B}},\boldsymbol{\gamma}_{\ell',i},\gamma_{\ell,i}=1,\boldsymbol{\eta}|\boldsymbol{X},\theta) \right) \right)\\
&+
E_{\ell,i'\neq i}\left(\log \left(P(\boldsymbol{Y},\widetilde{\boldsymbol{B}},\boldsymbol{\gamma}_{\ell,i'},\gamma_{\ell,i}=1,\boldsymbol{\eta}|\boldsymbol{X},\theta) \right) \right),\notag
\end{align}
we can obtain the best approximation $q(\widetilde{B}_{\ell,i,i}|\gamma_{\ell,i}=1)$. Again, the cases for the given $\gamma_{\ell,i}=0$, $\eta_{\ell,i,k}=1$, and $\eta_{\ell,i,k}=0$ can be derived with the same procedure. Since both $\gamma_{\ell,i}$ and $\eta_{\ell,i,k}$ are from the independent Bernoulli distribution, with Eq. \eqref{Estep_q}, we can add some variational parameters regarding $q(\gamma_{\ell,i})$ and $q(\eta_{\ell,i,k})$ and then derive the conditional distributiona of $\widetilde{B}_{\ell,i,i}$ given $\gamma_{\ell,i}$ and $\widetilde{B}_{\ell,i,\widetilde{s}_k}$ given $\eta_{\ell,i,k}$. Last, we optimize $L(q)$ to find the variational parameters.\\
First, we derive $q(\widetilde{B}_{\ell,i,i}|\gamma_{\ell,i})$ and $q(\widetilde{B}_{\ell,i,\widetilde{s}_k}|\eta_{\ell,i,k})$ where this involves the joint probability function. To find the optimal form of Eq. (\ref{Estep_q}), we rearrange the joint probability function to retain only the terms involving $\ell$, $i$, and $\widetilde{s}_k$ as follows: 
\begin{align}
\label{Eq:logP}
\log \left(P(\boldsymbol{Y},\widetilde{\boldsymbol{B}},\boldsymbol{\gamma},\boldsymbol{\eta}|\boldsymbol{X};\theta)\right)&=\frac{-Tm}{2}\log(2\pi)-\frac{T}{2}\log(det(\Sigma))\\
&-\frac{1}{2}\left( -\frac{1}{2}tr \left(\Sigma^{-1}(\boldsymbol{Y}-\boldsymbol{XB})'(\boldsymbol{Y}-\boldsymbol{XB})\right) \right)\notag\\
&- \frac{pm}{2}\left(\log(2\pi)+\log(\sigma_{B}^2)\right)-\frac{\sum_\ell\sum_i \widetilde{B}^2_{\ell,i,i}}{2\sigma_{B}^2}\notag\\
&-\frac{pm(m-1)}{2}\left(\log(2\pi)+\log(\sigma_{B}^2)\right)\notag\\
&-\sum_\ell\sum_i \sum_{|\widetilde{s}_k|}\frac{1}{2}tr\left((\sigma_{B}^2{I}_{|\widetilde{s}_k|})^{-1}\widetilde{B}_{\ell,i,\widetilde{s}_k}'\widetilde{B}_{\ell,i,\widetilde{s}_k}\right)\notag\\
&+\sum_\ell\sum_i\gamma_{\ell,i}\log(\pi_1)+\sum_{\ell}\sum_{i}(1-\gamma_{\ell,i})\log(1-\pi_1)\notag\\
&+\sum_\ell\sum_i\sum_k\eta_{\ell,i,k}\log(\pi_2)+\sum_{\ell}\sum_{i}\sum_k(1-\eta_{\ell,i,k})\log(1-\pi_2)\notag\\
&\propto -\frac{T}{2}\log(det(\Sigma))-\frac{1}{2}tr \left(\Sigma^{-1}(\boldsymbol{Y}-\boldsymbol{XB})'(\boldsymbol{Y}-\boldsymbol{XB})\right) \notag\\
&-\sum_\ell\sum_i\frac{1}{2}\left((1-\gamma_{\ell,i})+\gamma_{\ell,i}\right)\log(\sigma_{B}^2)-\frac{\sum_\ell\sum_i \left((1-\gamma_{\ell,i}+\gamma_{\ell,i})\right)\widetilde{B}^2_{\ell,i,i}}{2\sigma_{B}^2}\notag\\
&-\sum_\ell \sum_i \sum_k \left( (1-\eta_{\ell,i,k})+\eta_{\ell,i,k}\right)\log(det(\sigma_{B}^2{I}_{|\widetilde{s}_k|}))\notag\\
&-\frac{1}{2}tr\left((\sigma_{B}^2{I}_{|\widetilde{s}_k|})^{-1}\left((1-\eta_{\ell,i,k})+\eta_{\ell,i,k}\right)\widetilde{B}_{\ell,i,\widetilde{s}_k}'\widetilde{B}_{\ell,i,\widetilde{s}_k}\right)\notag\\
&+\sum_\ell\sum_i\gamma_{\ell,i}\log(\frac{\pi_1}{1-\pi_1})+\sum_\ell\sum_i\sum_k\eta_{\ell,i,k}\log(\frac{\pi_2}{1-\pi_2})\notag\\
&+pm\left(\log(1-\pi_1)\right)+pmg\left(\log(1-\pi_2)\right)+constant \notag .
\end{align}
We can derive $\log(q(\widetilde{B}_{\ell,i,i}|\gamma_{\ell,i}))$ by taking the expectation in Eq. \eqref{Estep_q}. When $\gamma_{\ell,i}=1$, we have
\begin{align*}
\log\left(q(\widetilde{B}_{\ell,i,i}|\gamma_{\ell,i}=1)\right)&=-\frac{1}{2}\left[(\Sigma)^{-1}_{i,i}(\boldsymbol{X}_\ell^{(i)})'\boldsymbol{X}_\ell^{(i)}+\frac{1}{\sigma_{B}^2}\right]\widetilde{B}_{\ell,i,i}^2\\
&-(\Sigma)^{-1}_{i,i}\widetilde{B}'_{\ell,i,i}\left[(\boldsymbol{X}_\ell^{(i)})'(\boldsymbol{Y}^{(i)}-\boldsymbol{X}_{-l}E(B_{-\ell}^{(i)})-\boldsymbol{X}_\ell^{(-i)}E(B_{\ell}^{(-i,i)}))\right]\\
&-E\left(tr((\Sigma)_{-i,i}^{-1}\widetilde{B}'_{\ell,i,i}((\boldsymbol{X}_\ell^{(i)})'(\boldsymbol{Y}^{(-i)}-(\boldsymbol{XB})^{(-i)})))\right)+constant.
\end{align*}
We can find that this is a quadratic form of $\widetilde{B}_{\ell,i,i}$, the posterior of $q(\widetilde{B}_{\ell,i,i}|\gamma_{\ell,i}=1)$ that follows a normal distribution in the form $N(\mu_{1,\ell,i,i},\Sigma_{B_{\ell,i,i}})$, where
\begin{align*}
\Sigma_{B_{\ell,i,i}}&=\left(\boldsymbol{X}_\ell^{(i)'}\boldsymbol{X}_\ell^{(i)}(\Sigma^{-1})_{i,i}+\sigma_{B}^2\right)^{-1},\\
\mu_{1,\ell,i,i}&=
\begin{aligned}[t]
&\Sigma_{B_{\ell,i}}\left((\Sigma^{-1})_{i,i}\boldsymbol{X}_\ell^{(i)'}\left(Y^{(i)}-\sum_{j \neq l}^p\boldsymbol{X}_jE(B_j^{(i)})-\boldsymbol{X}_{\ell}^{(-i)}E(B_{\ell}^{(-i,i)})\right)\right.\\
+&E\left( \left.tr\left((\Sigma^{-1})_{-i,i}\boldsymbol{X}_\ell^{(i)'}\left(Y^{(-i)}-\boldsymbol{X}B^{(-i)}\right)\right)\right)\right).
\end{aligned}
\end{align*}
Similarly, $log(q(\widetilde{B}_{\ell,i,\widetilde{s}_k}|\eta_{\ell,i,k}))$ takes the expectation in Eq. \eqref{Estep_q}; when $\eta_{\ell,i,k}=1$, we have
\begin{align*}
\log\left(q(\widetilde{B}_{\ell,i,\widetilde{s}_k}|\eta_{\ell,i,k}=1)\right)&=-\frac{1}{2}tr\left((\Sigma)^{-1}_{\widetilde{s}_k,\widetilde{s}_k}(\boldsymbol{X}_\ell^{(i)})'\boldsymbol{X}_\ell^{(i)}\widetilde{B}'_{\ell,i,\widetilde{s}_k}\widetilde{B}_{\ell,i,\widetilde{s}_k}\right)\\
&-\frac{1}{2}tr\left((\sigma_{B}{I}_{|\widetilde{s}_k|})^{-1}\widetilde{B}'_{\ell,i,\widetilde{s}_k}\widetilde{B}_{\ell,i,\widetilde{s}_k}\right)\\
&-tr\left((\boldsymbol{X}_\ell^{(i)})'(\boldsymbol{Y}^{(\widetilde{s}_k)}-\boldsymbol{X}_{-l}E(B_{-\ell}^{(\widetilde{s}_k)})-\boldsymbol{X}_\ell^{(-i)}E(B_{\ell}^{(-i,\widetilde{s}_k)}))(\Sigma^{-1})_{\widetilde{s}_k,\widetilde{s}_k}\right)\\
&-E\left(tr\left((\boldsymbol{X}_\ell^{(i)})'(\boldsymbol{Y}^{(-\widetilde{s}_k)}-(\boldsymbol{XB})^{(-\widetilde{s}_k)})(\Sigma^{-1})_{\widetilde{s}_k,i}\right)\right)+constant,
\end{align*}
where the posterior $q(\widetilde{B}_{\ell,i,\widetilde{s}_k}|\eta_{\ell,i,\widetilde{s}_k}=1)$ follows the $1\times |\widetilde{s}_k|$ multivariate normal distribution $N_{1\times|\widetilde{s}_k|} (\boldsymbol{\mu}_{2,\ell,i,\widetilde{s}_k},\boldsymbol{I},\Sigma_{B_{\ell,i,\widetilde{s}_k}})$, in which
\begin{align*}
\Sigma_{B_{\ell,i,\widetilde{s}_k}}&=\left((\boldsymbol{X}_\ell^{(i)})'\boldsymbol{X}_\ell^{(i)}(\Sigma^{-1})_{\widetilde{s}_k,\widetilde{s}_k}+\sigma_{B}^2{I}_{|\widetilde{s}_k|}\right)^{-1},\\
\boldsymbol{\mu}_{2,\ell,i,
\widetilde{s}_k}&=\begin{aligned}[t]
&\left((\boldsymbol{X}_\ell^{(i)})'\left(Y^{(\widetilde{s}_k)}-\sum_{j \neq l}^p\boldsymbol{X}_jE(B_j^{(\widetilde{s}_k)})-\boldsymbol{X}_{\ell}^{(-i)}E(B_{\ell}^{(-i,\widetilde{s}_k)})\right)(\Sigma^{-1})_{\widetilde{s}_k,\widetilde{s}_k}\right.\\
+&E\left(\left.tr\left((\boldsymbol{X}_\ell^{(i)})'\left(Y^{(-\widetilde{s}_k)}-\boldsymbol{X}B^{(-\widetilde{s}_k)}\right)(\Sigma^{-1})_{-\widetilde{s}_k,\widetilde{s}_k}\right)\right)\right)\Sigma_{B_{\ell,i,\widetilde{s}_k}}.
\end{aligned}\\
\end{align*}
According to the same process mentioned above, when $\gamma _{\ell,i}=0$ and $\eta_{\ell,i,k}=0$, we have
\begin{align*}
q(\widetilde{B}_{\ell,i,i}|\gamma_{\ell,i}=0) &\sim N(0,\sigma^2_{B}),\\
q(\widetilde{B}_{\ell,i,\widetilde{s}_k}|\eta_{\ell,i,\widetilde{s}_k}=0) &\sim N_{1\times|\widetilde{s}_k|}(\boldsymbol{0},\boldsymbol{I},\sigma^2_{B}{I}_{|\widetilde{s}_k|}).
\end{align*}
Therefore, $\phi_{1,\ell,i} $ and $\phi_{2,\ell,i,k}$ are the probabilities of $\gamma_{\ell,i}=1$ and $\eta_{\ell,i,k}=1$, respectively, and we have
\begin{align*}
q(\boldsymbol{\eta},\boldsymbol{\gamma},\widetilde{B})&=\begin{aligned}[t]
&\prod_l^p\prod_i^m\left(\phi_{1,\ell,i}N\left(\mu_{1,\ell,i,i},\Sigma_{B_{\ell,i,i}}\right)\right)^{\gamma_{\ell,i}}\left(\left(1-\phi_{1,\ell,i}\right)N(0,\sigma^2_{B})\right)^{(1-\gamma_{\ell,i})}\\
&\prod_k^g\left(\phi_{2,\ell,i,k}N_{1 \times |\widetilde{s}_k|}(\mu_{2,\ell,i,\widetilde{s}_k},\boldsymbol{I},\Sigma_{B_{\ell,i,\widetilde{s}_k}})\right)^{\eta_{\ell,i,k}}\left(\left(1-\phi_{2,\ell,i,k}\right)N_{1 \times |\widetilde{s}_k|}(0,\boldsymbol{I},\sigma_{B}^2{I}_{|\widetilde{s}_k|})\right)^{(1-\eta_{\ell,i,k})},
\end{aligned}
\end{align*}
where $\phi_{1,\ell,i}$ and $\phi_{2,\ell,i,k}$ are
\begin{align*}
\phi_{1,\ell,i}&= Inv-logit\left\{logit(\pi_1)-\frac{1}{2}log(\sigma^2_{B})+\frac{1}{2}log(det(\Sigma_{B_{\ell,i,i}}))+\frac{(\Sigma_{B_{\ell,i,i}})^{-1}\mu_{1,\ell,i,i}^2}{2}\right\}\text{ and}\\
\phi_{2,\ell,i,k}&= Inv-logit\left \{ logit(\pi_2)-\frac{1}{2}log(det(\sigma^2_\beta{I}_{|\widetilde{s}_k|}))+\frac{1}{2}log(det(\Sigma_{B_{\ell,i,k}} ))\right.\\
&+\left.\frac{1}{2}tr\left((\Sigma_{B_{\ell,i,\widetilde{s}_k}})^{-1}\boldsymbol{\mu}'_{2,\ell,i,\widetilde{s}_k}\boldsymbol{\mu}_{2,\ell,i,\widetilde{s}_k}\right)\right\}.
\end{align*}
\subsubsection {M-Step}
During the M-step, we update the parameters $\theta=\left\{\pi_1,\pi_2,\Sigma,\sigma^2_B\right\}$ with $\frac{L(q)}{\partial\theta}=0$. Considering $\pi_1$ and $\pi_2$, by setting $\frac{L(q)}{\partial\pi_1}=0$ and $\frac{L(q)}{\partial\pi_2}=0$, we obtain
\begin{align*}
 \pi_1&=\frac{\sum_\ell^p\sum_i^m\phi_{1,\ell,i}}{pm},\\
 \pi_2&=\frac{\sum_\ell^p \sum_i^m \sum_k^g \phi_{2,\ell,i,k}}{\sum_k^g |\widetilde{s}_k|pm}.
\end{align*}
For $\Sigma$ and $\sigma^2_{\beta}$, setting $\frac{L(q)}{\partial\Sigma}=0$ and $\frac{L(q)}{\partial\sigma^2_{\beta}}=0$, we obtain
\begin{align*}
\Sigma&=\frac{E\left((\boldsymbol{Y}-\boldsymbol{XB})'(\boldsymbol{Y}-\boldsymbol{XB})\right)}{T},\\
\sigma^2_{B}&=\frac{\sum_\ell^p\sum_i^m\phi_{1,\ell,i}(\Sigma_{B,\ell,i,i}+\mu^2_{1,\ell,i})+\sum_\ell^p\sum_i^m\sum_k^g\phi_{2,\ell,i,k}tr(\Sigma_{B,l,i,\widetilde{s}_k}+\boldsymbol{\mu}_{2,\ell,i,\widetilde{s}_k}'\boldsymbol{\mu}_{2,\ell,i,\widetilde{s}_k})}{\sum_\ell^p\sum_i^m\left(\phi_{1,\ell,i}+\sum_k^{g}|\widetilde{s}_k|\phi_{2,\ell,i,k}\right)}.
\end{align*}

\end{document}